\documentclass[twocolumn]{aastex61}
\usepackage{CJK} % adding Chinese

\usepackage{amsmath}
\usepackage{amssymb}
\usepackage{epsfig}
\usepackage{apjfonts}
\usepackage{natbib}
\usepackage{hyperref}
\usepackage{epstopdf}
\usepackage{verbatim}
\usepackage{enumitem}
\usepackage{color}
\setlist[enumerate]{itemsep=-1mm}
\usepackage{soul, xcolor}
\setstcolor{blue}

\usepackage{mathtools}

\usepackage{bigints}

\usepackage{tikz}

\makeatletter
\global\let\tikz@ensure@dollar@catcode=\relax
\makeatother

\received{July 31, 2020; Revised March 21, 2021; Accepted April 14, 2021}
\submitjournal{ApJ}

\begin{document}

%%%%%%%%%%
%%% TEXT  %%%
%%%%%%%%%%

\begin{CJK*}{UTF8}{gbsn}

\title{Uniform Forward-Modeling Analysis of Ultracool Dwarfs. \\I. Methodology and Benchmarking}

\author[0000-0002-3726-4881]{Zhoujian Zhang (张周健)}
\affiliation{Institute for Astronomy, University of Hawaii at Manoa, Honolulu, HI 96822, USA}
\affiliation{Visiting Astronomer at the Infrared Telescope Facility, which is operated by the University of Hawaii under contract 80HQTR19D0030 with the National Aeronautics and Space Administration.}

\author[0000-0003-2232-7664]{Michael C. Liu}
\affiliation{Institute for Astronomy, University of Hawaii at Manoa, Honolulu, HI 96822, USA}

\author[0000-0002-5251-2943]{Mark S. Marley}
\affiliation{NASA Ames Research Center, Mail Stop 245-3, Moffett Field, CA 94035, USA}
\affiliation{The University of Arizona, Tuscon AZ 85721, USA}

\author[0000-0002-2338-476X]{Michael R. Line}
\affiliation{School of Earth \& Space Exploration, Arizona State University, Tempe AZ 85287, USA}

\author[0000-0003-0562-1511]{William M. J. Best}
\affiliation{Department of Astronomy, University of Texas at Austin, Austin, Texas 78712, USA}

\begin{abstract}
We present a forward-modeling framework using the Bayesian inference tool Starfish and cloudless Sonora-Bobcat model atmospheres to analyze low-resolution ($R\approx80-250$) near-infrared ($1.0-2.5$~$\mu$m) spectra of T dwarfs. Our approach infers effective temperatures, surface gravities, metallicities, radii, and masses, and by accounting for uncertainties from model interpolation and correlated residuals due to instrumental effects and modeling systematics, produces more realistic parameter posteriors than traditional ($\chi^2$-based) spectral-fitting analyses. We validate our framework by fitting the model atmospheres themselves and finding negligible offsets between derived and input parameters. We apply our methodology to three well-known benchmark late-T dwarfs, HD~3651B, GJ~570D, and Ross~458C, using both solar and non-solar metallicity atmospheric models. We also derive these benchmarks' physical properties using their bolometric luminosities, their primary stars' ages and metallicities, and Sonora-Bobcat evolutionary models. Assuming the evolutionary-based parameters are more robust, we find our atmospheric-based, forward-modeling analysis produces two outcomes. For HD~3615B and GJ~570D, spectral fits provide accurate $T_{\rm eff}$ and $R$ but underestimated $\log{g}$ (by $\approx1.2$~dex) and $Z$ (by $\approx0.35$~dex), likely due to the systematics from modeling the potassium line profiles. For Ross~458C, spectral fits provide accurate $\log{g}$ and $Z$ but overestimated $T_{\rm eff}$ (by $\approx120$~K) and underestimated $R$ (by $\approx1.6\times$), likely because our model atmospheres lack clouds, reduced vertical temperature gradients, or disequilibrium processes. Finally, the spectroscopically inferred masses of these benchmarks are all considerably underestimated.
\end{abstract}

\section{Introduction}
\label{sec:introduction}
The past two decades have witnessed a wealth of exoplanet discoveries. Characterizing such a large and diverse population is essential to understanding their formation pathways, evolutionary stages, and habitability. Emission \citep[e.g.,][]{2016AJ....152..203L, 2017A&A...603A..57S}, transmission \citep[e.g.,][]{2013ApJ...774...95D, 2014Natur.505...69K}, and reflection \citep[e.g.,][]{2016ApJ...828...22P, 2018ApJ...858...69M} spectra and photometry are all useful for characterizing the thermal structure, cloud properties, and composition of exoplanet atmospheres, which are crucial in shaping the appearance and evolution of exoplanets \citep[e.g.,][]{2001RvMP...73..719B, 2008ApJ...683.1104F, 2015ARA&A..53..279M}. However, robust spectroscopic analyses of exoplanets are often hampered by low signal-to-noise (S/N) data due to the planets' faintness and small planet-to-star area ratios, as well as by data with fairly narrow and/or non-contiguous wavelength coverage. Fortunately, non-irradiated, self-luminous brown dwarfs \citep[$\approx 13-70$~M$_{\rm Jup}$;][]{2011ApJ...727...57S, 2017ApJS..231...15D} overlap in temperature and surface gravity to imaged and transiting planets and are readily viable for high-quality emission spectroscopy. Investigating the properties of brown dwarf atmospheres can therefore provide insight into the similar atmospheric chemical and physical processes operating in exoplanets.

Studying atmospheres of brown dwarfs has commonly relied on forward modeling (e.g., \citealt{2000ApJ...541..374S, 2001ApJ...556..373G, 2006ApJ...639.1095B, 2008ApJ...678.1372C, 2011ApJ...743...50C, 2009ApJ...702..154S, 2011ApJ...740..108L, 2014A&A...562A.127B, 2018A&A...618A..63B, 2016MNRAS.455.1341M}), i.e., comparing the observed spectra with a grid of theoretical model spectra, a.k.a. ``grid models'', which are pre-computed with a few free parameters (e.g., effective temperature, surface gravity, and metallicity) and self-consistent assumptions (including radiative-convective equilibrium, [dis]equilibrium chemistry, and simplified cloud properties). Such models are essential for characterizing the properties of brown dwarfs and exoplanets, developing our understanding of ultracool atmospheres, and planning ground- and space-based observations. 

Traditional forward-modeling analysis compares the observed spectrum with the model spectrum at each grid point and then typically derives best-fit parameters with a least-squares approach \citep[e.g.,][]{2008ApJ...678.1372C, 2010ApJS..186...63R, 2020ApJ...891..171Z}. This process usually assumes that flux measurements follow Gaussian statistics and that the (data$-$model) residuals between different wavelengths are independent. These assumptions lead to a diagonal covariance matrix when evaluating free parameters, with inverse weights (usually squared flux uncertainties) placed along the diagonal axis. In addition, since model grids are coarsely created, interpolation of models is required to evaluate parameters in between the grid points. 

However, this traditional method has two limitations. First, model interpolation is an important source of parameter uncertainties, but this is usually ignored. A typical approach is to use linear interpolation to determine the model spectrum for physical parameters in between model grid points, which returns a single spectrum with no uncertainty. This can result in artificially small uncertainties for inferred physical parameters and may bias parameter posteriors toward the model grid points \citep[e.g.,][]{2014ApJ...794..125C, 2015ApJ...812..128C, 2020ApJ...891..171Z}. Second, residuals from the data-model comparison are correlated between different wavelengths, given that (1) spectrographs are designed to over-sample the instrumental line spread function, so fluxes recorded on adjacent pixels are correlated, and (2) the number of free parameters for grid models is usually too small to comprehensively describe real data (as many other free parameters that describe, e.g., cloud sedimentation efficiency and vertical mixing of chemical species, are often pre-assumed within the grid models). Failure to account for interpolation uncertainties and correlated residuals can cause the resulting derived parameters to have underestimated errors, which in practice are often masked simply by adopting a fraction of the model grid spacing for the final parameter uncertainties \citep[e.g.,][]{2007ApJ...667..537L, 2008ApJ...678.1372C, 2009ApJ...702..154S, 2020ApJ...891..171Z}.

\citeauthor{2015ApJ...812..128C} (\citeyear{2015ApJ...812..128C}; C15) have developed a new Bayesian framework for forward-modeling analysis of stellar spectroscopy, dubbed Starfish,\footnote{\url{https://github.com/iancze/Starfish}.} which can mitigate limitations of traditional methods. To synthesize spectra in between grid points, Starfish is equipped with a ``spectral emulator'', which generates a probability distribution of spectra at the desired grid parameters and propagates the associated interpolation uncertainties into the resulting inferred parameters. Also, when evaluating physical parameters, Starfish constructs a covariance matrix with off-diagonal components to account for correlated residuals caused by the instrumental effect and model imperfections. Therefore, Starfish derives larger but more realistic parameter uncertainties than the traditional spectral-fitting technique.

In this work, we extend the Starfish methodology to the brown dwarf regime for the first time and study spectra of three benchmark objects, HD~3651B \citep[T7.5;][]{2006MNRAS.373L..31M}, GJ~570D \citep[T7.5;][]{2000ApJ...531L..57B}, and Ross~458C \citep[T8;][]{2010MNRAS.405.1140G}, using the cloudless Sonora-Bobcat model grids (\citealt{2017AAS...23031507M}; Marley et al., submitted) with both solar and non-solar metallicities. These 3 objects are common proper-motion companions to stars at wide orbital separations. Hence, their distances, metallicities, and ages can be independently obtained from their primary hosts. Being late-T dwarfs, these objects are believed to be free of optically thick clouds in the atmospheric extent probed by near-infrared wavelengths \citep[although optically thin clouds might exist; e.g.,][]{2012ApJ...756..172M}. Therefore, they constitute an optimal sample for testing cloud-free models, which serve as the starting point for the more complex cloudy models representative of directly imaged exoplanets \citep[e.g.,][]{2008Sci...322.1348M, 2017AJ....153...18B, 2017A&A...605L...9C}.

We start with a brief review of the three benchmarks and observations (Section~\ref{sec:companions}) and the cloudless Sonora-Bobcat models (Section~\ref{sec:model}). We then construct and validate our forward-modeling framework using the Sonora-Bobcat atmospheric models, infer physical properties of our sample, and assess the systematics of the model atmospheres (Section~\ref{sec:atm}). We also derive the objects' physical properties using their ages and metallicities and the Sonora-Bobcat evolutionary models (Section~\ref{sec:evo}), and we compare these properties with atmospheric-based spectral fits to test the cloudless model predictions (Section~\ref{sec:benchmarking}). Finally, we provide a summary and an outlook of future work (Section~\ref{sec:summary}).

\section{Benchmark Companions and Observations}
\label{sec:companions}
\subsection{HD~3651B}
\label{subsec:hd3651b}
HD~3651B (T7.5) is the first directly-imaged brown dwarf companion to an exoplanet host, found independently by \cite{2006MNRAS.373L..31M} and \cite{2007ApJ...654..570L}, with spectroscopic characterization also by \cite{2007ApJ...660.1507L} and \cite{2007ApJ...658..617B}. The companion is at $43''$ ($479$~au) projected separation to HD~3651A (K0V), which has a {\it Gaia}~DR2 distance of $11.134 \pm 0.007$~pc \citep[][]{2018AJ....156...58B} and hosts a Saturn-mass planet ($M\sin{i} = 0.2$~M$_{\rm Jup}$) on a close ($a = 0.3$~au), eccentric orbit \citep[$e=0.63$;][]{2003ApJ...590.1081F}. 

Table~\ref{tab:HD3651} summarizes the metallicity and age of HD~3651A based on the literature and our own calculations \citep[also see][]{2007ApJ...660.1507L}. The metallicity of HD~3651A is slightly super-solar based on high-resolution spectroscopy ($R \approx 5 \times 10^{4}$), and here we adopt $Z = 0.1-0.2$~dex. The age of HD~3651A has been estimated using various techniques. Isochrone fitting from the literature suggests an age of $3-13$~Gyr based on a variety of stellar parameters and model sets. Gyrochronology suggests an age of $6.1-7.3$~Gyr, with the lower limit computed by \cite{2007ApJ...669.1167B} using this object's rotation period of $44$~days \citep[][]{1996ApJ...457L..99B} and the upper limit computed by us using the same rotation period but the \cite{2008ApJ...687.1264M} gyrochrones. The stellar-activity age of HD~3651A has been estimated based on the chromospheric activity index $R^{\prime}_{\rm HK}$ or the X-ray emission index $\log{R_{X}} \equiv \log{L_{X}/L_{\rm bol}}$ using various activity-age relations \citep[e.g.,][]{1993PhDT.........3D, 1991ApJ...375..722S, 1998PASP..110.1259G, 2008ApJ...687.1264M}. We re-compute ages from these measured activity indices using the \cite{2008ApJ...687.1264M} $R^{\prime}_{\rm HK}$-age and $R_{X}$-age relations, leading to an age of $2.7-8.3$~Gyr. We equally weight the age estimates from the stellar-isochrone fitting, gyrochronology, and stellar activity and generate $10^{5}$ draws from each age range assuming a uniform distribution. We then combine these draws, compute their 16th and 84th percentiles, and adopt an age of $4.5-8.3$~Gyr for the HD~3651 system.

\subsection{GJ~570D}
\label{subsec:gj570d}
GJ~570D (T7.5) was found by \cite{2000ApJ...531L..57B}  in a hierarchical quadruple system with GJ~570A (K4V) and a M1.5V$+$M3V close binary GJ~570BC \citep[][]{1988A&A...200..135D}. This system is located at $5.884 \pm 0.003$~pc \citep[][]{2018AJ....156...58B}. The projected separations of A--BC, B--C, and A--D components are $24.7''$ (145~au), $0.15''$ (0.9~au), and $258.3''$ ($1520$~au), respectively. 

Table~\ref{tab:GJ570} summarizes the metallicity and age of GJ~570A \citep[also see][]{2007ApJ...660.1507L}. We adopt a metallicity of $Z = -0.05$ to $0.05$~dex based on high-resolution spectroscopy ($R \approx 5 \times 10^{4}$). The age of GJ~570A has been estimated as $1-8$~Gyr by the stellar-isochrone fitting and $3.7-5.3$~Gyr by gyrochronology. Similar to HD~3651A, we re-compute the object's stellar-activity age using the measured activity indices and the \cite{2008ApJ...687.1264M} $R^{\prime}_{\rm HK}$-age and $R_{X}$-age relations, leading to an age of $0.6-2.5$~Gyr. We combine these estimates as for HD~3651A and adopt an age of $1.4-5.2$~Gyr for the GJ~570 system.

%%%%%%%%%%%%%%%%%%%%%%%
%------------- Synthetic Spectra --------------------
%%%%%%%%%%%%%%%%%%%%%%%
\begin{figure*}[t]
\begin{center}
\includegraphics[height=5in]{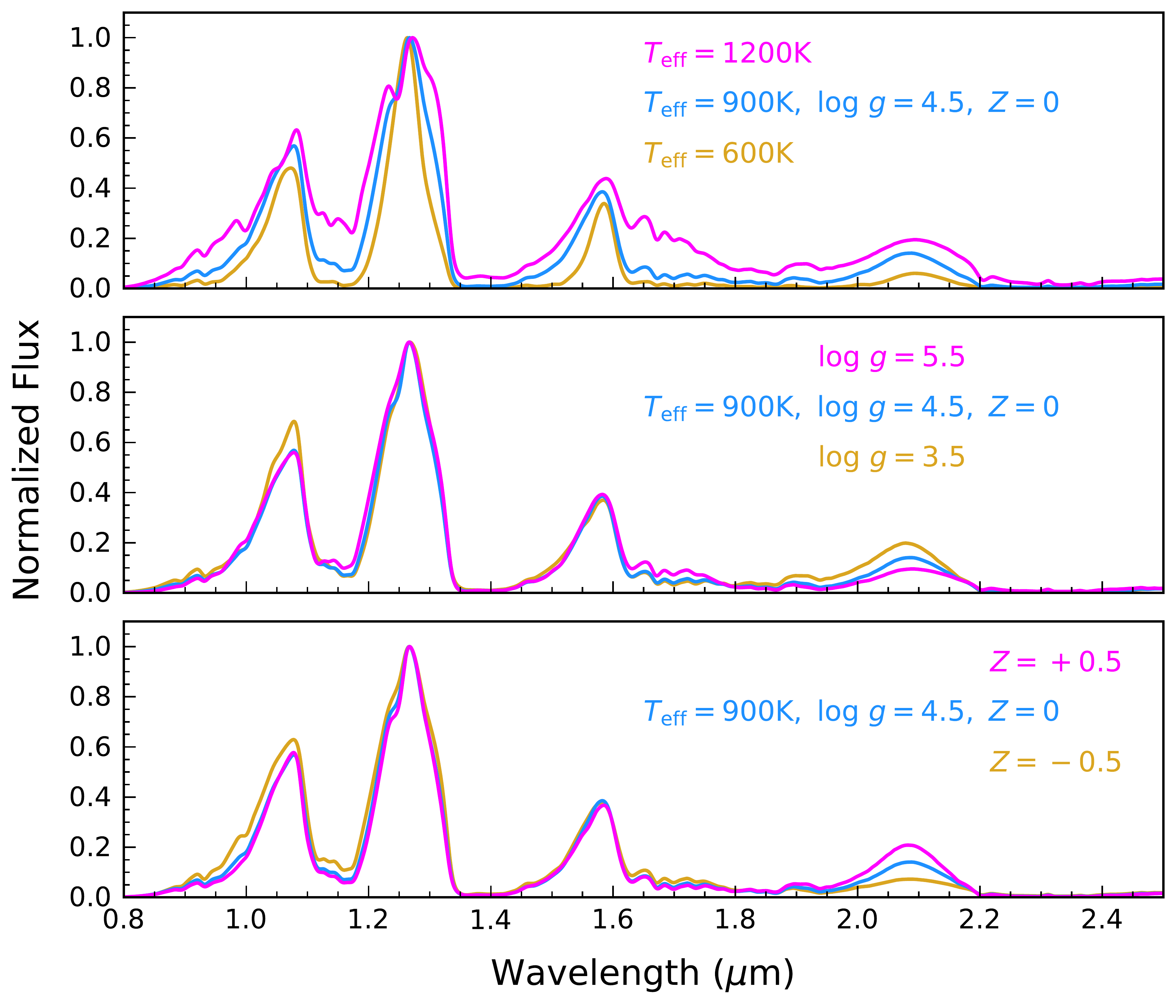}
\caption{Comparisons of cloudless Sonora-Bobcat model spectra with varying effective temperature $T_{\rm eff}$, surface gravity $\log{g}$, and metallicity $Z$. The upper panel compares spectra with $T_{\rm} = 1200$~K (magenta), $900$~K (blue), and $600$~K (orange), and fixed $\log{g}=4.5$ and $Z=0$. The middle panel compares spectra with $\log{g} = 5.5$ (magenta), $4.5$ (blue), and $3.5$ (orange), and fixed $T_{\rm eff} = 900$~K and $Z=0$. The lower panel compares spectra with $Z = +0.5$ (magenta), $0$ (blue), and $-0.5$ (orange), and fixed $T_{\rm eff} = 900$~K and $\log{g} = 4.5$. All model spectra have been smoothed to match the wavelength-dependent resolution of the $0.5''$ slit of SpeX prism ($R \approx 80-250$) and normalized by their peak $J$-band fluxes.  }
\label{fig:mod_spec}
\end{center}
\end{figure*}
%------------figure end-----------------

\subsection{Ross~458C}
\label{subsec:ross458c}
Ross~458C (T8) was found by \cite{2010MNRAS.405.1140G} and \cite{2010A&A...515A..92S} as a $102''$ ($1175$~au) companion to the M0.5Ve$+$M7Ve binary Ross~458AB \citep[separated by $6$~au;][]{2004A&A...425..997B}, located at $11.509 \pm 0.020$~pc\footnote{We note the {\it Gaia} DR2 astrometry of Ross~458AB is consistent with {\it Gaia}'s single-star model and unlikely affected by the orbital motion of the close binary, given its renormalized unit weight error (RUWE $=1.09$) is smaller than 1.4 \citep[the quality cut suggested by][]{Lindegren2018} and its astrometric excess noise is 0~mas. } \citep[][]{2018AJ....156...58B}. \cite{2001MNRAS.328...45M} noted the space motion of Ross~458AB is close to the Hyades cluster, but the object's Hyades membership was later rejected by \cite{2010ApJ...725.1405B}, who found a significant difference between the updated space motions of Ross~458AB and the Hyades. Using BANYAN~$\Sigma$ \citep[version 1.0;][]{2018ApJ...856...23G} and the {\it Gaia} DR2 proper motion, parallax, and radial velocity, we find Ross~458AB is a field dwarf with a $99.9\%$ probability.

The metallicity and age of the Ross~458 system were summarized by \cite{2010ApJ...725.1405B}, which we briefly describe here. Based on $V$- and $K$-band photometry, \cite{2010ApJ...725.1405B} derived a photometric metallicity of $0.2-0.3$~dex for Ross~458A, which is consistent with the more recent measurement of $0.25 \pm 0.08$~dex by \cite{2014ApJ...791...54G} using the moderate-resolution ($R \approx 2000$) near-infrared spectroscopy. We thus adopt a super-solar metallicity of $Z = 0.17-0.33$~dex. The fast rotation of Ross~458A \citep[$1.5-3$~days;][]{2003AandA...397..147P, 2007AcA....57..149K} suggests an age of $<1$~Gyr \citep[][]{2004A&A...425..997B}, and its strong stellar activity, indicated by H$\alpha$ emission and a very bright X-ray luminosity, suggests an age of $<0.8$~Gyr \citep[][]{2003ApJ...586..464B}. The absence of youth or low-gravity indicators (e.g., Li I absorption or weak K I doublets) in optical spectroscopy suggests an age of $>0.15$~Gyr. We therefore follow \cite{2010ApJ...725.1405B} and adopt an age of $0.15-0.8$~Gyr for the Ross~458 system.  

\subsection{Observations}
\label{subsec:obs}
The near-infrared spectra of the three benchmarks were all observed using the SpeX spectrograph \citep{2003PASP..115..362R} mounted on the NASA Infrared Telescope Facility (IRTF). We obtain spectra of HD~3651B and GJ~570D from the SpeX Prism Library \citep{2014ASInC..11....7B} as observed by \cite{2007ApJ...658..617B} and \cite{2004AJ....127.2856B}, respectively. We observed Ross~458C using IRTF/SpeX in prism mode on 2015 July 07 (UT). We took 30 exposures with 120~seconds each for the target in a ABBA pattern and contemporaneously observed a nearby A0V standard star HD~116960 for telluric correction. We reduced the data using version 4.1 of the Spextool software package \citep[][]{2004PASP..116..362C}. 

Spectra of all these objects were observed with the SpeX $0.5'' \times 15''$ slit and thus have same spectral resolution ($R \approx 80-250$). We flux-calibrate these spectra using the objects' $H_{\rm MKO}$ magnitudes and the WFCAM $H$-band filter response \citep{2006MNRAS.367..454H} and the zero-point flux \citep{2007MNRAS.379.1599L}. Astrometry and photometry of the three benchmarks are listed in Table~\ref{tab:astrom_phot}.

%%%%%%%%%%%%%%%%%%%%%%%
%------------- Spectral Derivative --------------------
%%%%%%%%%%%%%%%%%%%%%%%
\begin{figure*}[t]
\begin{center}
\includegraphics[height=4.5in]{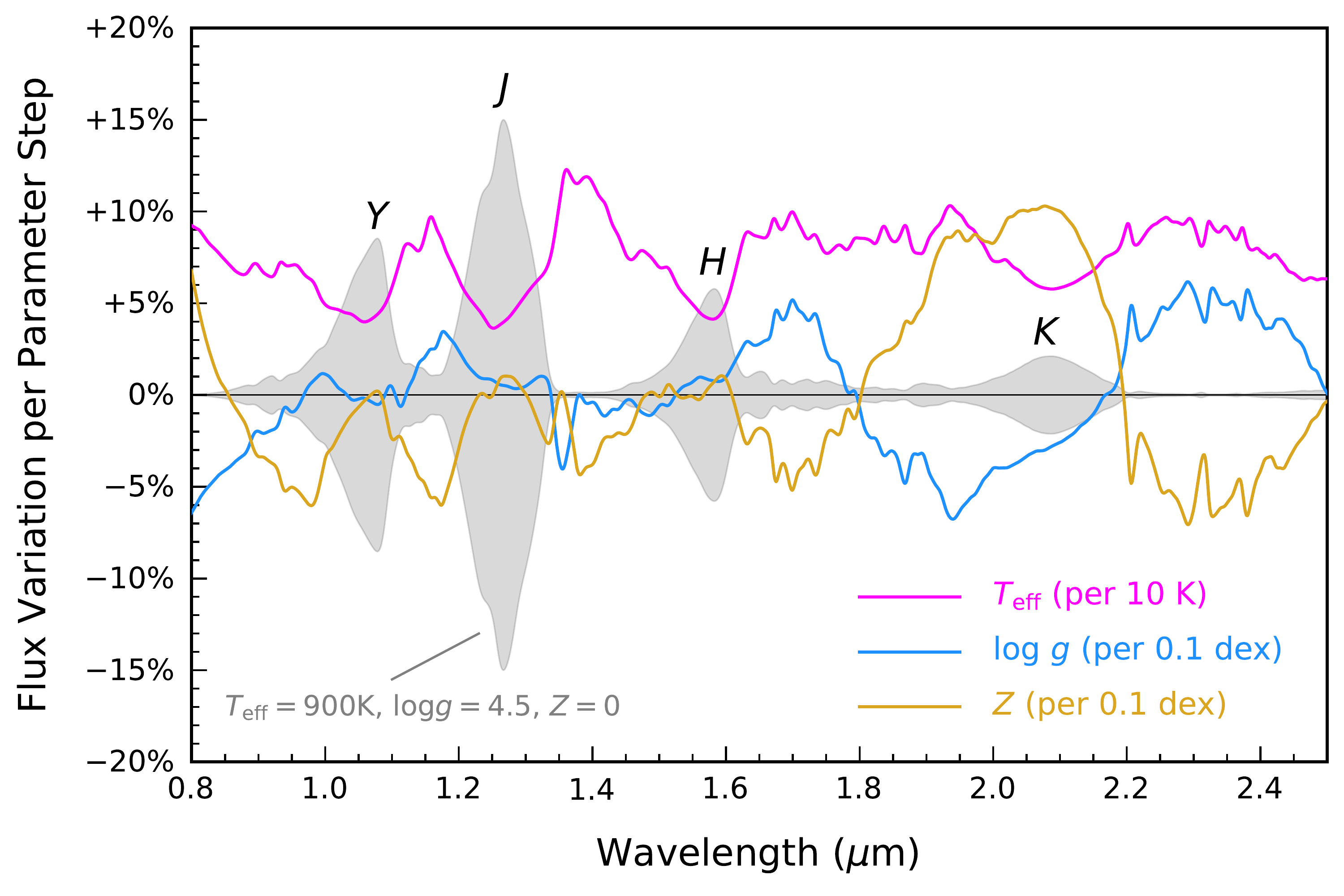}
\caption{The average flux increase of model spectra relative to a reference spectrum at $\{T_{\rm eff} = 900\, {\rm K},\ \log{g} = 4.5,\ Z = 0\}$, when $T_{\rm eff}$ is increased by $10$~K (magenta), $\log{g}$ is increased by $0.1$ (blue), and $Z$ is increased by $0.1$ (orange), respectively. We compute the spectral derivatives of $T_{\rm eff}$, $\log{g}$, and $Z$ by using Sonora-Bobcat model spectra with varying $T_{\rm eff}$ of $800, 850, 900, 950, 1000$~K, $\log{g}$ of $3.5, 4.0, 4.5, 5.0, 5.5$~dex, and $Z$ of $-0.5, 0, +0.5$~dex, respectively, with the other two grid parameters fixed at those of the reference spectrum. As a guide to the location of prominent spectral features, the upper and lower boundary of the background grey shadow correspond to the reference spectrum, scaled with a positive and an equally negative factor.}
\label{fig:diff_spec}
\end{center}
\end{figure*}
%------------figure end-----------------

\section{The Cloudless Sonora-Bobcat Models}
\label{sec:model}

\subsection{Model Description}
\label{subsec:description}
Throughout this work, we use the cloudless Sonora-Bobcat models (\citealt{2017AAS...23031507M}; Marley et al., submitted). The models are 1D radiative-convective equilibrium atmosphere models that iteratively solve for a self-consistent atmospheric structure given a specified gravity and effective temperature. The model atmosphere code was originally developed for modeling Titan's atmosphere \citep[][]{1989Icar...80...23M} and was subsequently modified by \cite{1999Icar..138..268M} to study the atmosphere of Uranus. It has since been applied to the study of brown dwarfs \citep[e.g.,][]{1996Sci...272.1919M, 1997ApJ...491..856B, 2008ApJ...678.1372C, 2008ApJ...689.1327S}, and solar and extrasolar giant planets \citep[e.g.,][]{1999ApJ...513..879M, 2002ApJ...568..335M, 2012ApJ...754..135M, 2005ApJ...627L..69F, 2008ApJ...678.1419F, 2013ApJ...775...80F}. The modeling framework is described in \cite{2015ARA&A..53..279M}. Chemical equilibrium is computed by interpolation of pre-computed tables using a modified version of the NASA CEA Gibbs minimization code \citep[see][]{gordon94}, computed accounting for rainout based upon prior thermochemical models of substellar atmospheres \citep[e.g.,][]{1999ApJ...519..793L, 2010ApJ...716.1060V, 2013ApJ...763...25M}. After a thermal structure solution in chemical and radiative-convective equilibrium has been found, the profile can be post-processed to yield model spectra at arbitrary spectral resolution. The evolutionary models are discussed in detail by \cite{2008ApJ...689.1327S}, with several updates described in Marley et al. (submitted). These models assume the objects have fully-convective adiabatic interiors and start their evolution with high entropy (``hot start''). The resulting models span a parameter space of $200-2400$~K in effective temperature ($T_{\rm eff}$), $3.25-5.5$~dex in surface gravity ($\log{g}$), and $-0.5$, $0$, and $0.5$~dex in metallicity ($Z$).

\subsection{Synthetic Spectra}
\label{subsec:syn_spec}
In order to understand the effects of physical parameters on the resulting spectra, we compare three sets of synthetic spectra with varying $T_{\rm eff}$ from $600$~K to $1200$~K, $\log{g}$ from $3.5$~dex to $5.5$~dex, and $Z$ from $-0.5$~dex to $+0.5$~dex (Figure~\ref{fig:mod_spec}). We smooth each model spectrum using a Gaussian kernel with varying width to match the wavelength-dependent resolution of the $0.5''$ slit of the SpeX prism \citep[][]{2003PASP..115..362R}. We then normalize all model spectra using their peak $J$-band flux to highlight the variations in spectral shape.

To compare the spectral variations among different grid parameters, we differentiate the synthetic spectra (which have been already smoothed and normalized by their peak $J$-band fluxes) with respect to the physical parameters, by computing the average flux increase at each wavelength relative to a spectrum at $\{T_{\rm eff} = 900\, {\rm K},\ \log{g}=4.5,\ Z=0\}$, when $T_{\rm eff}$ is increased by $10$~K, $\log{g}$ by $0.1$~dex, and $Z$ by $0.1$~dex, respectively (Figure~\ref{fig:diff_spec}). The computed flux increase per parameter step, i.e., the derivative of synthetic spectra, illustrates the wavelengths where a given parameter significantly affects the spectral appearance. Comparing such spectral derivatives among different parameters allows us to understand which parameter has the most effect at a given wavelength.

As shown in the upper panel of Figure~\ref{fig:mod_spec}, decreasing temperatures causes the absorption bands of H$_{2}$O and CH$_{4}$ to become broader and deeper, due to the increasing column densities of absorbers \citep[e.g.,][]{2014ApJ...797...41Z}. This can also been seen from Figure~\ref{fig:diff_spec}, where the spectral derivative in terms of $T_{\rm eff}$ is positive across almost the entire near-infrared wavelength range. In general, the cooling of atmospheres leads to a series of flux peaks at $Y$, $J$, $H$, and $K$ bands, which symbolize the brown dwarf evolution from earlier to later spectral types.

As shown in the middle panel of Figure~\ref{fig:mod_spec}, changing surface gravity causes strong spectral variations at $K$ band and the $Y$-band peak relative to the $J$-band peak. The $K$-band flux is suppressed at high gravity, due to enhanced absorption from collisions between H$_{2}$ molecules and the H and He atoms \citep[e.g.,][]{1991ApJS...76..759L}. In $Y$ band, the spectral variation due to gravity is not significant for $\log{g} \gtrsim 4.5$, in agreement with the spectra of benchmark substellar companions whose surface gravities are independently derived from ages of their primary stars \citep[e.g., HD~3651B and GJ~570D;][]{2007ApJ...660.1507L, 2007ApJ...667..537L}. Figure~\ref{fig:mod_spec} shows that low-gravity objects with $\log{g} \approx 3.5$ have synthetic spectra with bluer $Y-J$ colors than higher-gravity objects ($\log{g} \gtrsim 4.5$). However, based on the cloudless Sonora-Bobcat evolutionary models, $\log{g} = 3.5$ corresponds to objects with either (1) an age of $\lesssim 0.5$~Myr for a mass of $5-70$~M$_{\rm Jup}$, which are much younger than current search efforts, or (2) a mass of $1-2$~M$_{\rm Jup}$ for a field age of $1-10$~Gyr, which are too faint to be directly imaged with existing instruments. More discoveries of young, low-mass ultracool dwarfs are thereby needed to validate such blue $Y-J$ colors at very low surface gravities as predicted by the cloudless Sonora-Bobcat models.

As shown in the lower panel of Figure~\ref{fig:mod_spec}, lower metallicity has qualitatively similar effects as higher surface gravity (except for the blue wing of $Y$ band). A metal-poor atmosphere has relatively more abundant H and He atoms, and is less opaque, with the photosphere residing at lower-altitude, higher-pressure layers. These lead to enhanced collision-induced H$_{2}$ absorption (CIA) relative to other molecular absorption, thereby suppressing the flux in $K$ band. While such $K$-band flux suppression also occurs for objects with high surface gravities, the amount of the suppression has a different dependence on $\log{g}$ and $Z$ \citep[e.g.,][]{2006ApJ...640.1063B, 2007ApJ...667..537L, 2007ApJ...660.1507L, 2009MNRAS.395.1237B}. Quantitatively, our Figure~\ref{fig:diff_spec} shows that the $\log{g}$ should be increased/decreased by a factor of $\approx 3$ more than the amount that $Z$ should be decreased/increased, in order to achieve comparable $K$-band flux changes in the cloudless Sonora-Bobcat models. In addition, higher metallicity results in the suppressed flux at the blue wing of $Y$ band and at $\approx 1.24-1.25$~$\mu$m, primarily because the enhancement of Na and K abundances leads to larger opacity in the optical resonance lines of these species, whose pressure-broadened red wings extend to near-infrared \citep[e.g.,][]{1999ApJ...520L.119T, 2003ApJ...583..985B, 2007A&A...474L..21A}.

\section{Atmospheric Model Analysis}
\label{sec:atm}
Here we use the cloudless Sonora-Bobcat models and the Starfish methodology (C15; as updated by \citealt{2017ApJ...836..200G}) to study atmospheric properties of HD~3651B, GJ~570D, and Ross~458C. We first construct and validate our forward-modeling framework using Sonora-Bobcat models (Section~\ref{subsec:starfish}). Then we present results of our atmospheric model analysis and compare them to those derived from the traditional forward-modeling approach and previous spectroscopic analyses (Section~\ref{subsec:results}). Finally, we use the Starfish hyper-parameters to quantify the systematic difference between data and models (Section~\ref{subsec:assess_model_systematics}).

\subsection{Forward-Modeling Analysis with Starfish}
\label{subsec:starfish}

%%%%%%%%%%%%%%%%%%%%%%%%%%%%
%------------- Model Reconstruction PCA --------------------
%%%%%%%%%%%%%%%%%%%%%%%%%%%%
\begin{figure*}[t]
\begin{center}
\includegraphics[height=4.3in]{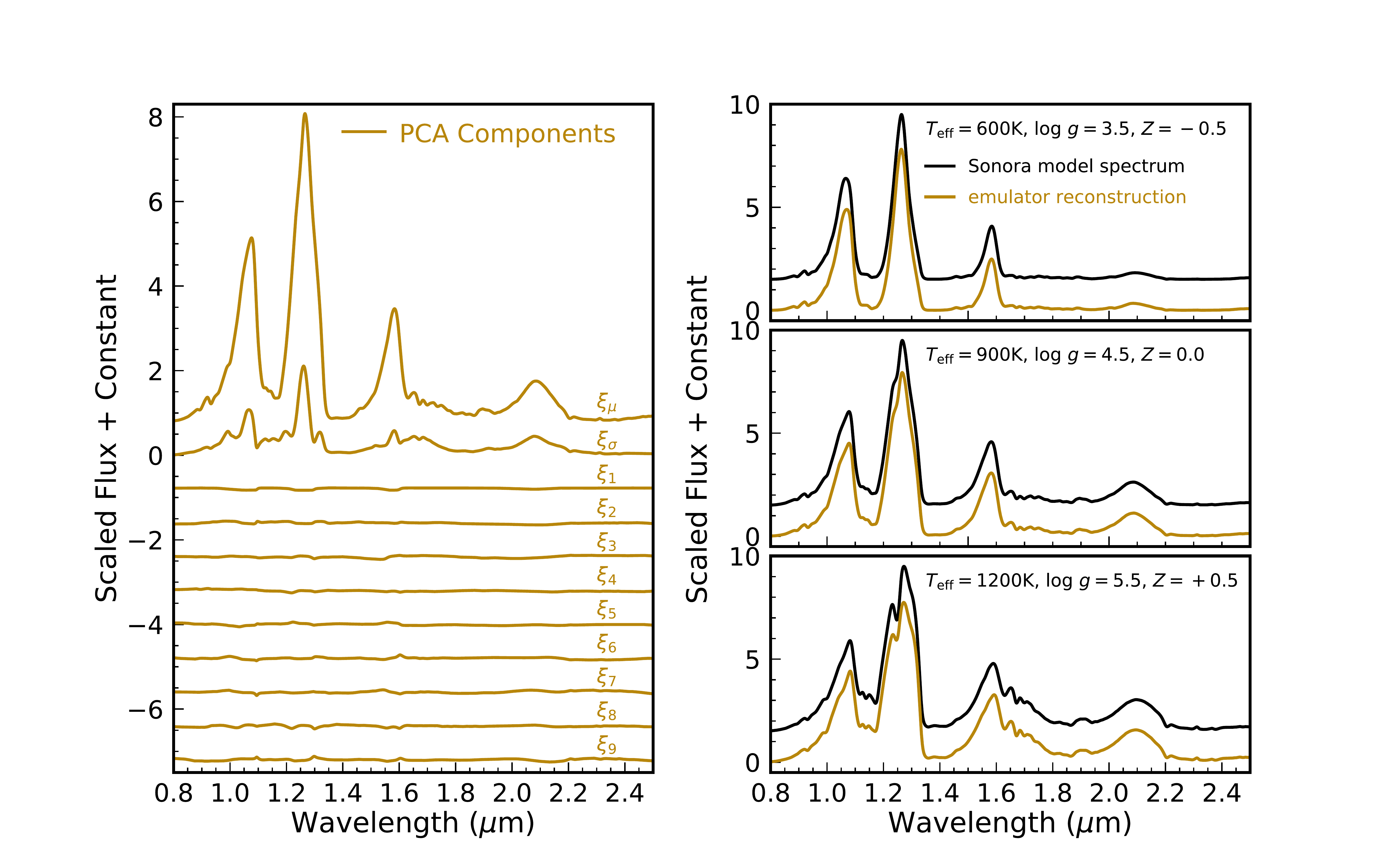}
\caption{Left: The mean spectrum $\xi_{\mu}$, standard deviation spectrum $\xi_{\sigma}$, and 9 eigenspectra $\xi_{k}$ of the 198 cloudless Sonora-Bobcat models $f_{\lambda}\left(\{\theta_{\star}\}^{\rm train}\right)$ used to train the spectral emulator (Section~\ref{subsubsec:spec_emu}). These spectra are scaled with the same scale factor and then offset by constants. Right: Comparisons between the original Sonora-Bobcat model spectra (offset by a constant; black) and the linear combination of $\xi_{\mu}$, $\xi_{\sigma}$, and $\xi_{k}$ (orange) at $\{T_{\rm eff},\ \log{g},\ Z\} =$ $\{600\, {\rm K},\ 3.5,\ -0.5\}$ (top), $\{900\, {\rm K},\ 4.5,\ 0.0\}$ (middle), and $\{1200\, {\rm K},\ 5.5,\ +0.5\}$ (bottom). We scale the flux of each Sonora-Bobcat model spectrum and use the same scale factor for the corresponding reconstructed spectra. The reconstructed spectra by PCA components well-approximate the original models. }
\label{fig:recon_pca}
\end{center}
\end{figure*}
%------------figure end-----------------

\subsubsection{Spectral Emulator Construction}
\label{subsubsec:spec_emu}
We start our forward-modeling analysis by constructing Starfish's spectral emulator, which can generate the cloudless Sonora-Bobcat model spectra with an arbitrary set of grid parameters. (1) Following the Appendix of C15 (also see \citealt{2007PhRvD..76h3503H} and \citealt{2009ApJ...705..156H}), we first trim the cloudless Sonora-Bobcat models to wavelengths of $0.8-2.5$~$\mu$m over the parameter space relevant to late-T dwarfs: $[600, 1200]$~K in $T_{\rm eff}$, $[3.25, 5.5]$~dex in $\log{g}$, and $[-0.5, +0.5]$~dex in $Z$, encompassing 286 grid points. We then pick all 198 grid points where the Sonora-Bobcat models cover the same $\{T_{\rm eff},\ \log{g}\}$ values among the three metallicity points $Z=\{+0.5, 0, -0.5\}$ (see Figure~\ref{fig:bias} for the parameter space of models). The selected 198 grid points are used to train the spectral emulator and are denoted as $\{\theta_{\star}\}^{\rm train}$, where $\theta_{\star} = \{T_{\rm eff}, \log{g}, Z\}$. Model spectra from these training grid points, $f_{\lambda}\left(\{\theta_{\star}\}^{\rm train}\right)$, have four dimensions (i.e., wavelength, $T_{\rm eff}$,\ $\log{g}$,\ and $Z$) with a spacing of $50$~K and $100$~K in $T_{\rm eff}$, $0.25$~dex and $0.5$~dex in $\log{g}$, and $0.5$~dex in $Z$.\footnote{The $T_{\rm eff}$ spacing is $50$~K for $[600, 1000]$~K and $100$~K for $[1000, 1200]$~K. The $\log{g}$ spacing is $0.25$~dex for $[3.25, 3.5]$~dex and $0.5$~dex for $[3.5 ,5.5]$~dex.} The remaining 88 ($= 286 - 198$) grid points, denoted as $\{\theta_{\star}\}^{\rm test}$, and their corresponding model spectra, $f_{\lambda}\left(\{\theta_{\star}\}^{\rm test}\right)$, will be used to validate the performance of our spectral emulator (Section~\ref{subsubsec:recons}). 

(2) We downgrade the resolution of the model spectra using a Gaussian kernel corresponding to the $0.5''$ slit of SpeX. Our convolution process accounts for the wavelength-dependent spectral resolution of the SpeX prism mode \citep[][]{2003PASP..115..362R}. We therefore end up with convolved models at both the training ($\{\theta_{\star}\}^{\rm train}$) and the testing ($\{\theta_{\star}\}^{\rm test}$) grid points that have the same spectral resolution as the data.

(3) We perform principal component analysis (PCA) of the model spectra from the training grid points $f_{\lambda}\left(\{\theta_{\star}\}^{\rm train}\right)$. We standardize $f_{\lambda}\left(\{\theta_{\star}\}^{\rm train}\right)$ by subtracting their mean spectrum $\xi_{\mu}$ and then dividing by their standard deviation spectrum $\xi_{\sigma}$. Then we decompose the standardized model spectra into a few principal eigenspectra \citep[e.g.,][]{2014sdmm.book.....I}. We obtain 9 eigenspectra $\xi_{k}\, (k = 1, 2, \dots, 9)$ with a PCA threshold value of $0.99$ (Figure~\ref{fig:recon_pca}), meaning that the $99\%$ summative variance of the standardized model spectra can be modeled with these their corresponding eigenspectra. The model spectrum at each grid point can be thereby approximated using the linear combination of $\xi_{\mu}$, $\xi_{\sigma}$, and 9 eigenspectra $\xi_{k}$ (Figure~\ref{fig:recon_pca}; also see Equation~23 of C15),
\begin{equation} \label{eq:spec_emu}
\begin{aligned} 
f_{\lambda}\left(\{\theta_{\star}\}^{\rm train}\right) &= f_{\lambda}^{\rm rec}\left(\{\theta_{\star}\}^{\rm train}\right) + \epsilon_{\rm emu}\left(\{\theta_{\star}\}^{\rm train}\right) \\
&\equiv \xi_{\mu} + \xi_{\sigma} \sum_{k=1}^{9} w_{k}\left(\{\theta_{\star}\}^{\rm train}\right) \xi_{k} + \epsilon_{\rm emu}\left(\{\theta_{\star}\}^{\rm train}\right) 
\end{aligned} 
\end{equation} 
Here $f_{\lambda}^{\rm rec}\left(\{\theta_{\star}\}^{\rm train}\right)$ is the reconstructed model spectra at the training grid points, and $\epsilon_{\rm emu}$ is the reconstruction error, i.e., the flux difference between the original and reconstructed models. The standardized weights of eigenspectra, $w_{k}\left(\{\theta_{\star}\}^{\rm train}\right)$, is a function of grid parameters. 

(4) We train a Gaussian process on the standardized weights $w_{k}\left(\{\theta_{\star}\}^{\rm train}\right)$ for the eigenspectra at different grid parameters. For a given set of parameters $\theta_{\star} = \{T_{\rm eff},\ \log{g},\ Z\}$, this Gaussian process provides a probability distribution of $w_{k}\left(\theta_{\star}\right)$, thereby leading to a distribution of the interpolated spectra (Equation~\ref{eq:spec_emu}) as well as the propagation of the interpolation uncertainties into the resulting inferred parameters. 

%%%%%%%%%%%%%%%%%%%%%%%%%%%%%%%%%%%
%------------- Model Reconstruction emulator (training) --------------------
%%%%%%%%%%%%%%%%%%%%%%%%%%%%%%%%%%%
\begin{figure*}[t]
\begin{center}
\includegraphics[height=5.2in]{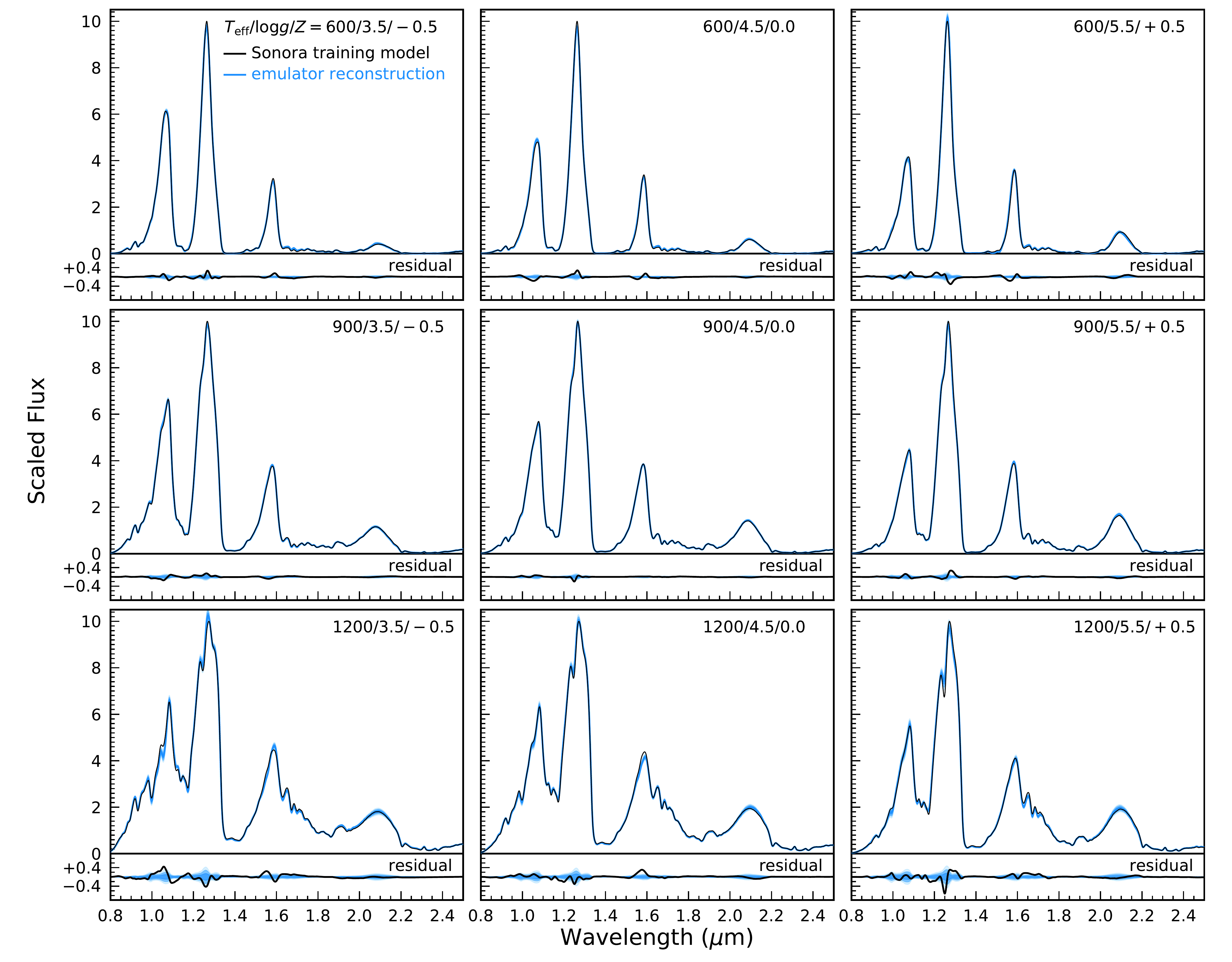}
\caption{Each panel compares a Sonora-Bobcat model spectrum (black) used to define the spectral emulator (a.k.a., the training models; Section~\ref{subsubsec:spec_emu}) with a spectra distribution generated from the emulator (blue) at a given grid point (with $\{T_{\rm eff},\ \log{g},\ Z\}$ values listed at the upper corner). We normalize the flux of each Sonora-Bobcat model spectrum to the same $J$-band peak and then use the same scale factor for the reconstructed ones from the spectral emulator. Residuals between the original and reconstructed models (black) are shown at the bottom of each panel (with the same vertical scale as the spectra shown at the top), overlaid with the $1\sigma$, $2\sigma$, and $3\sigma$ dispersions (blue background shadows) of the spectra distribution from the emulator. Our emulator can generate spectra nearly identical to the Sonora-Bobcat models. Small residuals can be found near the $YJHK$ flux peaks, which mostly constitute $<0.5\%$ of the peak $J$-band flux of the original models, smaller than the modeling systematics of the cloudless Sonora-Bobcat models (Section~\ref{subsec:assess_model_systematics}). }
\label{fig:recon_emu_train}
\end{center}
\end{figure*}
%------------figure end-----------------

%%%%%%%%%%%%%%%%%%%%%%%%%%%%%%%%%%%
%------------- Model Reconstruction emulator (testing) --------------------
%%%%%%%%%%%%%%%%%%%%%%%%%%%%%%%%%%%
\begin{figure*}[t]
\begin{center}
\includegraphics[height=5.2in]{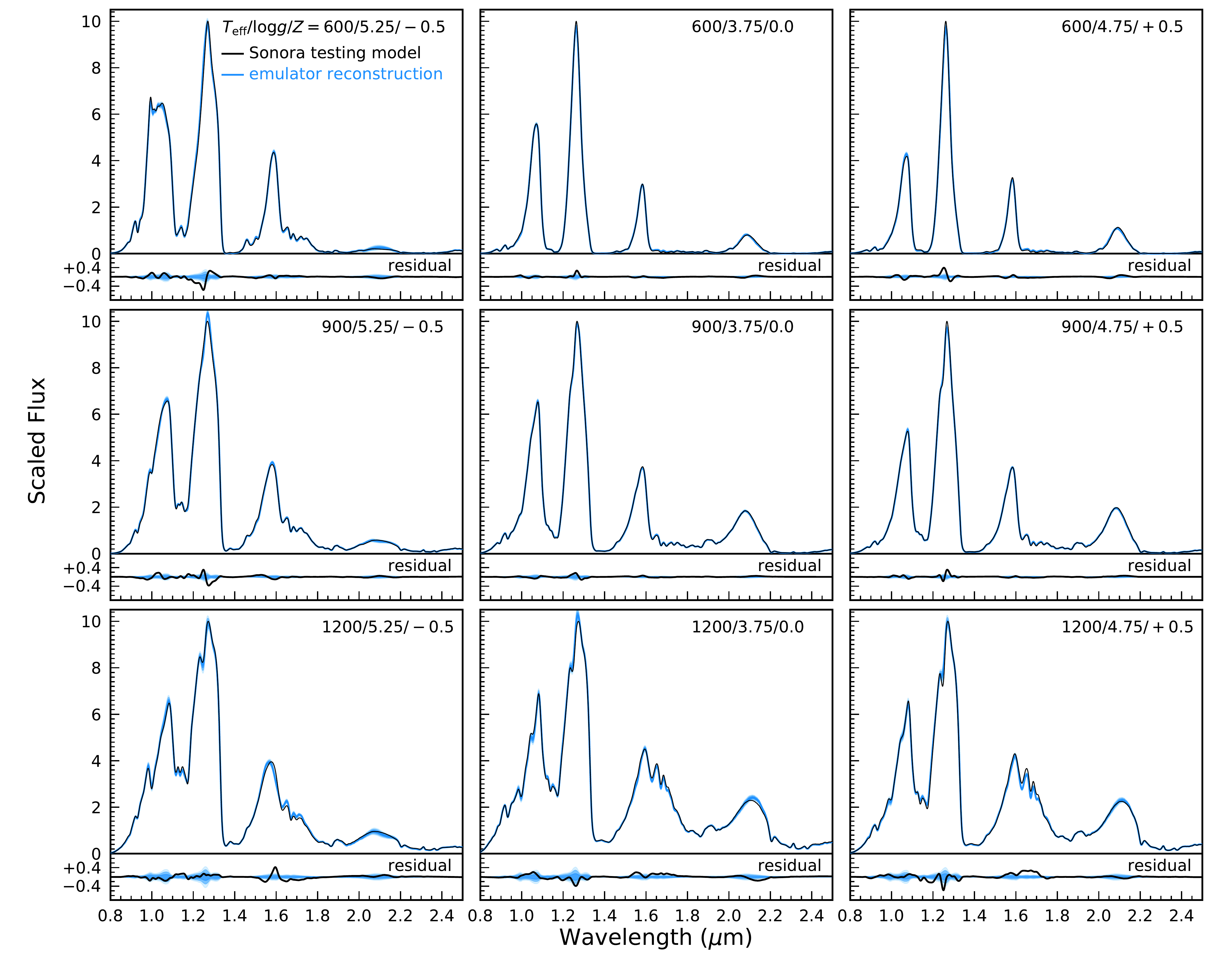}
\caption{Comparisons between the Sonora-Bobcat model spectra (black) not used to define the spectral emulator (a.k.a. the testing models) and the spectra distribution generated from the spectral emulator (blue), with the same format as Figure~\ref{fig:recon_emu_train}. The testing models can be well reconstructed even though they are not used to construct the spectral emulator. }
\label{fig:recon_emu_test}
\end{center}
\end{figure*}
%------------figure end-----------------

Starfish uses a squared exponential kernel to describe the Gaussian process covariance matrix for the standardized weights of each eigenspectrum. This kernel is characterized by four hyper-parameters, $\phi_{{\rm int}, k} = \{a_{{\rm int}, k}, \ell_{T_{\rm eff}}, \ell_{\log{g}}, \ell_{Z}\}$. The parameter $a_{{\rm int},k}$ describes the covariance along the diagonal axis, and $\ell_{T_{\rm eff}}, \ell_{\log{g}}, \ell_{Z}$ describe the length scales in grid parameters over which the covariance decreases in an exponential manner. We follow the suggestions of C15 and assume a $\Gamma$ function prior for each length scale, so that each $\Gamma$ function reaches its peak probability at a value ($\approx 100$~K in $T_{\rm eff}$, $\approx 1.0$~dex in $\log{g}$, and $\approx 1.0$~dex in $Z$) which is one or two times the grid spacing of each parameter. 

When the PCA process decomposes grid models into 9 principal components, the flux information from the remaining less important components is truncated, leading to residuals $\epsilon_{\rm emu}$ between the original training grid models and the reconstructed ones (Equation~\ref{eq:spec_emu}). C15 accounts for this reconstruction error by including an additional hyper-parameter $\lambda_{\xi}$ into the likelihood function of the Gaussian process. We assume a $\Gamma$ function prior for $\lambda_{\xi}$ and compute its rate and shape following Equation~43$-$44 of C15.

(5) We determine the posteriors of all 37 hyper-parameters (4 $\phi_{{\rm int}, k}$ parameters for each of 9 eigenspectra and one $\lambda_{\xi}$) using the Markov chain Monte Carlo (MCMC) algorithm {\it emcee} \citep{2013PASP..125..306F} with 148 walkers. We use the medians of the hyper-parameter posteriors to construct the spectral emulator. 

These 37 hyper-parameters constitute the core of the spectral emulator, allowing computation of the probability distribution of the 9 eigenspectra' standardized weights $w_{k}\left(\theta_{\star}\right)$ at a given set of grid parameters $\theta_{\star}$ (Equation~$47-51$\footnote{The component ``$(\lambda_{\xi} \Phi^{T} \Phi)$'' in Equation~47$-$48 of C15 should be ``$(\lambda_{\xi} \Phi^{T} \Phi)^{-1}$'', though the Starfish code itself does not have this typo.} of C15). Combining $\xi_{\mu}$, $\xi_{\sigma}$, $\xi_{k}$, and the distribution of their standardized weights, Equation~\ref{eq:spec_emu} allows us to compute the distribution of interpolated model spectra (Equation~$52$ of C15). Starfish also uses these hyper-parameters to define an ``emulator covariance matrix'' $\mathcal{K}^{E}$, which describes the covariance of the interpolated spectral fluxes between different wavelengths (Equation~59 of C15). We include $\mathcal{K}^{E}$ into the final covariance matrix for the subsequent model evaluation, so that the interpolation uncertainties are propagated into the inferred physical parameters (Section~\ref{subsubsec:eval}).

\subsubsection{Model Reconstruction}
\label{subsubsec:recons}
Before inferring physical parameters of objects, we verify that our spectral emulator can reconstruct both the training and the testing Sonora-Bobcat models within an acceptable level. At each grid point, we generate a probability distribution of spectra using the spectral emulator and then compare them with the original model spectrum. Representative examples of the reconstructed training and testing models are shown in Figures~\ref{fig:recon_emu_train} and \ref{fig:recon_emu_test}, respectively. 

We note the spectra generated from our spectral emulator are nearly identical to the Sonora-Bobcat models over all grid points, confirming that our emulator is a robust model interpolator. At several grid points and in a certain wavelength ranges (primarily near the flux peaks in $YJHK$ bands), there are noticeable flux differences between the original and reconstructed model spectra. These differences mostly constitutes $<6\%$ of the local flux and $<0.5\%$ of the peak $J$-band flux in the original models, much smaller than the systematic difference between the cloudless Sonora-Bobcat models and late-T dwarf spectra ($\approx 2\%-4\%$ of the peak $J$-band flux; Section~\ref{subsec:assess_model_systematics}). In Section~\ref{subsubsec:systematics}, we quantify the systematics in the derived physical parameters due to imperfect model reconstruction.

\subsubsection{Physical Parameters and Priors}
\label{subsubsec:phys_param}
There are six physical parameters to be determined: effective temperature $T_{\rm eff}$, logarithmic surface gravity $\log{g}$, metallicity $Z$, radial velocity $v_{r}$, projected rotational velocity $v\sin{i}$, and logarithmic solid angle $\log{\Omega} = \log\ (R/d)^{2}$, where $i$, $R$, and $d$ is the inclination of the rotation axis, the radius, and the distance, respectively. For a given set of these parameters, we use our spectral emulator to generate a probability distribution of Sonora-Bobcat model spectra at $\{T_{\rm eff},\ \log{g},\ Z\}$, convolve the interpolated spectra with kernels that apply the Doppler shift by radial velocity $v_{r}$ and the broadening by rotation $v\sin{i}$ \citep[e.g.,][]{2008oasp.book.....G}, and then scale the spectra using $\log{\Omega}$. The resulting model spectra are thereby ready to be compared with data to infer physical parameters.

We assume uniform priors for $T_{\rm eff}$, $\log{g}$, and $Z$ within the parameter space where our spectral emulator is constructed, i.e., $\left[600, 1200\right]$~K in $T_{\rm eff}$, $\left[3.25, 5.5\right]$~dex in $\log{g}$, and $\left[-0.5, +0.5\right]$~dex in $Z$. We assume a uniform prior of $(-\infty, +\infty)$ for both $v_{r}$ and $\log{\Omega}$.

We assume a uniform prior of $\left[0, v_{\rm max}\right]$ for $v\sin{i}$ and determine the $v_{\rm max}$ value using an object's oblateness. We follow \cite{2020ApJ...891..171Z} and express the rotationally induced oblateness as $f = 2Cv_{\rm rot}^{2} / 3gd\sqrt{\Omega}$ \citep[e.g.,][]{2003ApJ...588..545B}, where $d$ is the distance, $v_{\rm rot}$ is the equatorial rotational velocity, and $C = 0.9669$ corresponds to a polytropic index of $n=1.5$ \citep[][]{1939isss.book.....C}, which well approximates fully-convective brown dwarfs \citep[e.g.,][]{1993RvMP...65..301B}. To be rotationally stable, the oblateness of a $n=1.5$ polytrope should be within a critical value of $f_{\rm crit} = 0.385$ \citep{1964ApJ...140..552J}. Altogether, we derive the following constraints for $v\sin i$:
\begin{equation} \label{eq:vmax_with_d}
\begin{aligned} 
    &0 \leqslant v\sin{i} \leqslant v_{\rm rot} \leqslant \Omega^{1/4} \left(\frac{3f_{\rm crit}gd}{2C}\right)^{1/2}  \\
    \Rightarrow\quad &v\sin{i} \in \left[0,\ v_{\rm max} \equiv \Omega^{1/4} \left(\frac{3f_{\rm crit}gd}{2C}\right)^{1/2}\right]
\end{aligned} 
\end{equation}
For the three benchmarks, we follow Equation~\ref{eq:vmax_with_d} and compute $v_{\rm max}$ using their $d$, $\log{g}$, and $\Omega$ in each step of the spectral-fitting process.

\subsubsection{Covariance Hyper-Parameters and Priors}
\label{subsubsec:covhyper_param}
In addition to physical parameters, Starfish uses three hyper-parameters $\{a_{N},\ a_{G},\ \ell\}$\footnote{C15 originally used the letter ``$b$'' for the hyper-parameter ``$a_{N}$'' that we describe here.} to describe the covariance matrix for subsequent model evaluation (Section~\ref{subsubsec:eval}). The hyper-parameter $a_{N}$ characterizes a ``noise covariance matrix'' $\mathcal{K}^{N}$, which describes the covariance in (data$-$model) residuals between wavelength pixels $i$ and $j$, namely $\mathcal{K}_{ij}^{N}= a_{N} \delta_{ij} \sigma_{i} \sigma_{j}$, where $\delta_{ij}$ is the Kronecker delta and $\sigma_{i}, \sigma_{j}$ is observed flux uncertainty in wavelength pixels $i, j$, respectively. This diagonal $\mathcal{K}^{N}$ assumes the residuals from different wavelengths are independent and is usually adopted as the full covariance matrix by traditional forward-modeling analyses (Section~\ref{subsubsec:compare_traditional}). We assume a Gaussian prior on $a_{N}$ with mean of $1.0$, standard deviation of $0.25$, and a range of $[0, +\infty)$.

The other two hyper-parameters $a_{G}$ and $\ell$ characterize a ``global covariance matrix'' $\mathcal{K}^{G}$, which describes the covariance in residuals between two wavelength pixels using a customized Mat{\'e}rn kernel. The hyper-parameter $a_{G}$ describes the covariance along the diagonal axis of the matrix and $\ell$ describes the wavelength scale over which the covariance between two wavelengths decreases in nearly an exponential manner (see Appendix~\ref{app:ell} for details). This $\mathcal{K}^{G}$ matrix has off-diagonal components and can account for the important correlation in residuals caused by (1) the over-sampled instrumental line spread function, and (2) systematics in model atmospheres. We assume a uniform prior of $(-\infty, +\infty)$ in $\log{a_{G}}$ and assume a uniform prior on $\ell$ based on the prism-mode line spread function (Appendix~\ref{app:ell}). 

%%%%%%%%%%%%%%%%%%%%%%%%%%%
%------------- Parameter Systematics --------------------
%%%%%%%%%%%%%%%%%%%%%%%%%%%
\begin{figure*}[t]
\begin{center}
\includegraphics[height=2.in]{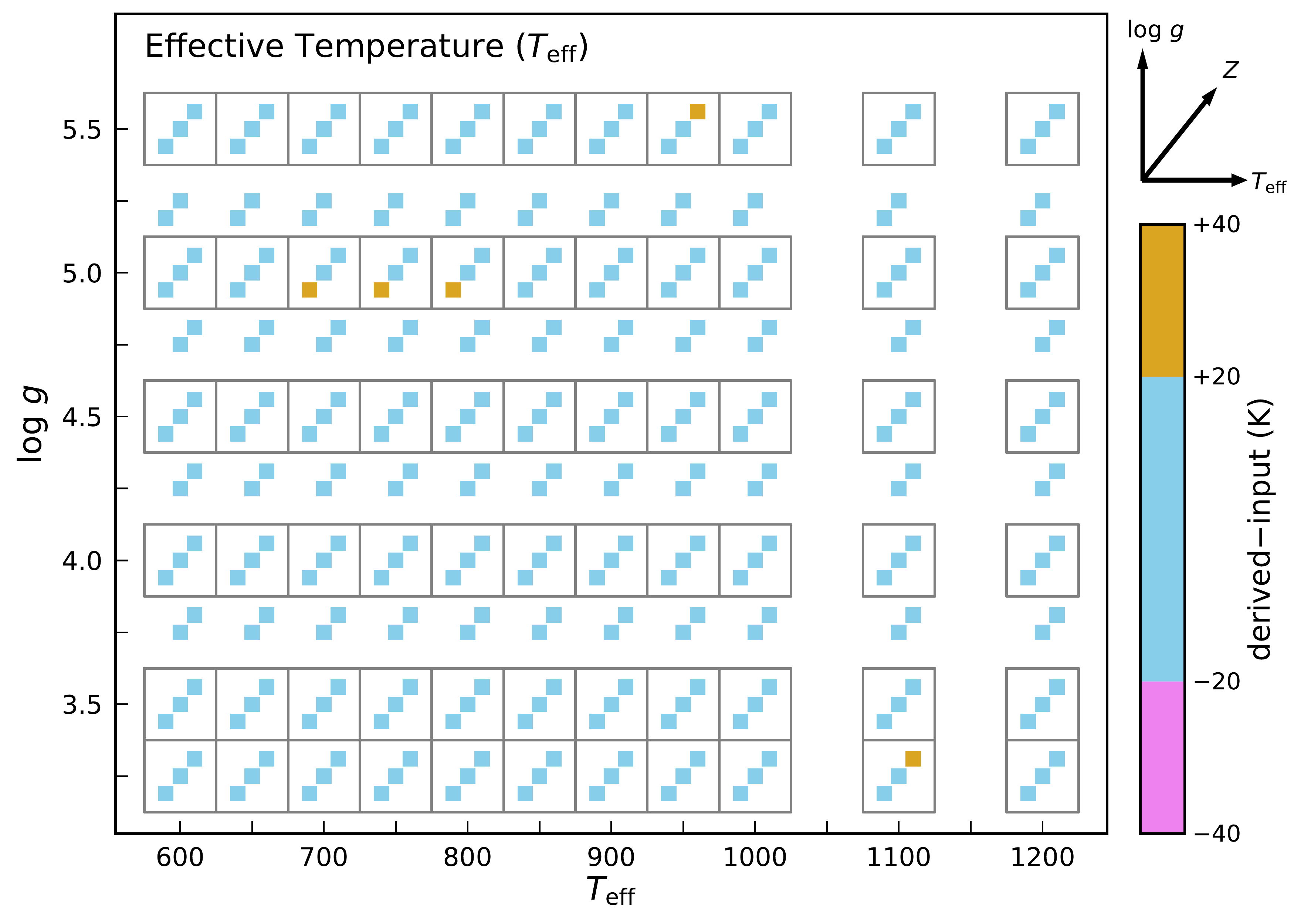}
\includegraphics[height=2.in]{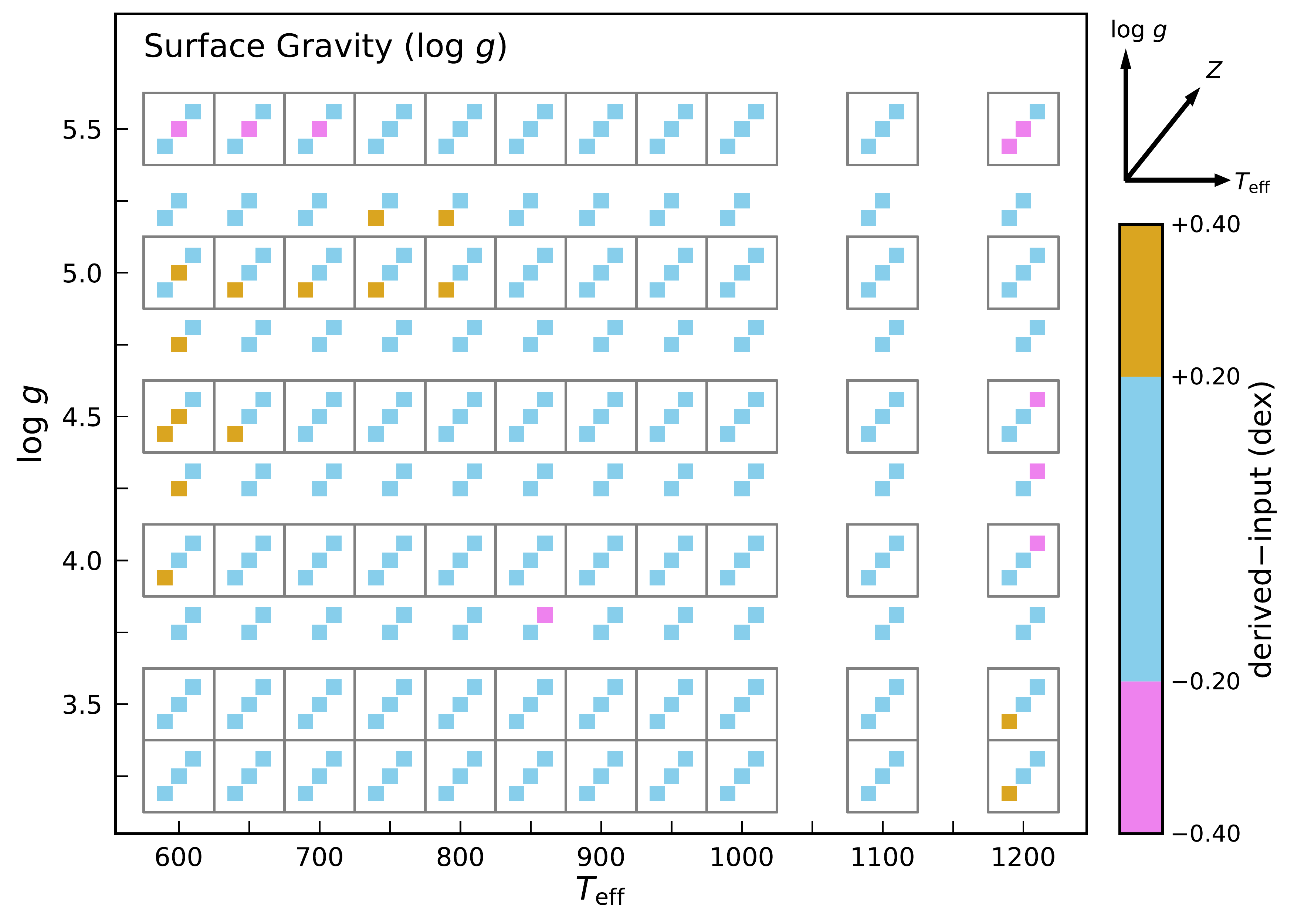}
\includegraphics[height=2.in]{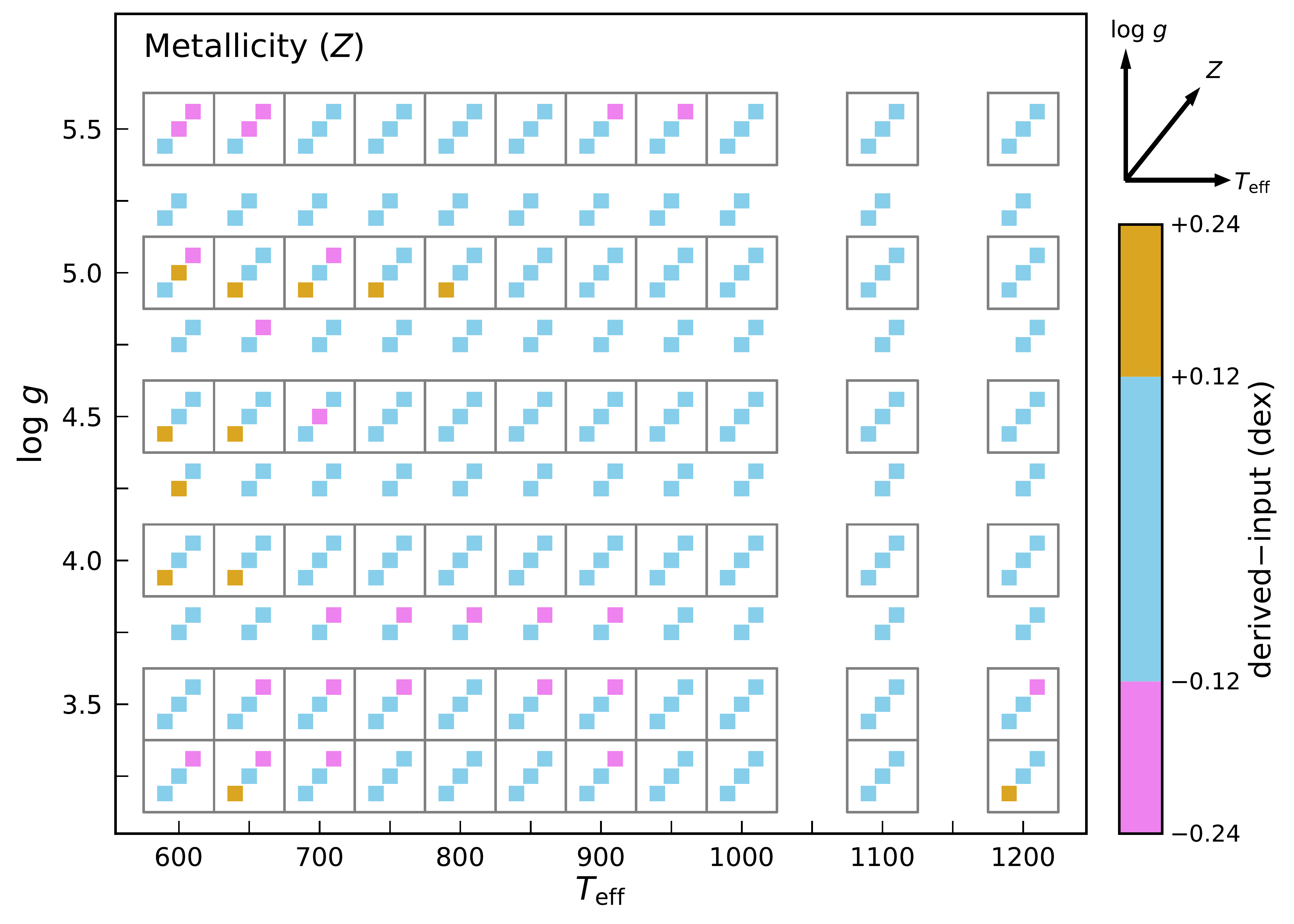}
\includegraphics[height=2.in]{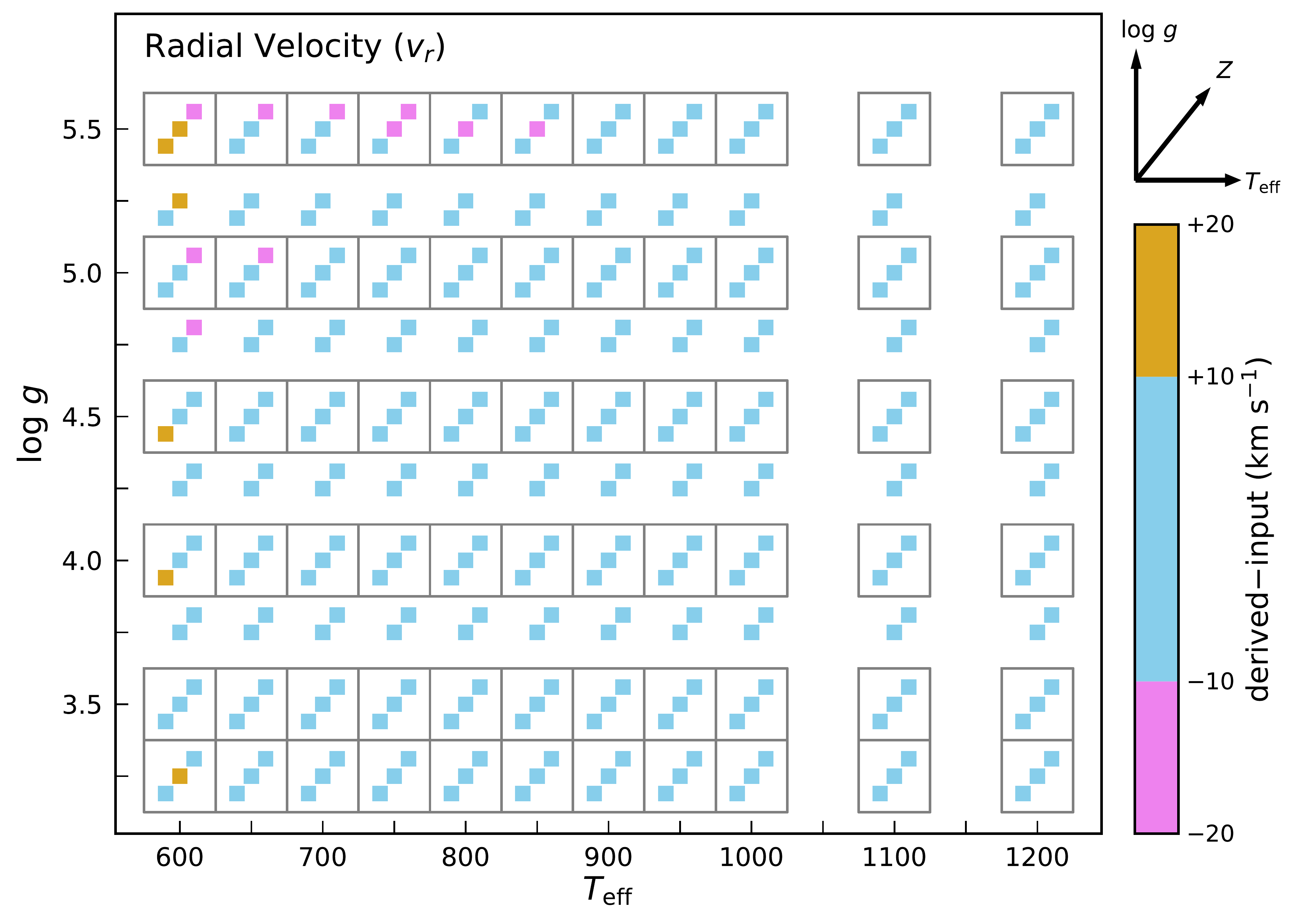}
\includegraphics[height=2.in]{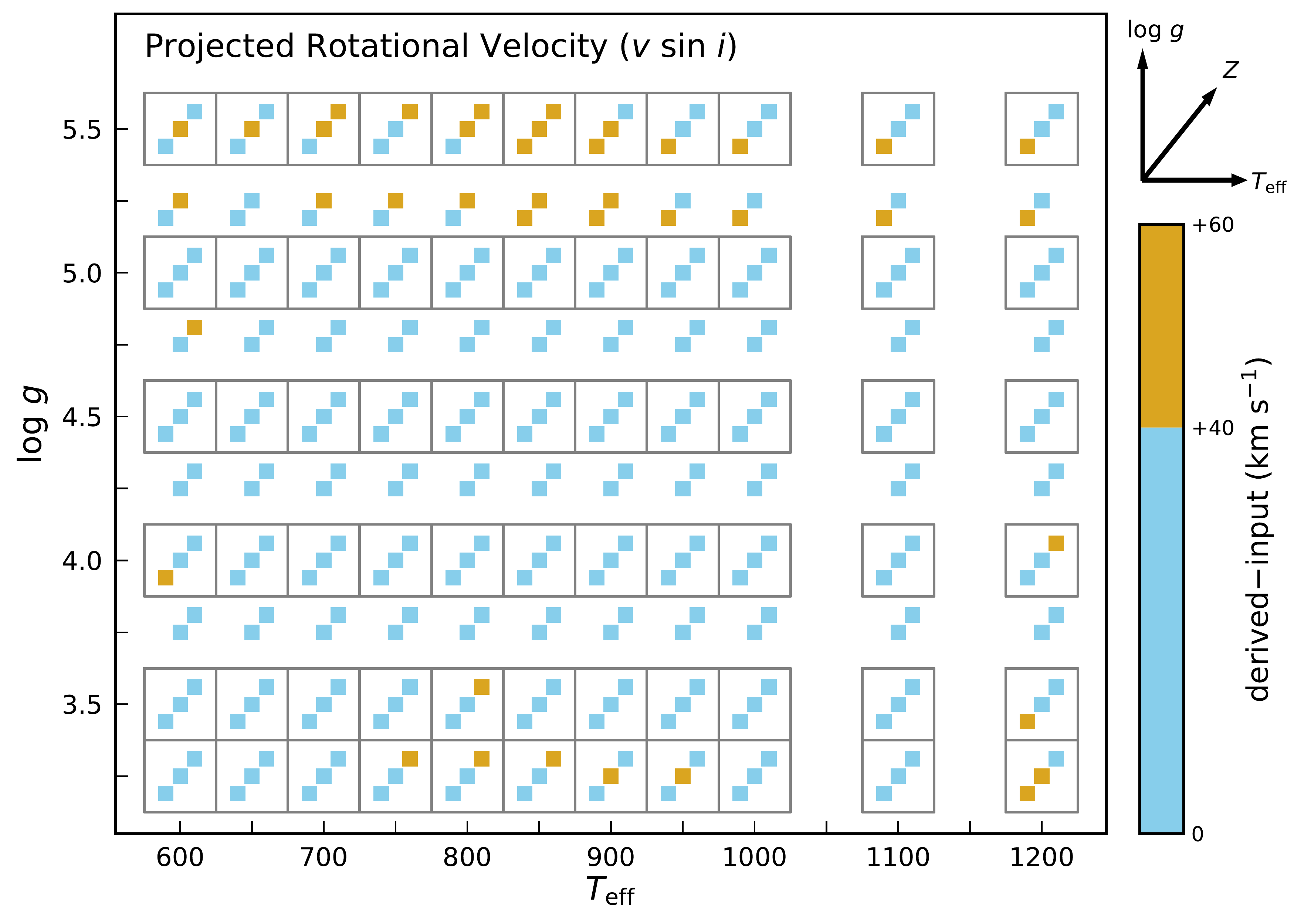}
\includegraphics[height=2.in]{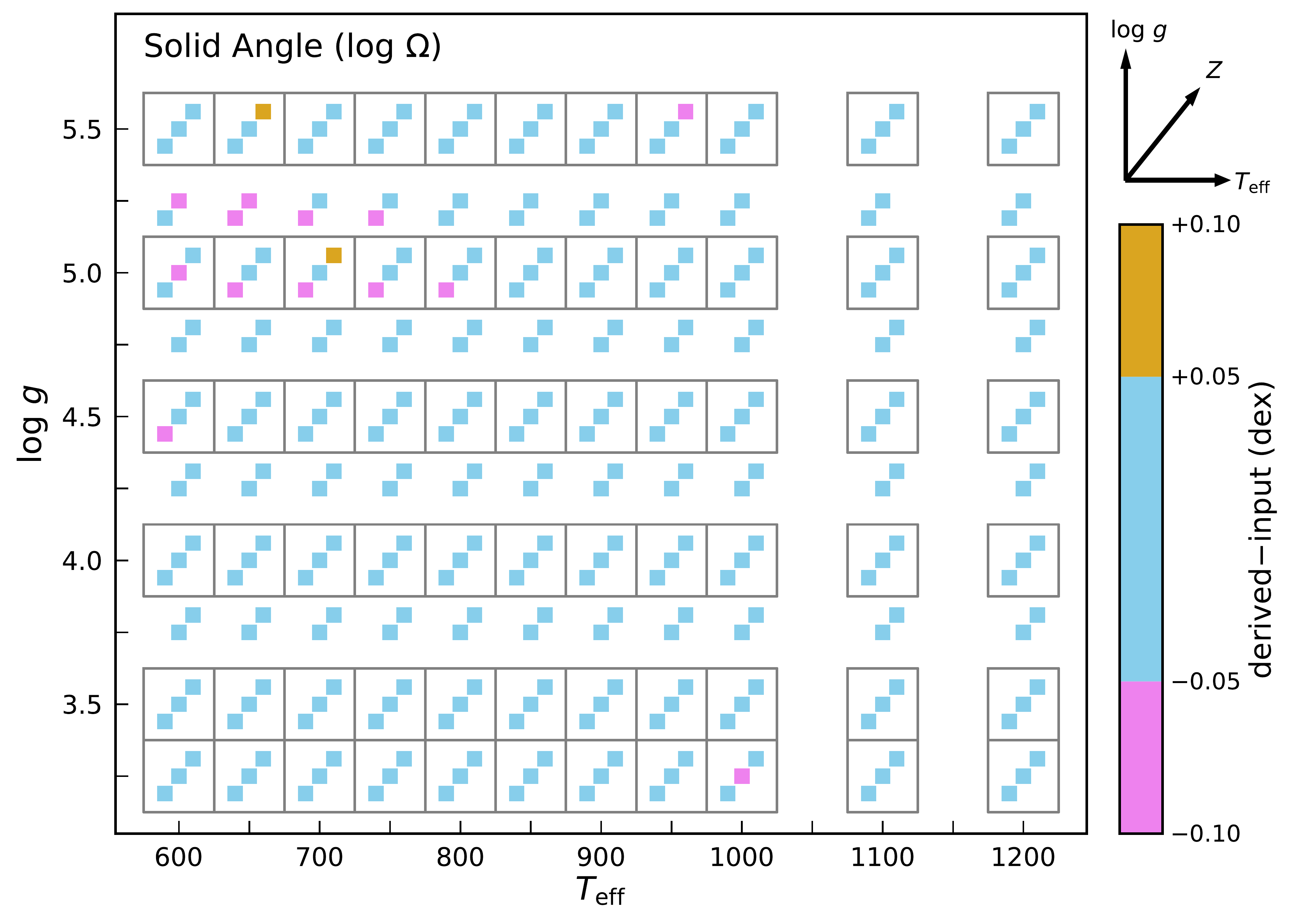}
\caption{Test of the systematic errors (arising from the spectral emulator) derived by fitting the original Sonora-Bobcat models using the Starfish analysis. Each panel shows one of the six physical parameters $\{T_{\rm eff},\ \log{g},\ Z,\ v_{r},\ v \sin{i},\ \log{\Omega}\}$ with all the 286 Sonora-Bobcat grid points, where the small squares located at the $\{T_{\rm eff}, \log{g}\}$ grid points correspond to solar metallicity ($Z=0$), and the two neighboring points correspond to the sub-solar metallicity ($Z = -0.5$; lower left) and super-solar metallicity ($Z = +0.5$; upper right). The layout of grid points is also indicated by the compass shown at the upper right in each panel. We use large grey squares to mark the training grid points used to construct the spectral emulator and do not mark the remaining testing grid points (Section~\ref{subsubsec:spec_emu}).  We use three different colors to map the derived-minus-input bias (caused by the spectral emulator) of the corresponding parameter, with boundary values defined by our adopted systematic errors (Equation~\ref{eq:six_systematics}). We note that the parameter bias over most grid points are well within our adopted systematics (light blue), and such bias has no significant difference between the training and testing grid points. }
\label{fig:bias}
\end{center}
\end{figure*}
%------------figure end-----------------

\subsubsection{Model Evaluation}
\label{subsubsec:eval}
There are nine free parameters $\{T_{\rm eff},\ \log{g},\ Z,\ v_{r},\ v\sin{i},\ \\\log{\Omega},\ a_{N},\ a_{G},\ \ell\}$ to be determined by our forward-modeling framework. For a given spectrum, we first generate model spectra $\mathbf{M}$ using the spectral emulator with physical parameters incorporated. We then compare $\mathbf{M}$ with the data $\mathbf{D}$ to compute the residual spectra,
\begin{equation} \label{eq:residual}
\mathbf{R} = \mathbf{D} - \mathbf{M}(T_{\rm eff},\ \log{g},\ Z,\ v_{r},\ v\sin{i},\ \log{\Omega})
\end{equation}
We evaluate the free parameters using the following covariance matrix $\mathbf{C_{S}}$ and log-likelihood function $\mathcal{L}_{\rm S}$,
\begin{equation} \label{eq:covlike_emu}
\begin{aligned} 
    \mathbf{C_{S}} &= \mathcal{K}^{E}(\phi_{{\rm int},k},\ \lambda_{\xi}) + \mathcal{K}^{N}(a_{N}) + \mathcal{K}^{G}(a_{G},\ \ell) \\
    \mathcal{L}_{\rm S} &= - \frac{1}{2} \left(\mathbf{R^{T}} \mathbf{C_{S}^{-1}} \mathbf{R} + \ln\lvert\mathbf{C_{S}}\rvert \right) -\frac{n}{2} \ln\left(2\pi\right) 
\end{aligned} 
\end{equation}
where $n$ is number of wavelength pixels in the spectra. The $\mathbf{C_{S}}$ matrix has three components: the emulator covariance matrix $\mathcal{K}^{E}$ (Section~\ref{subsubsec:spec_emu}), the noise covariance matrix $\mathcal{K}^{N}$, and the global covariance matrix $\mathcal{K}^{G}$ (Section~\ref{subsubsec:covhyper_param}). The $\mathcal{K}^{E}$ matrix is already determined from the spectral emulator construction and therefore is fixed for a given slit width of the data. The $\mathcal{K}^{N}$ and $\mathcal{K}^{G}$ matrices (described by the three hyper-parameters) are determined as part of the analysis process.

We use {\it emcee} \citep{2013PASP..125..306F} to fit our $1.0-2.5$~$\mu$m spectra with 48 walkers. During the fitting the process, we remove the first 5000 iterations as the burn-in phase and then iteratively estimate the chains' average autocorrelation length using both the \cite{2010CAMCS...5...65G} method and the revised version suggested by Daniel Foreman-Mackey.\footnote{\url{https://emcee.readthedocs.io/en/stable/tutorials/autocorr/}.} We terminate the fitting process once the number of iteration exceeds 50 times the autocorrelation length of all parameters \citep[e.g.,][]{2020ApJ...891..171Z}. The MCMC analysis for all objects converges with $1 \times 10^{5}$ iterations, and we use the resulting chains to derive the parameter posteriors for each object.

\subsubsection{Systematics of Inferred Physical Parameters}
\label{subsubsec:systematics}
Here we validate our forward-modeling framework by investigating the systematics of the inferred physical parameters caused by the Starfish machinery and data reduction. 

As described in Section~\ref{subsubsec:recons}, Starfish's spectral emulator decomposes the Sonora-Bobcat grid models into eigenspectra, leading to small flux differences between the original model spectra and those reconstructed by the spectral emulator (e.g., Figures~\ref{fig:recon_emu_train} and \ref{fig:recon_emu_test}). While such reconstruction error is generally insignificant over the entire grid, it may introduce systematic errors to the inferred physical parameters. These ``emulator-induced systematics'' are not fully accounted\footnote{When constructing the spectral emulator, Starfish uses the hyper-parameter $\lambda_{\xi}$ to account for the reconstruction error (Section~\ref{subsubsec:spec_emu}). However, $\lambda_{\xi}$ only characterizes an average level of such error among all the training models across the entire $0.8-2.5$~$\mu$m wavelengths (also see Equation~33 of C15). Therefore, this single, constant $\lambda_{\xi}$ parameter is not enough to absorb all reconstruction errors that occur at specific grid points and/or in short wavelength ranges, leading to occasionally noticeable flux differences between the original and reconstructed model spectra as seen in Figures~\ref{fig:recon_emu_train} and \ref{fig:recon_emu_test}.} for when evaluating model parameters. Here we quantify them by running our forward-modeling analysis directly on the original Sonora-Bobcat model spectra and comparing the resulting parameter posteriors to the input model properties.

We use all 286 Sonora-Bobcat model spectra with their spectral resolution downgraded to the $0.5''$ slit. We treat them the same as the on-sky data and use the aforementioned spectral emulator, parameter priors, covariance matrix, and likelihood function to infer the properties of each model spectrum. We adopt zero flux uncertainties for the original model spectra and thus do not include the noise covariance matrix $\mathcal{K}^{N}$. We then compare the resulting parameters to the input parameters of each model spectrum $\{T_{\rm eff}^{\rm grid},\ \log{g}^{\rm grid},\ Z^{\rm grid},\ v_{r} \equiv 0,\ v\sin i \equiv 0,\ \log{\Omega} \equiv 0\}$ to estimate the emulator-induced systematics.

Figure~\ref{fig:bias} summarizes each physical parameter over all 286 grid points. We find the median and $1\sigma$ confidence intervals of the ``derived-minus-input'' parameter bias is $+2^{+8}_{-7}$~K in $T_{\rm eff}$, $0.00^{+0.10}_{-0.09}$~dex in $\log{g}$, $-0.01 \pm 0.09$~dex in $Z$, $+1^{+4}_{-5}$~km~s$^{-1}$ in $v_{r}$, $+15^{+24}_{-8}$~km~s$^{-1}$ in $v\sin i$, and $-0.005^{+0.017}_{-0.025}$~dex in $\log{\Omega}$. The spread of the derived-minus-input bias of each parameter (except $v\sin{i}$) is smaller than or comparable to the formal parameter uncertainties for our late-T dwarfs from the Starfish analysis (Section~\ref{subsec:results}; Table~\ref{tab:results}), illustrating that such emulator-induced systematics do not dominate the errors and our forward-modeling analysis is robust. The derived derived-minus-input bias in $v\sin{i}$ is notably larger than zero at many grid points, and this occurs because $v\sin{i}$ is defined to be non-negative and the rotational broadening cannot be well-constrained by spectra with such low spectral resolution (see Section~\ref{subsec:results}). 

We adopt the following systematic errors for the six physical parameters to account for the emulator-induced systematics:
\begin{equation} \label{eq:six_systematics}
\begin{aligned}
&\sigma_{T_{\rm eff},\,\,{\rm emu}} &=&\quad 20~{\rm K} \\
&\sigma_{\log{g},\,\,{\rm emu}} &=&\quad  0.2~{\rm dex} \\
&\sigma_{Z,\,\,{\rm emu}} &=&\quad  0.12~{\rm dex} \\
&\sigma_{v_{r},\,\,{\rm emu}} &=&\quad  10~{\rm km~s}^{-1} \\
&\sigma_{v\sin{i},\,\,{\rm emu}} &=&\quad  40~{\rm km~s}^{-1} \\
&\sigma_{\log{\Omega},\,\,{\rm emu}} &=&\quad  0.05~{\rm dex} 
\end{aligned}
\end{equation}
These values are chosen conservatively to encompass the derived-minus-input values of parameters over most of the grid points (Figure~\ref{fig:bias}). 

In addition to these errors, the objects' radial velocities $v_{r}$ has another source of systematic error due to the uncertainty in the wavelength calibration of the SpeX data. This uncertainty is  $5.9$~\AA\ (M.~Cushing, private communication), independent of the slit width and equivalent to a radial velocity of $180 - 70$~km~s$^{-1}$ over the $1.0-2.5$~$\mu$m wavelength range. We thus adopt an additional error of $\sigma_{v_{r},\,\,{\rm wcal}} =  180~{\rm km~s}^{-1}$ for the inferred radial velocity. The total systematic error of $v_{r}$ is $(\sigma_{v_{r},\,\,{\rm emu}}^{2} + \sigma_{v_{r},\,\,{\rm wcal}}^{2})^{1/2} \approx 180~{\rm km~s}^{-1}$. 

Given that we flux-calibrate the objects' spectra using $H_{\rm MKO}$, the associated photometric uncertainties contribute to the total systematic error of $\log{\Omega}$, which we compute as $[\sigma_{\log{\Omega},\,\,{\rm emu}}^{2} + (0.4\sigma_{H_{\rm MKO}})^{2}]^{1/2}$, where $\sigma_{H_{\rm MKO}}$ is the photometric uncertainty in magnitudes. Calibrating the objects' spectra using photometry in different bands will alter our resulting $\log{\Omega}$ posteriors \citep[e.g., Figure~8][]{2015ApJ...807..183L}. The three benchmarks studied in this work have similar photometric uncertainties among $J$, $H$, and $K$ bands, therefore, their $\log{\Omega}$ posteriors will have different median values but similar uncertainties if we calibrate their spectra using $J$- or $K$-band magnitudes. Specifically, conducting the flux calibration using $J_{\rm MKO}$ ($K_{\rm MKO}$) magnitudes would cause the inferred $\log{\Omega}$ values to be higher by $0.05$~dex (lower by $0.03$~dex) for HD~3651B, higher by $0.01$~dex (higher by 0.01~dex) for GJ~570D, and higher by $0.07$~dex (lower by $0.14$~dex) for Ross~458C. We do not incorporate these small shifts into the systematic error of $\log{\Omega}$, so our inferred spectroscopic parameters are tied to the objects' $H$-band photometry.

We incorporate these systematic errors for each object by modifying the MCMC chains obtained from the spectral-fitting process (Section~\ref{subsubsec:eval}). For each of the six physical parameters $\{T_{\rm eff},\ \log{g},\ Z,\ v_{r},\ v\sin{i},\ \log{\Omega}\}$, we draw errors from a normal distribution centered at zero with a standard deviation being the adopted systematic error, and we make the number of draws the same as the number of MCMC samples. We then add these errors to the corresponding chain values to produce the final parameter posteriors. For the four parameters $\{T_{\rm eff},\ \log{g},\ Z,\ v_{r}\}$, if any of their chain values are outside the range of $[600, 1200]$~K for $T_{\rm eff}$, $[3.25, 5.5]$ for $\log{g}$, $[-0.5, +0.5]$ for $Z$, and $[0, v_{\rm max}]$ for $v\sin i$ ($v_{\rm max}$ is computed from Equation~\ref{eq:vmax_with_d}), we force those values to be at the lower or upper boundary. The resulting medians of all the final parameter posteriors are nearly unchanged from those without the inclusion of these emulator-induced systematics.

%%%%%%%%%%%%%%%%%%%%%%%%%%%%%%%%
%------------- Forward-Modeling Results (Starfish) --------------------
%%%%%%%%%%%%%%%%%%%%%%%%%%%%%%%%
\begin{figure*}[t]
\begin{center}
\includegraphics[height=2.in]{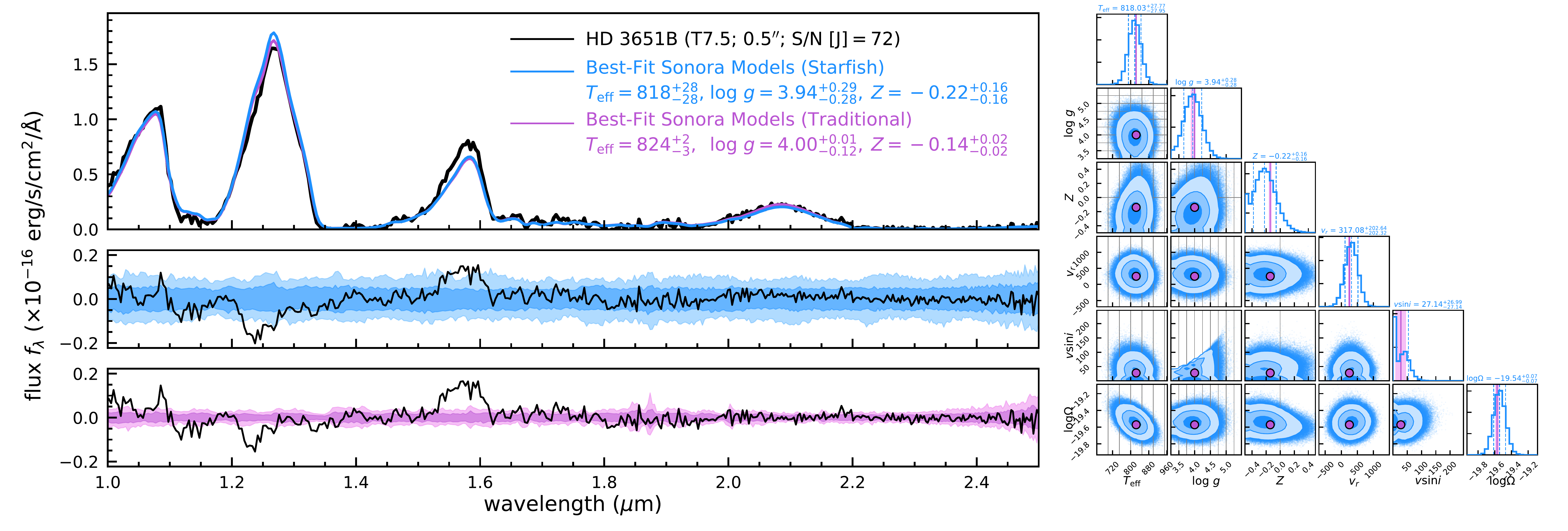}
\includegraphics[height=2.in]{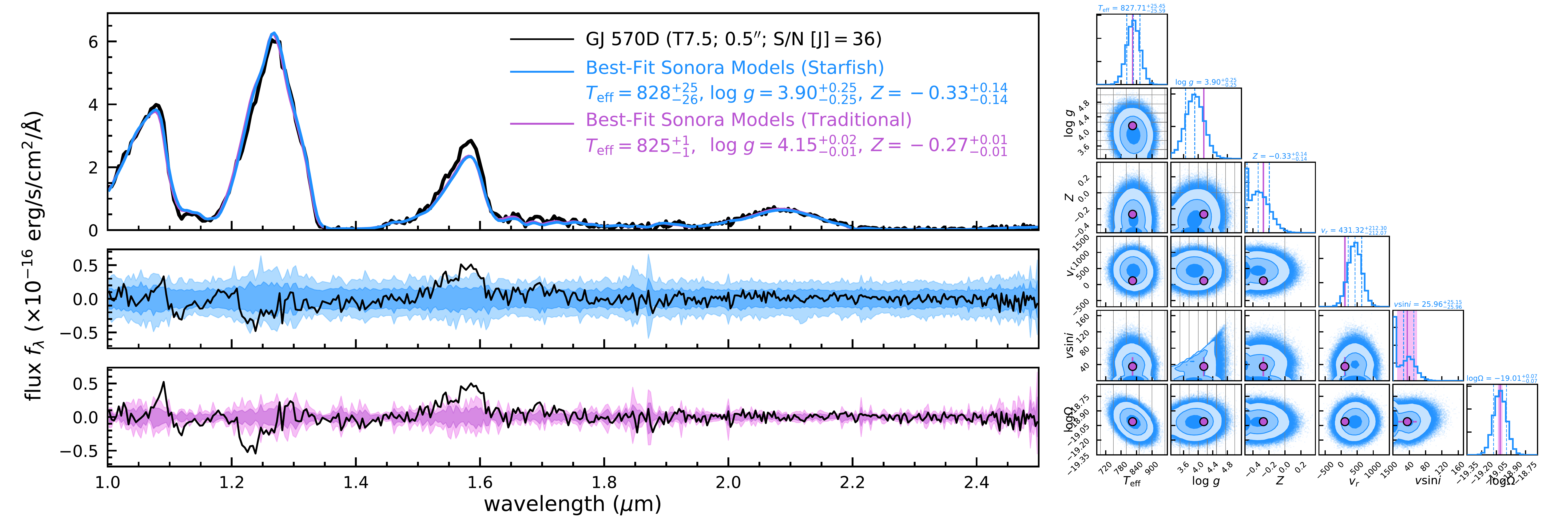}
\includegraphics[height=2.in]{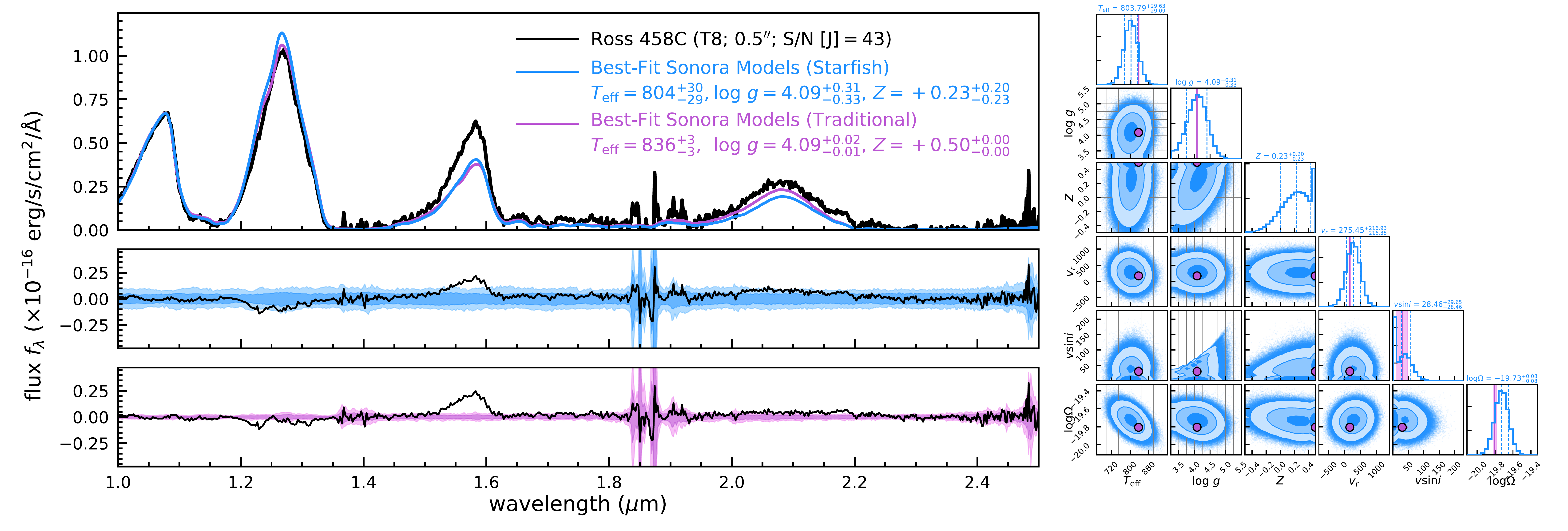}
\caption{Forward-modeling results of HD~3651B, GJ~570D, and Ross~458C, using the Starfish (blue) and the traditional approach (purple). Left: The upper panel for each object shows the observed spectrum (black) and the median Sonora-Bobcat model spectra of those interpolated at parameters drawn from the MCMC chains based on the Starfish (blue) and traditional (purple) methods. The object's name, spectral type, the slit width and $J$-band S/N of its spectrum, and inferred physical parameters are in the upper right corner.  The middle and lower panel for each object shows the residual of each method (data$-$model; black), with the blue and purple shadows being $1\sigma$ and $2\sigma$ dispersions of $5 \times 10^{4}$ draws from the Starfish and traditional covariance matrix, respectively. Right: Posteriors of the six physical parameters $\{T_{\rm eff},\ \log{g},\ Z,\ v_{r},\ v\sin{i},\ \log{\Omega}\}$ derived from the Starfish-based forward-modeling analysis (blue). We overlay the median values and uncertainties from the traditional method (purple), shown as vertical lines and shadows in the 1-D histograms and as circles and error bars in the 2-D histograms. We use grey vertical and horizontal lines to mark the $\{T_{\rm eff},\ \log{g},\ Z\}$ grids points of the cloudless Sonora-Bobcat models. }
\label{fig:emulin_results}
\end{center}
\end{figure*}
%------------figure end-----------------

\subsection{Results}
\label{subsec:results}
Figure~\ref{fig:emulin_results} presents the resulting parameter posteriors of our three benchmarks and also compares the observed data with the fitted Sonora-Bobcat model spectra, which are interpolated at the physical parameters drawn from the MCMC samples (with systematic errors incorporated). The fitted Sonora-Bobcat models match the observations well, but large residuals are found in $J$ and especially $H$ bands, suggesting some physical and chemical processes of late-T dwarf atmospheres are inadequately modeled or missing in the cloudless Sonora-Bobcat models. In $J$ band ($\approx 1.2-1.3$~$\mu$m), the fitted models over-predict spectra of the three benchmarks. This discrepancy plausibly arises from clouds \citep[e.g.,][]{2012ApJ...756..172M} and/or reductions in the vertical temperature gradient \citep[e.g.,][]{2015ApJ...804L..17T}, which can redistribute fluxes from shorter to longer wavelengths and thereby produce spectra that could better match the observations. In $H$ band ($\approx 1.5-1.6$~$\mu$m), the fitted models under-predict the emergent flux. This discrepancy might result from the spectral-fitting process itself aiming to balance the residuals from over-estimated fluxes in $J$ band. Also, it might be related to the disequilibrium abundance of NH$_{3}$, which would be less than the amount computed by the equilibrium chemistry of Sonora-Bobcat models and thereby could weaken absorption in the blue wing of $H$ band to better match the data \citep[e.g.,][]{1994Icar..110..117F, 2006ApJ...647..552S, 2014ApJ...797...41Z}. A detailed discussion of the cloudless Sonora-Bobcat models is in Section~\ref{sec:benchmarking}.

We further use the fitted parameter posteriors and the objects' parallaxes (from their primary stars) to compute their radii ($R$) and masses ($M$). For each object, we draw parallaxes from a normal distribution corresponding to the measured parallax and uncertainty, truncated within $(0, +\infty)$. We make the number of draws the same as their MCMC samples ($=$ 48 walkers $\times$ $10^{5}$ iterations) and combine these parallaxes and the $\log{\Omega}$ posterior to compute the $R$ posterior. We then use $\log{g}$ and $R$ to determine the $M$ posterior. 

We summarize the inferred physical parameters and covariance hyper-parameters of the three benchmarks in Table~\ref{tab:results} and \ref{tab:hyper_params}, respectively. Our derived uncertainties in $T_{\rm eff}$, $\log{g}$, and $Z$ are close to $1/3-1/2$ of the Sonora-Bobcat model grid spacing, suggesting that adopting a fraction of grid spacing as final parameter uncertainties is reasonable, as is usually done in traditional forward-modeling analysis \citep[e.g.,][]{2007ApJ...667..537L, 2008ApJ...678.1372C, 2011ApJ...726...30M, 2020ApJ...891..171Z}. In addition, we note that our forward-modeling analysis cannot constrain the radial velocity and the projected rotational velocity given the low-resolution prism spectra ($R \approx 80-250$). These two parameters might thus behave like nuisance parameters in our fits, aiming for the maximum likelihood without conserving any physical meaning. Our inferred $v_{r}$ of the three benchmark companions range from $+200$~km~s$^{-1}$ to $+500$~km~s$^{-1}$, which are not consistent with their primary stars' {\it Gaia}~DR2 radial velocities and are even comparable with the local Galactic escape speed \citep[$\approx 540$~km~s$^{-1}$;][]{2007MNRAS.379..755S, 2014A&A...562A..91P}. However, the inferred $v_{r}$ are mostly insignificant from zero ($<2\sigma$) and well within one resolution pixel ($1200-3750$~km~s$^{-1}$) of the prism spectra. Also, the inferred $v\sin{i}$ of these objects are only at $1\sigma$ significance. Finally, we note that while both $v_{r}$ and $v\sin{i}$ are coupled with modeling systematics and cannot be well-constrained from the data, their posteriors have at most weak correlations with (and thereby negligible impact on) the posteriors of the other physical parameters.

\subsubsection{Comparison with Traditional Forward-Modeling Approach}
\label{subsubsec:compare_traditional}
We have also conducted a forward-modeling analysis for the three benchmarks following the traditional approach \citep[e.g.,][]{2008ApJ...678.1372C, 2009ApJ...702..154S, 2010ApJS..186...63R, 2020ApJ...891..171Z}, where we use linear interpolation to generate the model spectrum in between grid points and adopt a diagonal covariance matrix (defined by observed flux uncertainties) for model evaluation. 

We use the cloudless Sonora-Bobcat model atmospheres over their entire parameter space of $[200, 2400]$~K in $T_{\rm eff}$, $[3.25, 5.5]$~dex in $\log{g}$, and $[-0.5, +0.5]$~dex in $Z$, amounting to 1014 grid points. The grid spacing is $25$~K, $50$~K, and $100$~K in $T_{\rm eff}$, $0.25$~dex and $0.5$~dex in $\log{g}$, and $0.5$~dex in $Z$.\footnote{The $T_{\rm eff}$ spacing is $25$~K for $[200, 600]$~K, $50$~K for $[600, 1000]$~K, and $100$~K for $[1000,2400]$~K. The $\log{g}$ spacing varies with metallicity. At $Z=-0.5$, the $\log{g}$ spacing is $0.5$~dex for $[3.5, 5.0]$~dex and $0.25$~dex for $[3.25, 3.5]$~dex. At $Z=0$, the $\log{g}$ spacing is $0.25$~dex. At $Z=+0.5$~dex, the $\log{g}$ spacing is $0.5$~dex for $[5.0, 5.5]$~dex and $0.25$~dex among the rest of points.} Then we determine the six physical parameters $\{T_{\rm eff},\ \log{g},\ Z,\ v_{r},\ v\sin{i},\ \log{\Omega}\}$ with same priors as our Starfish analysis (Section~\ref{subsec:starfish}). At a given set of $\{T_{\rm eff},\ \log{g},\ Z\}$, we use linear interpolation of the flux (conducted in logarithmic units for $T_{\rm eff}$) to generate a single model spectrum. Compared to the spectral emulator of Starfish, linear interpolation provides the exact model spectrum at each grid point but does not provide interpolation uncertainties for spectra in between grid points. We apply $\{v_{r},\ v\sin{i},\ \log{\Omega}\}$ values to the interpolated spectra with the same approach as in Starfish and then compare data to models to compute the residual spectra $\mathbf{R}$ (Equation~\ref{eq:residual}). We then estimate parameters using the following standard covariance matrix $\mathbf{C_{T}}$ and log-likelihood function $\mathcal{L}_{\rm T}$,
\begin{equation} \label{eq:covlike_lin}
\begin{aligned} 
    \mathbf{C_{T}} &= \mathcal{K}^{N}(a_{N} \equiv 1) \\
    \mathcal{L}_{\rm T} &= - \frac{1}{2} \left(\mathbf{R^{T}} \mathbf{C_{T}^{-1}} \mathbf{R} + \ln\lvert\mathbf{C_{T}}\rvert \right) -\frac{n}{2} \ln\left(2\pi\right)  
\end{aligned} 
\end{equation}
Here we construct the covariance matrix $\mathbf{C_{T}}$ with solely the diagonal noise covariance matrix and with $a_{N}$ fixed at $1$. We do not include other covariance hyper-parameters as adopted by Starfish. 

We use {\it emcee} to fit the objects' $1.0-2.5$~$\mu$m spectra with 24 walkers, remove the first 5000 iterations as the burn-in phase, and terminate the fitting process with $3 \times 10^{4}$ iterations so these MCMC chains converge. Following the same approach as in our Starfish-based analysis (Section~\ref{subsubsec:systematics}), we incorporate the systematic error of $180$~${\rm km~s}^{-1}$ into the inferred $v_{r}$ to account for the uncertainty in the wavelength calibration of the SpeX prism data. We also incorporate the systematic error of $0.4\sigma_{H_{\rm MKO}}$ into the inferred $\log{\Omega}$ to account for the uncertainty in flux calibration, where $\sigma_{H_{\rm MKO}}$ is the photometric error in the objects' $H$-band magnitudes. We compute the objects' radii and masses using their parallaxes and $\log{g}$ and $\log{\Omega}$ posteriors, and summarize all inferred parameters and their uncertainties in Table~\ref{tab:results}. 

As shown in Figure~\ref{fig:emulin_results}, the inferred parameters and fitted models derived from our Starfish-based and the traditional methods are well consistent. However, the traditional approach produces artificially small parameter errors, smaller than those of Starfish by factors of $2-15$ in $\{T_{\rm eff},\ \log{g},\ Z,\ R,\ M\}$. We note the larger parameter uncertainties derived from Starfish are not due to the incorporation of the aforementioned emulator-induced systematic errors (which are not applied to the traditional approach; Section~\ref{subsubsec:systematics}), but rather because Starfish propagates the model interpolation uncertainties and the correlated residuals among adjacent wavelengths to the inferred parameters (which are ignored by the traditional approach). Therefore, our Starfish-based forward-modeling analysis produces more realistic error estimates.

\subsubsection{Comparison with Previous Spectroscopic Analyses}
\label{subsubsec:previous}
Table~\ref{tab:results} compares our inferred physical parameters of the three benchmarks to those from previous spectroscopic analyses, which used grid models or the retrieval method\footnote{The compilation of literature values for HD~3651B and GJ~570D is mostly from \cite{2017ApJ...848...83L}, to which we have added a few new studies and more details of the grid models used in literature.}. For HD~3651B and GJ~570D, our fitted $T_{\rm eff}$ are consistent with literature values, especially those using grid models with cloudless and chemical-equilibrium atmospheres. However, our $\log{g}$ and $Z$ are smaller by $\approx 1.2$~dex and $\approx 0.35$~dex, respectively. Surface gravity and metallicity have similar effect on the spectral morphology, as either a high (low) $\log{g}$ or a low (high) $Z$ leads to the same suppressed (enhanced) $K$-band flux in late-T dwarf spectra (Section~\ref{subsec:syn_spec}). As a consequence, these two parameters are degenerate in the spectroscopic analysis (e.g., \citealt{2006ApJ...639.1095B, 2007ApJ...667..537L, 2007ApJ...660.1507L}). As shown in Table~\ref{tab:results}, most past studies of HD~3651B and GJ~570D used grid models with a fixed metallicity at solar abundance or that of their primary stars, and their fitted $\log{g}$ are therefore anchored at a specific $Z$. In comparison, we determine $\log{g}$ and $Z$ simultaneously from the spectral fitting process and therefore our results represent a comprehensive evaluation of the model parameter space. According to the subsequent evolutionary model analysis (Section~\ref{sec:evo}), our fitted $\log{g}$ and $Z$ of these two objects are underestimated, indicating shortcomings of the adopted model atmospheres (see Section~\ref{sec:benchmarking}). Moreover, our analysis implies that using the cloudless Sonora-Bobcat models to analyze late-T dwarf spectra whose metallicities are not known in priori can lead to inaccurate $\log{g}$ and $Z$. 

For Ross~458C, our fitted $\log{g}$ and $Z$ match the literature values and our $T_{\rm eff}$ are consistent with those using cloudless, chemical-equilibrium model atmospheres. However, we note that analyses using cloudy models can better match the data \citep[e.g.,][]{2010ApJ...725.1405B, 2011MNRAS.414.3590B} and they infer $\approx 100$~K cooler $T_{\rm eff}$, which is consistent with our evolutionary model analysis (Section~\ref{sec:evo}). While the specific condensation opacities of some cloud models are perhaps uncertain \citep[iron and silicate clouds versus sulfide clouds; see][]{2012ApJ...756..172M}, it is clear that cloudless model atmospheres cannot fully interpret the spectra of Ross~458C unless disequilibrium processes and reduced vertical temperature gradient are also considered \citep[e.g.,][]{2015ApJ...804L..17T}.

%%%%%%%%%%%%%%%%%%%%%%%%%%%%%%%%%
%------------- Evolutionary Model Analysis Results --------------------
%%%%%%%%%%%%%%%%%%%%%%%%%%%%%%%%%
\begin{figure*}[t]
\begin{center}
\includegraphics[height=6in]{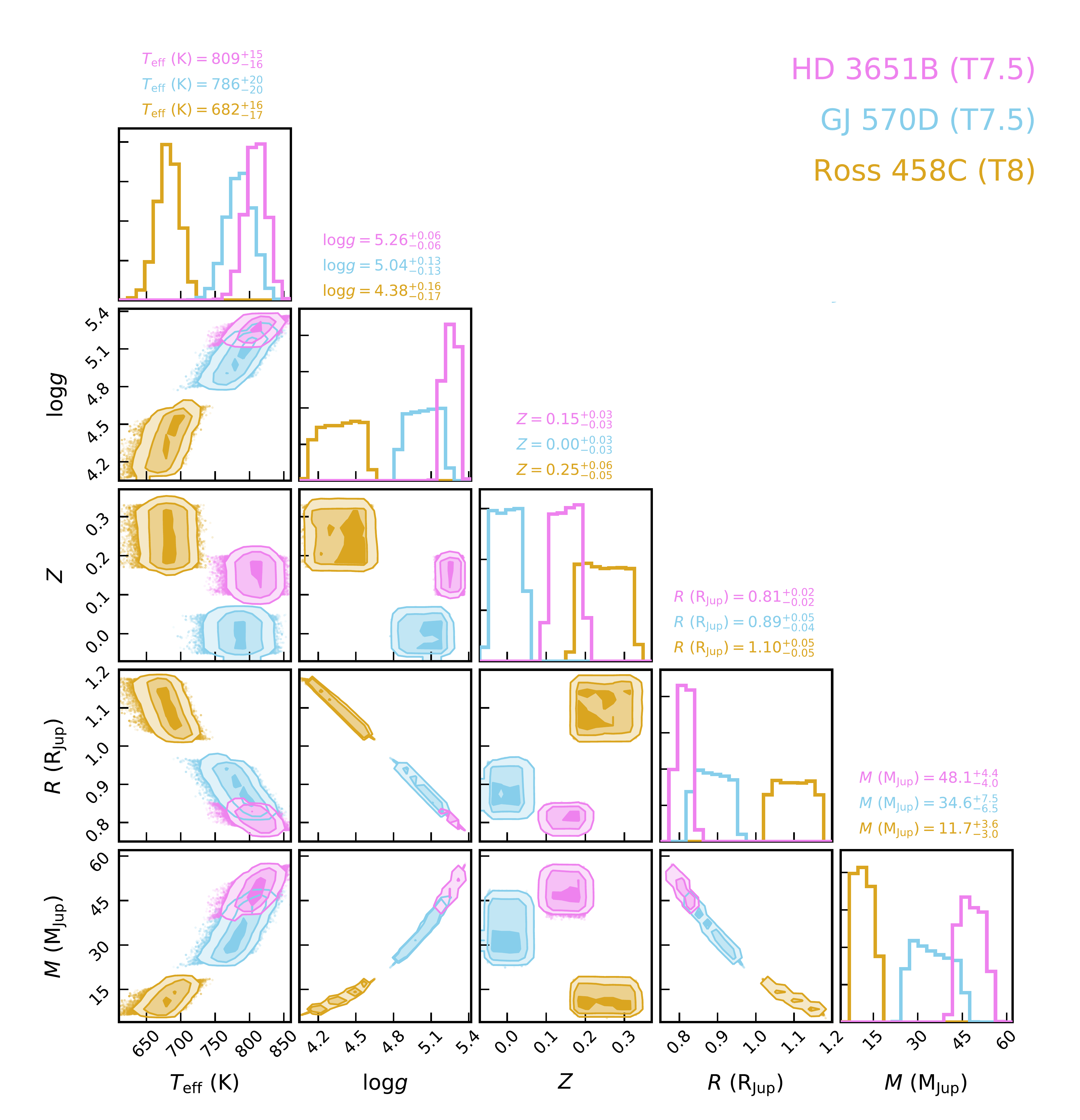}
\caption{Posterior distributions of evolutionary model parameters for HD~3651B (violet), GJ~570D (light blue), and Ross~458C (orange). The median and $1\sigma$ uncertainties of each parameter are labeled over each 1-D histogram, with colors indicating the corresponding objects. The parameters plotted here are effective temperature ($T_{\rm eff}$), logarithmic surface gravity ($\log{g}$), metallicity ($Z$), radius ($R$), and mass ($M$). The companions' $\log{g}$ and $Z$ posteriors are close to uniform distributions because we assume uniform distributions for their primary stars' age and metallicity. }
\label{fig:evo}
\end{center}
\end{figure*}
%------------figure end-----------------

\subsection{Assessment of Modeling Systematics using Starfish Hyper-Parameters}
\label{subsec:assess_model_systematics}

We can assess the modeling systematics using the derived covariance hyper-parameters of the three benchmarks listed in Table \ref{tab:hyper_params}. As described in Section~\ref{subsec:starfish}, the full covariance matrix of Starfish consists of three components, the emulator covariance matrix $\mathcal{K}^{E}$, the noise covariance matrix $\mathcal{K}^{N}$, and the global covariance matrix $\mathcal{K}^{G}$. While $\mathcal{K}^{E}$ has been already pre-determined during the spectral emulator construction, $\mathcal{K}^{N}$ and $\mathcal{K}^{G}$ are computed as part of our spectral-fitting process. The $\mathcal{K}^{N}$ matrix has one hyper-parameter $a_{N}$ and accounts for measurement uncertainties. The $\mathcal{K}^{G}$ matrix is characterized by two hyper-parameters, $\ell$ and $a_{G}$, with the latter describing the amplitude of $\mathcal{K}^{G}$'s normalization (Equation~\ref{eq:kG}). The $\ell$ parameter describes the auto-correlation wavelength scale of the residual and such correlated residuals are caused by (1) the over-sampled instrumental line spread function (LSF) and (2) the modeling systematics. If the latter effect dominates, then we can use these hyper-parameters to examine the performance of the models.

As shown in Table~\ref{tab:hyper_params}, the fitted $\ell$ of all the three benchmarks are $2700-8500$~km~s$^{-1}$ and exceed the expected range of $820-1840$~km~s$^{-1}$ from LSF (Appendix~\ref{app:ell}). Therefore, the correlated residual of our sample are caused by the modeling systematics rather than the instrumental effect. As a consequence, the global covariance matrix $\mathcal{K}^{G}$ mainly describes the model imperfections, and we can use its determinant to quantify such systematics. 

The hyper-parameter $a_{G}$ is a proxy for $\mathcal{K}^{G}$'s normalization. Ideally, if the models match the observations within the measurement uncertainties, then the need for including $\mathcal{K}^{G}$ into the spectral fitting is fairly low and thus the fitted $a_{G}$ should be very small. A larger fitted $a_{G}$ indicates measurement uncertainties cannot solely explain the difference between data and models, and thus Starfish needs $\mathcal{K}^{G}$ with higher values to account for this data-model discrepancy. The meaning of $a_{G}$ can be understood from the definition of the noise covariance $\mathcal{K}^{N}$ (Section~\ref{subsubsec:covhyper_param}), which is a diagonal matrix with values of $a_{N}\sigma_{i}^{2}$, where $\sigma_{i}$ is the flux uncertainty at the $i$-th wavelength pixel. Similarly, the global covariance $\mathcal{K}^{G}$ is a band matrix, with $a_{G}$ on its main diagonal and with much smaller non-zero values along diagonals above or below (Equation~\ref{eq:kG}). Following $\mathcal{K}^{N}$, we can express $a_{G} \equiv a_{N} \sigma_{m}^{2}$ and regard $\sigma_{m}$ as an equivalent flux that characterizes ``model uncertainties''. We can then normalize $\sigma_{m}$ of the object by its observed peak $J$-band flux and define
\begin{equation} \label{eq:epsilon_j}
\epsilon_{J} = \frac{\sqrt{a_{G} / a_{N}}}{{\rm max}\left(f_{{\rm obs}, J}\right) }
\end{equation}
The $\epsilon_{J}$ parameter quantifies the modeling systematics, with higher values suggesting more significant data-model discrepancy. We compute $\epsilon_{J}$ values for the three benchmarks (Table~\ref{tab:hyper_params}) and find that the systematic difference between the cloudless Sonora-Bobcat models and spectra of the three benchmarks comprises $2\%-4\%$ of the observed peak $J$-band fluxes, equivalent to a S/N of $50-25$. This implies that the model uncertainties can exceed measurement uncertainties when fitting the cloudless Sonora-Bobcat models to low-resolution spectra with high S/N ($>50$ in $J$ band) and increasing the S/N of data does not promise enhanced precision of fitted physical parameters.

\section{Evolutionary Model Analysis}
\label{sec:evo}
We also derive physical properties of HD~3651B, GJ~570D, and Ross~458C using cloudless Sonora-Bobcat evolutionary models. We extend the \cite{2006ApJ...647..552S} method from solar- to multi-metallicity evolutionary models. Specifically, we compute $\{T_{\rm eff},\ \log{g},\ Z\}$ at which (1) the companions' interpolated bolometric luminosities from the evolutionary models match the measured luminosities from their near-infrared spectra, their primary stars' distances, and the atmospheric model-based ratio between near-infrared and bolometric fluxes, and (2) the companions' metallicity and the interpolated age from evolutionary models match those of their primary stars (Section~\ref{sec:companions}). We implement this method in a Bayesian fashion following \cite{2020ApJ...891..171Z}.

For each benchmark companion, we first integrate its SpeX spectrum in a wavelength range of $1.0-2.5$~$\mu$m and use its primary star's {\it Gaia}~DR2 distance to derive a near-infrared luminosity, namely $L_{\rm 1.0-2.5\mu m}$, with uncertainties in spectra and distance propagated accordingly. We then execute a MCMC routine with free parameters being $T_{\rm eff}$, $\log{g}$, and $Z$. Given a set of these parameters, we use linear interpolation of the cloudless Sonora-Bobcat models to construct the corresponding synthetic spectrum, from which we compute a ratio of the integrated flux in $1.0-2.5$~$\mu$m to that in $0.4-50$~$\mu$m. We note such ratio is $0-0.3$ for models of $T_{\rm eff} = 200 - 600$~K, $0.3-0.6$ for $T_{\rm eff} = 600-1200$~K, and $0.6-0.7$ for $1200-2400$~K, with an average change per $\{T_{\rm eff},\ \log{g},\ Z\}$ grid spacing of $<0.03$, $<0.04$, and $<0.01$ in each of the above $T_{\rm eff}$ ranges, respectively. We therefore adopt a conservative uncertainty of $0.04$. We apply this ratio to the measured near-infrared luminosity to derive the companion's atmospheric-based bolometric luminosity $L_{\rm bol, atm}$.\footnote{As discussed in Section~\ref{sec:benchmarking}, Sonora-Bobcat models adopt the potassium line shape theory by \cite{2007A&A...465.1085A} and these authors have recently updated their potassium opacities in \cite{2016A&A...589A..21A}. Different treatments of potassium line profiles will lead to different model fluxes in the optical, $Y$, and $J$ bands, potentially impacting the model-based flux ratio and thereby $L_{\rm bol, atm}$ computed in this work. In order to quantify such impact, we examine the ATMO atmospheric models developed by \cite{2020A&A...637A..38P}. Two sets of ATMO model spectra are available (M. Phillips, private communication), which we denote as ``ATMO-A07'' and ``ATMO-A16'' and are computed using the \cite{2007A&A...465.1085A} and \cite{2016A&A...589A..21A} K I profiles, respectively. These models are generated with $T_{\rm eff}$ in $[200, 3000]$~K, $\log{g}$ in $[2.5, 5.5]$~dex, and a solar metallicity. For each grid point of each model set, we compute a ratio of the integrated flux in $1.0-2.5$~$\mu$m to that in $0.4-50$~$\mu$m. For both ATMO-A07 and ATMO-A16 model sets, we find such flux ratio is $0-0.35$ for $T_{\rm eff} = 200-600$~K, $0.35-0.6$ for $T_{\rm eff} = 600-1200$~K, and $0.6-0.75$ for $T_{\rm eff} = 1200-3000$~K, with an average change per $\{T_{\rm eff}, \log{g}\}$ grid spacing of $<0.04$, $<0.04$, and $<0.01$ in each of the above $T_{\rm eff}$ ranges, respectively. Following our analysis using Sonora-Bobcat models, we would adopt a conservative uncertainty of $0.04$ for such flux ratio if we use the ATMO models for computing the $L_{\rm bol, atm}$. In comparison, the absolute difference of flux ratios between the ATMO-A07 and the ATMO-A16 models is $<0.005$ over the entire model grid, significantly smaller than the notional $0.04$ uncertainty. Therefore, the different treatments of potassium line profiles between \cite{2007A&A...465.1085A} and \cite{2016A&A...589A..21A} have a negligible impact to our computed $L_{\rm bol, atm}$. } The uncertainties in $L_{\rm 1.0-2.5\mu m}$ and the model-based flux ratio are both propagated accordingly.

For each set of $\{T_{\rm eff},\ \log{g},\ Z\}$ in the MCMC chains, we also interpolate the Sonora-Bobcat evolutionary models (in logarithmic scales for $T_{\rm eff}$ and age) and compute the companion's evolutionary-based bolometric luminosity $L_{\rm bol, evo}$ and age $t_{\rm evo}$. Altogether, we derive $\{T_{\rm eff},\ \log{g},\ Z\}$ using the following likelihood function,
\begin{equation}
\mathcal{L} = p(L_{\rm bol,atm} \mid L_{\rm bol, evo}) \times p(Z_{\rm \star} \mid Z) \times p(t_{\rm \star} \mid t_{\rm evo}) \times p(T_{\rm eff}) \times p(\log{g}) \times p(Z) 
\end{equation}
where $Z_{\star}$ and $t_{\star}$ stand for the metallicity and age of primary stars summarized in Section~\ref{sec:companions}. We assume the priors, $p(T_{\rm eff})$, $p(\log{g})$, and $p(Z)$ are all uniform distributions within the parameter space of the Sonora-Bobcat models, i.e., $[200, 2400]$~K for $T_{\rm eff}$, $[3.25, 5.5]$~dex for $\log{g}$, and $[-0.5, +0.5]$~dex for $Z$. To compute $p(L_{\rm bol,atm} \mid L_{\rm bol, evo})$, we assume the bolometric luminosity follows a normal distribution corresponding to the value and uncertainty of $L_{\rm bol,atm}$. To compute $p(Z_{\rm \star} \mid Z)$ and $p(t_{\rm \star} \mid t_{\rm evo})$, we assume the metallicity and age of the companion follow the uniform distributions constrained by those of their primary stars. 

We use {\it emcee} to derive $\{T_{\rm eff},\ \log{g},\ Z\}$ values of each companion with 30 walkers, removing the first 1000 iterations as the burn-in phase and terminating with $1.0 \times 10^{4}$ iterations so that our MCMC analysis converges. We use the companions' resulting $\{T_{\rm eff},\ \log{g},\ Z\}$ posteriors and the interpolated cloudless Sonora-Bobcat evolutionary models to derive their radii, masses, and bolometric luminosities. We present the posteriors of their evolutionary model parameters in Figure~\ref{fig:evo} and summarize their values in Table~\ref{tab:comp_evo}.

%%%%%%%%%%%%%%%%%%%%%%%%%%%%%%%%%%%%%%%
%------------- Atmospheric - Evolutionary Parameter Posteriors  --------------------
%%%%%%%%%%%%%%%%%%%%%%%%%%%%%%%%%%%%%%%
\begin{figure*}[t]
\begin{center}
\includegraphics[height=6in]{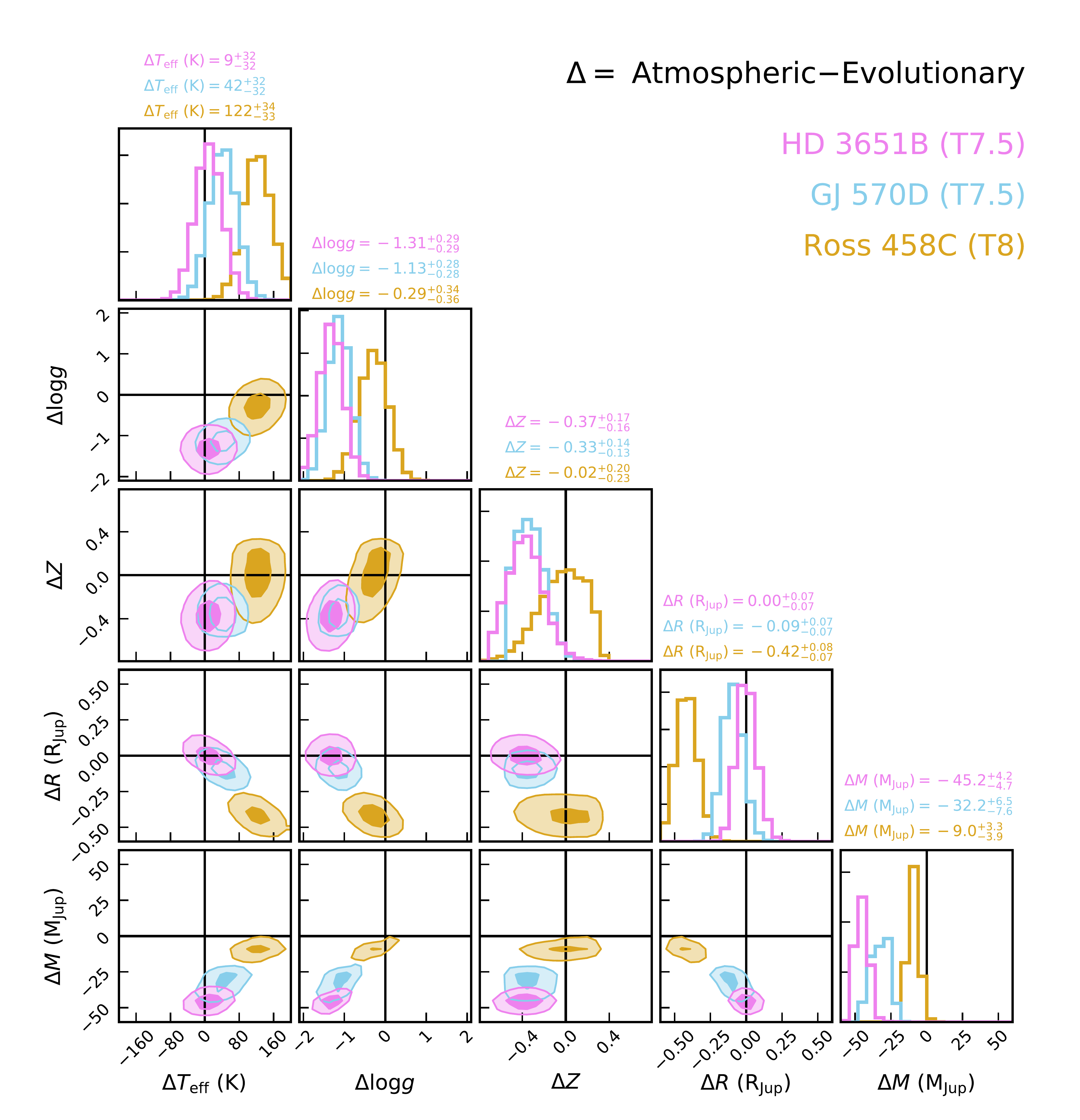}
\caption{Posterior distributions of the atmospheric$-$evolutionary model parameter differences for HD~3651B (violet), GJ~570D (light blue), and Ross~458C (orange) as described in Section~\ref{sec:benchmarking}. We generate these posteriors by subtracting the evolutionary-based MCMC chains from the atmospheric-based chains. The median and $1\sigma$ uncertainties of the parameter differences are labeled over each 1-D histogram, with colors indicating the corresponding objects. The positions of these posteriors suggest the accuracy of model assumption has two outcomes. At $T_{\rm eff} \approx 780-810$~K and $\log{g} \approx 5.0-5.3$~dex (corresponding to HD 3651B and GJ 570D), spectral fits produce robust $T_{\rm eff}$ and $R$ but underestimated $\log{g}$ (by $\approx 1.1-1.3$~dex) and $Z$ (by $\approx 0.3-0.4$~dex). At $T_{\rm eff} \approx 700$~K and $\log{g} \approx 4.4$~dex, spectral fits provide robust $\log{g}$ and $Z$ but overestimated $T_{\rm eff}$ (by $\approx 120$~K) and underestimated $R$ (by $\approx 0.4$~R$_{\rm Jup}$ or $\approx 1.6\times$).}
\label{fig:atm_minus_evo}
\end{center}
\end{figure*}
%------------figure end-----------------

\section{Benchmarking}
\label{sec:benchmarking}
The three benchmarks now have the same set of physical parameters $\{T_{\rm eff},\ \log{g},\ Z,\ R,\ M\}$ derived from both atmospheric models (Section~\ref{sec:atm}) and evolutionary models (Section~\ref{sec:evo}). Ideally, the results from these two approaches would be consistent, but they are not. The evolutionary model parameters are expected to be more robust \citep[though not totally immune from the systematics; e.g.,][]{2009ApJ...692..729D, 2014ApJ...790..133D, 2018AJ....156..168B, 2020AJ....160..196B}, given that the consistency between observations and evolutionary model predictions has been found in many systems including ultracool dwarfs with dynamical masses \citep[e.g.,][]{2017ApJS..231...15D, 2019AJ....158..140B} and transiting brown dwarfs with directly measured radii \citep[e.g.,][]{2016ApJ...822L...6M, 2020AJ....160...53C, 2020NatAs...4..650T}. Also, the atmospheric model predictions have well-noticed shortcomings \citep[e.g.,][]{2008ApJ...678.1372C, 2009ApJ...702..154S, 2017ApJ...842..118L}. As a consequence, comparing physical parameters derived from these two model sets, a.k.a. the ``benchmarking'' process \citep[e.g.,][]{2006MNRAS.368.1281P, 2008ApJ...689..436L}, can calibrate our inferred atmospheric model parameters and shed light on modeling systematics.

To compare the results from atmospheric and evolutionary models, we subtract the values of each object's evolutionary-based MCMC chains from those of its atmospheric-based chains\footnote{Given that the number of MCMC samples of our atmospheric model analysis ($48$ walkers $\times 10^{5}$ iterations) is 16 times larger than that of our evolutionary model analysis ($30$ walkers $\times 10^{4}$ iterations), we simply repeat the chains of the latter 16 times and then subtract the two to generate the posteriors. We also generated posteriors of parameter differences using $1/16$ of the atmospheric-based MCMC chains and obtained consistent results.}. Distributions of the resulting parameter differences are shown in Figure~\ref{fig:atm_minus_evo}, with values summarized in Table~\ref{tab:comp_evo}.  

HD~3651B has the highest evolutionary-based $T_{\rm eff}$ ($809^{+15}_{-16}$~K) and $\log{g}$ ($5.26 \pm 0.06$~dex) among the three benchmarks. Its $T_{\rm eff}$ and $R$ derived from the atmospheric and evolutionary models are consistent within uncertainties, but its spectroscopically inferred $\log{g}$ and $Z$ are lower by $1.31 \pm 0.29$~dex and $0.37^{+0.17}_{-0.16}$~dex, respectively. The inaccurate atmospheric $\log{g}$ also leads to the object's very small inferred mass ($2.3^{+2.3}_{-1.1}$~M$_{\rm Jup}$).

GJ~570D has a slightly lower evolutionary-based $T_{\rm eff}$ ($786 \pm 20$~K) and $\log{g}$ ($5.04 \pm 0.13$~dex) than HD~3651B. The spectroscopically inferred $\log{g}$ and $Z$ are underestimated by $1.13 \pm 0.28$~dex and $0.33^{+0.14}_{-0.13}$~dex, respectively. In addition, our atmospheric model analysis slightly overestimates this object's $T_{\rm eff}$ by $42 \pm 32$~K and underestimates $R$ by $0.09 \pm 0.07$~R$_{\rm Jup}$ (or a factor of $1.12^{+0.10}_{-0.09}$). The inaccurate atmospheric $\log{g}$ and $R$ together lead to its small inferred mass ($2.0^{+1.7}_{-0.9}$~M$_{\rm Jup}$).

Ross~458C has the lowest evolutionary-based $T_{\rm eff}$ ($682^{+16}_{-17}$~K) and $\log{g}$ ($4.37^{+0.16}_{-0.17}$~dex). This object has consistent $\log{g}$ and $Z$ values from the two sets of models. However, its spectroscopically inferred $T_{\rm eff}$ is overestimated by $122^{+34}_{-33}$~K and $R$ is underestimated by $0.42^{+0.08}_{-0.07}$~R$_{\rm Jup}$ (or a factor of $1.61^{+0.16}_{-0.15}$), which lead to its very small inferred mass ($2.3^{+2.3}_{-1.2}$~M$_{\rm Jup}$).

To summarize, our atmospheric model analysis are in line with the evolutionary model analysis, but some parameters of the former appear to be under- or over-estimated. We find the parameter difference between these two sets of models exhibits two outcomes. At $T_{\rm eff} \approx 780-810$~K and $\log{g} \approx 5.0-5.3$~dex (corresponding to HD~3651B and GJ~570D), spectral fits produce robust $T_{\rm eff}$ and $R$, but underestimate $\log{g}$ and $Z$ by $\approx 1.1-1.3$~dex and $\approx 0.3-0.4$~dex, respectively. Going toward a cooler $T_{\rm eff} \approx 700$~K and a lower $\log{g} \approx 4.4$~dex (corresponding to Ross~458C), spectral fits produce robust $\log{g}$ and $Z$, but overestimate $T_{\rm eff}$ by $\approx 120$~K and underestimate $R$ by $\approx 0.4$~R$_{\rm Jup}$ (or a factor of $\approx 1.6$). For all three benchmarks, our spectroscopically inferred masses are thereby significantly underestimated. Analysis of more benchmark systems will allow us to further investigate the consistency between atmospheric and evolutionary models as a function of physical parameters, which will help quantify the specific modeling systematics and calibrate results of such forward-modeling analysis.  

While the evolutionary model parameters are considered more robust throughout our analysis, we first nevertheless examine whether the different atmospheric-based and evolutionary-based parameters of our sample can be solely explained by any systematics of the evolutionary models, which has been speculated for a small sample of ultracool benchmarks. In the T dwarf regime, recent work \citep[e.g.,][]{2018A&A...614A..16C, 2020AJ....160..196B} has found the measured dynamical masses of some brown dwarf companions to stars are more massive than evolutionary model predictions, derived based on the companions' bolometric luminosities and their host stars' ages. The reason of this discrepancy is unclear, as it might be caused by unknown missing components in the evolutionary models, inaccurate ages inferred from the host stars, or undetected additional companions in the system. Assuming the systematics of evolutionary models is the main reason and the three benchmarks in this work share similar systematics as those T dwarfs with dynamical masses, then our inferred evolutionary-based masses are all underestimated. Correcting for such an underestimation will further worsen the disagreement between the atmospheric-based and evolutionary-based masses for our benchmarks.  

In the L dwarf regime, \cite{2009ApJ...692..729D} and \cite{2014ApJ...790..133D} measured dynamical masses of two mid-L brown dwarf binaries which are both companions to solar-type stars. They found the dynamical masses of these L dwarfs are $10-25\%$ smaller than the evolutionary model predictions derived from the these objects' bolometric luminosities and their host stars' ages. \cite{2014ApJ...790..133D} speculated that such discrepancy might be resolved if the evolutionary models of L dwarfs could incorporate the onset of cloud clearing at appropriate time during the evolution. Our three late-T benchmarks are in a different physical regime compared to L dwarfs. However, even if we correct for a notional $10-25\%$ overestimation in the evolutionary-based masses of our benchmarks, we still find a significant tension of $6.8-8.3\sigma$, $3.1-3.8\sigma$, and $1.5-1.9\sigma$ between the atmospheric-based and evolutionary-based masses for HD~3651B, GJ~570D, and Ross~458C, respectively. Therefore, the different atmospheric-based and evolutionary-based physical parameters of our benchmark companions are very likely caused by the shortcomings of the atmospheric models.

%%%%%%%%%%%%%%%%%%%%%%%
%------------- NIR evo.+atm. vs. data --------------------
%%%%%%%%%%%%%%%%%%%%%%%
\begin{figure*}[t]
\begin{center}
\includegraphics[height=5in]{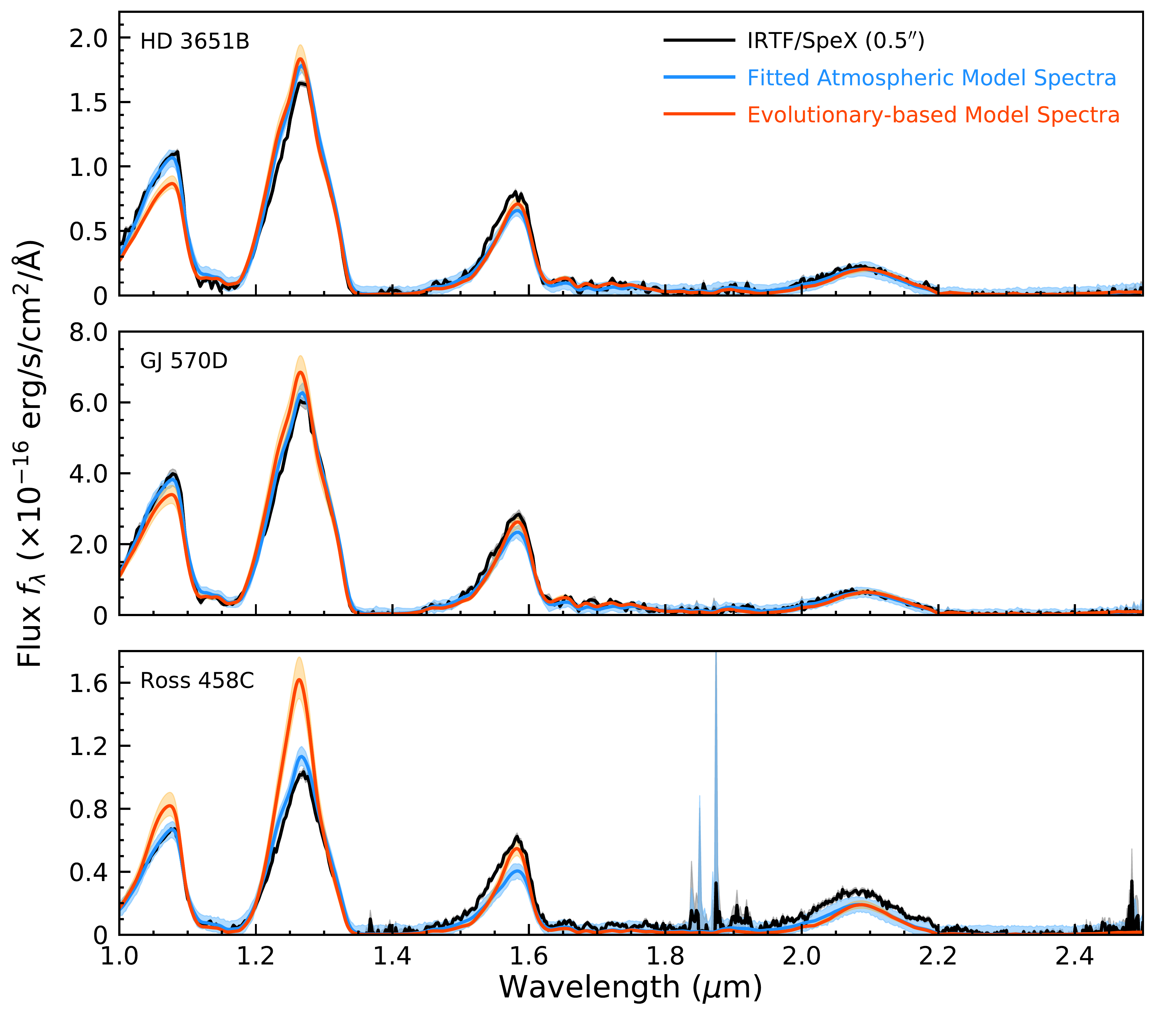}
\caption{The atmospheric model spectra (red) and their $1\sigma$ uncertainties (orange shadow) generated and scaled at the evolutionary-based $\{T_{\rm eff},\ \log{g},\ Z,\ R\}$ of three benchmark companions, HD~3651B (top), GJ~570D (middle), and Ross~458C (bottom). We also show the observed spectra (black) and fitted atmospheric model spectra (blue) as in Figure~\ref{fig:emulin_results}.  }
\label{fig:nir_evo_atm_data}
\end{center}
\end{figure*}
%------------figure end-----------------

%%%%%%%%%%%%%%%%%%%%%%%
%------------- FWL evo.+atm. vs. data --------------------
%%%%%%%%%%%%%%%%%%%%%%%
\begin{figure*}[t]
\begin{center}
\includegraphics[height=6.in]{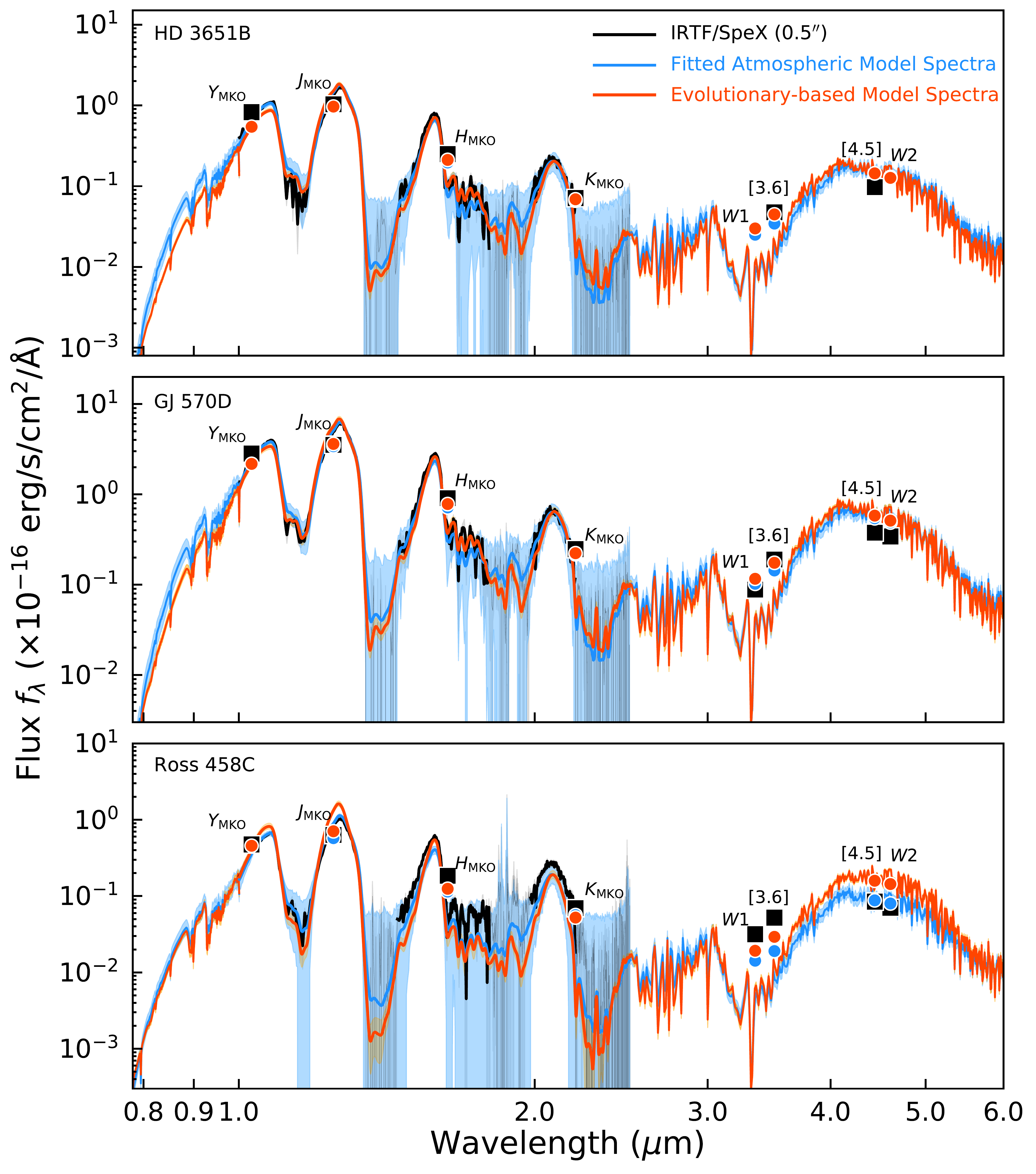}
\caption{The observed spectra (black), the fitted atmospheric model spectra (blue), and the evolutionary-based model spectra (red) of three benchmark companions, using the same format as Figure~\ref{fig:nir_evo_atm_data}. Thinner black lines are used to plot low-S/N regions of the observed spectra. Black squares show the observed fluxes derived from photometry, and red and blue circles show the synthetic fluxes from evolutionary-based and fitted atmospheric model spectra, respectively. We compute these fluxes using the filter responses and zero-point fluxes from \cite{2006MNRAS.367..454H} and \cite{2007MNRAS.379.1599L} for MKO photometry, \cite{2011ApJ...735..112J} for {\it WISE}, and the IRAC Instrument Handbook for {\it Spitzer}/IRAC. }
\label{fig:fwl_evo_atm_data}
\end{center}
\end{figure*}
%------------figure end-----------------

We next investigate the possible shortcomings of the Sonora-Bobcat atmospheric models. For each object, we use the Starfish spectral emulator to generate atmospheric model spectra from posteriors of its evolutionary-based $\{T_{\rm eff},\ \log{g},\ Z,\ R\}$ and its primary star's distance. If the atmospheric models are perfect, then these evolutionary-based model spectra should exactly match the observed spectrophotometry of each companion. In other words, inconsistencies between these two sets of spectra can inform the specific modeling systematics.  

Figure~\ref{fig:nir_evo_atm_data} compares the observed and evolutionary-based model spectra of three benchmarks at $1.0-2.5$~$\mu$m, and Figure~\ref{fig:fwl_evo_atm_data} extends such comparisons to $0.8-6.0$~$\mu$m\footnote{We use the spectral emulator to generate $0.8-2.5$~$\mu$m spectra and use the linear interpolation for $2.5-6.0$~$\mu$m, over which the spectral emulator is not defined. In $0.8-2.5$~$\mu$m, spectra of the three companions generated from the linear interpolation are consistent with those generated from our spectral emulator within uncertainties.}. For HD~3651B and GJ~570D, the model spectra are consistent with data in $JHK$ bands, but produce notably fainter fluxes in $Y$ band. This is likely related to the potassium resonance doublet at $0.77$~$\mu$m \citep{1999PhRvA..60.1021A, 2000ApJ...531..438B}, whose pressure-broadened wings can extend to $Y$ and $J$ bands. The cloudless Sonora-Bobcat models adopt the K line shape theory by \cite{2007A&A...465.1085A} and these authors have recently produced improved calculations of the K$-$H$_{2}$ potential \citep[][]{2016A&A...589A..21A}. As demonstrated by \cite{2020A&A...637A..38P}, model atmospheres with the \cite{2016A&A...589A..21A} prescription of K profiles can better match the observed $Y$-band spectra of GJ~570D than those of \cite{2007A&A...465.1085A}. In addition, the evolutionary-based model spectra of HD~3651B and GJ~570D have slightly brighter fluxes in $W2$ and $[4.5]$ bands, suggesting the disequilibrium abundance of CO in their atmospheres \citep[e.g.,][]{2014ApJ...797...41Z}.

The data-model mismatch of HD~3651B and GJ~570D might explain why their $\{\log{g},\ Z\}$ derived from our forward-modeling analysis are underestimated. Since the evolutionary-based model spectra have under-predicted $Y$-band fluxes, the fitting process favors a much lower $\log{g}$ in order to boost the $Y$-band emission (at a given $T_{\rm eff}$ and $Z$) to match the data (e.g., Figure~\ref{fig:mod_spec}). The decreased $\log{g}$ as such further causes a lower $Z$, given the $\log{g}-Z$ degeneracy (Section~\ref{subsubsec:previous}), to ensure the data and models remain consistent in other bands (e.g., $K$ band). 

For Ross~458C, the evolutionary-based model spectra have significantly brighter fluxes in $YJ$ bands and fainter fluxes in $K$ band, suggesting the models should probably include clouds to properly interpret the object's spectra \citep[e.g.,][]{2012ApJ...756..172M}. Alternatively, \cite{2015ApJ...804L..17T} proposed that the cloudless disequilibrium models with reduced vertical temperature gradient (as compared to the adiabatic thermal structure) can predict redder near-infrared spectra than the chemical equilibrium models and better match the observations. We note the evolutionary-based models of Ross~458C have fattener fluxes in the blue wing of $H$ band, likely due to the disequilibrium chemistry of NH$_{3}$ which is not included in the cloudless Sonora-Bobcat models. Convective mixing can disturb the chemical equilibrium between N$_{2}$ and NH$_{3}$ in the upper atmosphere, reducing the amount of NH3 and thereby weakening absorption in the blue wing of H band \citep[e.g.,][]{1994Icar..110..117F, 2006ApJ...647..552S, 2012ApJ...750...74S, 2011ApJ...743...50C, 2014ApJ...797...41Z, 2015ApJ...804L..17T}. Also, the brighter $W2$ and $[4.5]$ fluxes as predicted by model spectra are likely due to the disequilibrium mixing of CO.

Again, we can use the data-model mismatch of Ross~458C to explain why its $T_{\rm eff}$ and $R$ derived from our forward-modeling analysis are over- and under-estimated, respectively. The evolutionary-based model spectra have significantly bluer near-infrared colors (e.g., $Y-K$, $J-K$, $J-H$) than the data. Therefore, the spectral fitting is forced to choose models with redder colors to match the data. Among all three grid parameters, $T_{\rm eff}$ is the primary modulator of near-infrared colors for late-T dwarfs. Consequently, a higher $T_{\rm eff}$ than the evolutionary-based value is favored during the fitting process, which leads to a smaller $R$ to preserve the integrated flux of spectra.

\section{Summary and Future Work}
\label{sec:summary}
We have constructed a forward-modeling framework to analyze low-resolution ($R \approx 80 - 250$) near-infrared ($1.0-2.5$~$\mu$m) spectra of T dwarfs. We have extended the Bayesian inference tool Starfish \citep[][]{2015ApJ...812..128C} to the brown dwarf regime for the first time, using state-of-art, cloudless Sonora-Bobcat model atmospheres spanning $T_{\rm eff} = 600 - 1200$~K, $\log{g} = 3.25 - 5.5$~dex, and $Z = -0.5$, $0$, $+0.5$~dex. Starfish has two features that can lead to more robust parameters and more realistic error estimates than traditional ($\chi^{2}$-based) forward-modeling analyses. First, Starfish's spectral emulator generates a probability distribution of interpolated spectra at any model grid location, with the associated interpolation uncertainties then propagated into the resulting inferred parameters. Second, Starfish constructs a covariance matrix with off-diagonal components to account for correlated residuals caused by the instrumental effect and modeling systematics. Both of these two aspects are not accounted for in traditional analyses.

In our forward-modeling framework, we verified that the spectral emulator can reproduce the original Sonora-Bobcat models, with only slight flux differences that are smaller than the systematics of Sonora-Bobcat models. To further validate our approach, we treated the original model spectra as if they were on-sky data and used Starfish to infer the properties of each model spectrum. We found negligible offsets between derived and input physical parameters and thereby confirmed the emulator-induced flux differences from the original models do not bias our results.

We then applied our forward-modeling framework to three benchmark late-T dwarfs, HD~3651B, GJ~570D, and Ross~458C, which are wide-orbit companions to main-sequence stars. We derived these objects' effective temperatures ($T_{\rm eff}$), surface gravities ($\log{g}$), metallicities ($Z$), radii ($R$), and masses ($M$). Our fitted atmospheric models generally match the observed spectra of the three benchmarks. However, residuals are seen in $J$ and $H$ bands, likely arising from clouds and/or reductions of the vertical temperature gradient in the atmospheres, as neither of these two effects are incorporated in the cloudless Sonora-Bobcat models. The data-model discrepancy in the blue wing of $H$ band might also be related to the disequilibrium abundance of NH$_{3}$, which would be less than the amount assumed by the equilibrium chemistry of Sonora-Bobcat.

Our derived physical parameters are consistent with those derived from the traditional spectral-fitting approach and from previous spectroscopic studies. The parameter uncertainties from the traditional method are implausibly small, while the error estimates from our Starfish-based analysis are a factor of $2-15$ larger in $\{T_{\rm eff},\ \log{g},\ Z,\ R,\ M\}$ and more realistic. Our resulting uncertainties in $T_{\rm eff}$, $\log{g}$, and $Z$ are typically about $1/3-1/2$ of the Sonora-Bobcat model grid spacing. 

In addition, based on the fitted covariance hyper-parameters from Starfish, we find the systematic difference between the late-T dwarf spectra and the cloudless Sonora-Bobcat models comprises about $2\% - 4\%$ of the objects' observed peak $J$-band fluxes, equivalent to a S/N of $50-25$. Consequently, when using the cloudless Sonora-Bobcat models, increasing the S/N of observed spectra beyond 50 in $J$ band will not lead to enhanced precision of fitted physical parameters, given that the model uncertainties will exceed the measurement uncertainties in the forward-modeling analysis.

We also used the Sonora-Bobcat evolutionary models to derive these benchmarks' physical properties, based on their bolometric luminosities, and their primary stars' metallicities and ages. As a result, these benchmarks have $\{T_{\rm eff},\ \log{g},\ Z,\ R,\ M\}$ derived from both the atmospheric models (via our forward-modeling analysis) and the evolutionary models. Given the consistency between observations and evolutionary model predictions found for many ultracool dwarfs with dynamical masses and/or directly measured radii, we assume our evolutionary-based parameters are more robust and then used them to test the accuracy of spectral fits. We find the parameter difference between atmospheric and evolutionary models exhibits two outcomes. At $T_{\rm eff} \approx 780-810$~K and $\log{g} \approx 5.0-5.3$~dex (corresponding to HD~3651B and GJ~570D), spectral fits produce robust $T_{\rm eff}$ and $R$, but underestimate $\log{g}$ and $Z$ by $\approx 1.1-1.3$~dex and $\approx 0.3-0.4$~dex, respectively. Going toward a cooler $T_{\rm eff} \approx 700$~K and a lower $\log{g} \approx 4.4$~dex (corresponding to Ross~458C), spectral fits produce robust $\log{g}$ and $Z$, but overestimate $T_{\rm eff}$ by $\approx 120$~K and underestimate $R$ by $\approx 0.4$~R$_{\rm Jup}$ (or a factor of $\approx 1.6$). The spectroscopically inferred masses of these benchmarks are all underestimated.

In order to investigate the possible shortcomings of the cloudless Sonora-Bobcat models, we generated atmospheric model spectra at the evolutionary-based $\{T_{\rm eff},\ \log{g},\ Z,\ R\}$ of the three benchmarks and then compared with their observed spectrophotometry. For HD 3651B and GJ 570D, the evolutionary-based model spectra have fainter fluxes in $Y$ band and slightly brighter fluxes in $W2$ and $[4.5]$ bands, suggesting the modeling systematics mainly come from the uncertainties of potassium line profiles and the lack of disequilibrium abundance of CO. For Ross 458C, the evolutionary-based model spectra have brighter fluxes in $Y$, $J$, $W2$ and $[4.5]$ bands and fainter fluxes in the blue wing of $H$ band, $K$, $W1$, and $[3.6]$ bands, suggesting the modeling systematics mainly come from the lack of clouds, reductions of vertical temperature gradient, and disequilibrium chemistry of CO/CH$_{4}$ and N$_{2}$/NH$_{3}$.

In a companion paper, we apply our forward-modeling analysis to a much larger sample of late-T dwarf spectra and investigate their population properties. Unlike benchmark companions, metallicities and ages of most field dwarfs are unknown, but we can examine their spectral-fitting residuals as a function of wavelength and atmospheric properties to test model atmospheres. Also, thanks to the flexibility of Starfish, our forward-modeling framework can be conveniently extended to earlier/later spectral types, wider/narrower wavelength coverages, and different spectral resolutions, leading to a stronger understanding of the precision and accuracy of model assumption.

Finally, we emphasize the value of benchmark systems, including wide-orbit companions \citep[e.g.,][]{2006MNRAS.368.1281P, 2014ApJ...792..119D, 2020ApJ...889..176F, 2020ApJ...891..171Z}, members of nearby associations \citep[e.g.,][]{2009ApJ...703..399L, 2012ApJ...758...31L, 2013MNRAS.431.3222L, 2016ApJ...833...96L, 2016ApJS..225...10F, 2018ApJ...856...23G, 2018ApJ...858...41Z, 2021arXiv210205045Z, 2020ApJ...892..122J}, and substellar binaries with dynamical masses \citep[e.g.,][]{2008ApJ...689..436L, 2017ApJS..231...15D, 2019AJ....158..174D, 2018AJ....155..159B, 2019AJ....158..140B, 2020AJ....160..196B}. As exemplified by HD 3651B, GJ 570D, and Ross 458C, each benchmark can have its physical properties determined independently from spectral fitting and thus can be used to validate atmospheric models in a specific part of the parameter space. Continued discoveries and analyses of these systems with diverse ages, compositions, and masses will therefore test and help improve model atmospheres over an unprecedentedly large extent of physical parameters.

\acknowledgments
We thank Mark Phillips, Didier Saumon, Caroline Morley, Eugene Magnier, Paul Molli\`{e}re, Joe Zalesky, Trent Dupuy, Ehsan Gharib-Nezhad for insightful discussions and comments on this work. We thank Ian Czekala, Michael Gully-Santiago, and Miles Lucas for helpful discussions about Starfish. We also thank Michael Gully-Santiago for implementing Starfish for IRTF/SpeX prism data and sharing initial work on spectroscopic analysis for T dwarfs (\url{https://github.com/gully/jammer-Gl570D}). This work benefited from the 2017--2019 Exoplanet Summer Program in the Other Worlds Laboratory (OWL) at the University of California, Santa Cruz, a program funded by the Heising-Simons Foundation. M.C.L. acknowledges National Science Foundation (NSF) grant AST-1518339. The advanced computing resources from the University of Hawaii Information Technology Services -- Cyberinfrastructure are greatly acknowledged, and Z. Z. thanks the technical support received from Curt Dodds. This work presents results from the European Space Agency (ESA) space mission {\it Gaia}. {\it Gaia} data are being processed by the {\it Gaia} Data Processing and Analysis Consortium (DPAC). Funding for the DPAC is provided by national institutions, in particular the institutions participating in the {\it Gaia} MultiLateral Agreement (MLA). The {\it Gaia} mission website is https://www.cosmos.esa.int/gaia. The {\it Gaia} archive website is https://archives.esac.esa.int/gaia. This research was greatly facilitated by the TOPCAT software written by Mark Taylor (http://www.starlink.ac.uk/topcat/). Finally, the authors wish to recognize and acknowledge the very significant cultural role and reverence that the summit of Maunakea has always had within the indigenous Hawaiian community.  We are most fortunate to have the opportunity to conduct observations from this mountain.

\facilities{IRTF (SpeX)}

\software{Starfish \citep{2015ApJ...812..128C}, {\it emcee} \citep[][]{2013PASP..125..306F}, Spextool \citep[][]{2004PASP..116..362C}, TOPCAT \citep[][]{2005ASPC..347...29T}, Astropy \citep{2013A&A...558A..33A, 2018AJ....156..123A}, IPtyhon \citep{PER-GRA:2007}, Numpy \citep{numpy},  Scipy \citep{scipy}, Matplotlib \citep{Hunter:2007}.}

\appendix

\section{Notes on the Global Covariance Hyper-Parameters}
\label{app:ell}
In this Appendix, we describe the meaning of the global covariance hyper-parameters in Starfish and their implications for the systematics in model assumption.  As described in Section~\ref{subsubsec:covhyper_param}, Starfish uses the global covariance matrix $\mathcal{K}^{G}$ to characterize the correlation in residuals among adjacent pixels, as caused by (1) the over-sampled instrumental line spread function and (2) the systematics in model atmospheres. This covariance matrix has two hyper-parameters $\{a_{G},\ \ell\}$, with each matrix element expressed as follows (also see Equation~9$-$12 of C15)\footnote{The expression of $\mathcal{K}_{ij}^{G}$ in Equation~\ref{eq:kG} is adopted by Starfish~v0.2 (used in this work) and is slightly different from the original definition of C15 (their Equation~9$-$12) where
\begin{displaymath}
\begin{aligned} 
    r_{ij} &= \frac{c}{2} \bigg| \frac{\lambda_{i} - \lambda_{j}}{\lambda_{i} + \lambda_{j}} \bigg|  \\
    r_{0} &= 4\ell
\end{aligned} 
\end{displaymath}
These differences should have negligible impact for the inference of the objects' physical properties.}:
\begin{equation} \label{eq:kG}
\begin{aligned} 
    \mathcal{K}_{ij}^{G} &= w_{ij} a_{G} \left(1 + \frac{\sqrt{3} r_{ij}}{\ell}\right)\ {\rm exp} \left(- \frac{\sqrt{3} r_{ij}}{\ell}\right) \\
    \text{where} \quad r_{ij} &= \frac{c}{\lambda_{i}} \bigg|\lambda_{i} - \lambda_{j}\bigg| \\
                         w_{ij} &=
    \begin{cases}
      \left[1 + \cos \left(\pi r_{ij}/ r_{0}\right)\right]/2 & r_{ij} \leqslant r_{0}  \\
      0 & r_{ij} > r_{0}
    \end{cases}    \\
    r_{0} &= 6\ell
\end{aligned} 
\end{equation}
Here $c$ is the speed of light and $i,j \in \{1,2,3,\dots,n\}$, where $n$ is number of wavelength pixels in the spectra. Consequently, $\mathcal{K}^{G}$ is a band matrix, with $a_{G}$ on its main diagonal (where $i=j$) and with much smaller non-zero values located along diagonals above or below followed by a truncation at $r_{ij} = r_{0}$. The hyper-parameter $a_{G}$ is therefore a proxy of $\mathcal{K}^{G}$'s normalization or determinant. As described in Section~\ref{subsec:assess_model_systematics}, the fitted $a_{G}$ values from the forward-modeling analysis inform uncertainties of model atmospheres.

The other hyper-parameter $\ell$ describes the wavelength scale over which $\mathcal{K}^{G}$ decreases exponentially along anti-diagonal directions. $\ell$ has units of km~s$^{-1}$ (same as the $r_{ij}$ in Equation~\ref{eq:kG}) and describes such auto-correlation wavelength in an equivalent velocity space. In order to better understand the meaning of $\ell$ values as inferred from our forward-modeling analysis, here we derive a conversion between $\ell$ and the auto-correlation wavelength. 

We first convert the covariance matrix $\mathcal{K}^{G}$ into a correlation matrix $\mathcal{R}^{G}$, namely $\mathcal{R}_{ij}^{G} \equiv \mathcal{K}_{ij}^{G} / a_{G}$, which only depends on $\ell$. Then we compute the autocorrelation as a function of wavelength offset $\Delta\lambda$, 
\begin{equation} \label{eq:corr_G}
\mathcal{R}^{G}(\Delta\lambda; \ell) = \sum_{i=1}^{n} \mathcal{R}_{ij}^{G} (\ell)\quad \text{such that}\ \lambda_{i} - \lambda_{j} = \Delta\lambda,\ \ \text{for}\ j \in \{1,\dots,n\}
\end{equation}
Using the typical wavelength grid from the spectra in our late-T dwarf sample, we construct $\mathcal{K}^{G}$ using Equation~\ref{eq:kG} and compute $\mathcal{R}^{G}(\Delta\lambda; \ell)$ using Equation~\ref{eq:corr_G} for a set of different $\ell$ values and then derive the autocorrelation wavelength $\tau$ for each $\ell$ from the half width at half maximum (HWHM) of $\mathcal{R}^{G}(\Delta\lambda; \ell)$. We finally fit a line to the computed $\tau$ and their corresponding $\ell$ values and obtain
\begin{equation} \label{eq:tau_ell}
\frac{\tau}{1\ \textup{\AA}} = 0.0507 \times \frac{\ell}{1\ \rm km~s^{-1}} + 8.5329
\end{equation}
Using the Equation~\ref{eq:tau_ell}, we can thus converts the derived $\ell$ of each late-T dwarf from our forward-modeling analysis into an auto-correction wavelength.

Equation~\ref{eq:tau_ell} can also help determine the $\ell$ values that correspond to the prism mode line spread function. Based on the wavelength-dependent spectral resolution of SpeX's prism mode \citep[][]{2003PASP..115..362R}, we derive that the autocorrelation wavelength of the $0.5''$ slit from $1.0-2.5$~$\mu$m spans $50-102$~\AA\ (with a median of $83$~\AA), which is equivalent to $2-3$~wavelength pixels. We thus use Equation~\ref{eq:tau_ell} to convert such autocorrelation wavelength into $\ell = 820 - 1840$~km~s$^{-1}$ for the $0.5''$ slit.

In our forward-modeling analysis, we set the prior of $\ell$ based on the prism-mode line spread function. We assume a uniform prior of $[820,\ 1840 \times 5]$ in $\ell$ for spectra taken in the $0.5''$ slit and here we multiply 5 to the maximum $\ell$ as expected from the line spread function, in order to capture the correlation in residuals caused by modeling systematics. Therefore, comparing the resulting $\ell$ values of our late-T dwarfs to the range expected from the instrumental line spread function, we can examine whether the systematics in model atmospheres is playing a crucial role in interpreting the residuals (see Section~\ref{subsec:assess_model_systematics}).

\end{CJK*}

%%%%%%%%%%%%%%%
%%% BIBLIOGRAPHY %%%
%%%%%%%%%%%%%%%
\clearpage
\bibliographystyle{aasjournal}
\bibliography{ms}

\begin{thebibliography}{}
\expandafter\ifx\csname natexlab\endcsname\relax\def\natexlab#1{#1}\fi
\providecommand{\url}[1]{\href{#1}{#1}}
\providecommand{\dodoi}[1]{doi:~\href{http://doi.org/#1}{\nolinkurl{#1}}}
\providecommand{\doeprint}[1]{\href{http://ascl.net/#1}{\nolinkurl{http://ascl.net/#1}}}
\providecommand{\doarXiv}[1]{\href{https://arxiv.org/abs/#1}{\nolinkurl{https://arxiv.org/abs/#1}}}

\bibitem[{{Ackerman} \& {Marley}(2001)}]{2001ApJ...556..872A}
{Ackerman}, A.~S., \& {Marley}, M.~S. 2001, \apj, 556, 872,
  \dodoi{10.1086/321540}

\bibitem[{{Aguilera-G{\'o}mez} {et~al.}(2018){Aguilera-G{\'o}mez},
  {Ram{\'\i}rez}, \& {Chanam{\'e}}}]{2018AandA...614A..55A}
{Aguilera-G{\'o}mez}, C., {Ram{\'\i}rez}, I., \& {Chanam{\'e}}, J. 2018, \aap,
  614, A55, \dodoi{10.1051/0004-6361/201732209}

\bibitem[{{Allard} {et~al.}(2007{\natexlab{a}}){Allard}, {Allard}, {Homeier},
  {Kielkopf}, {McCaughrean}, \& {Spiegelman}}]{2007A&A...474L..21A}
{Allard}, F., {Allard}, N.~F., {Homeier}, D., {et~al.} 2007{\natexlab{a}},
  \aap, 474, L21, \dodoi{10.1051/0004-6361:20078362}

\bibitem[{{Allard} \& {Freytag}(2010)}]{2010HiA....15..756A}
{Allard}, F., \& {Freytag}, B. 2010, Highlights of Astronomy, 15, 756,
  \dodoi{10.1017/S1743921310011415}

\bibitem[{{Allard} {et~al.}(2001){Allard}, {Hauschildt}, {Alexander},
  {Tamanai}, \& {Schweitzer}}]{2001ApJ...556..357A}
{Allard}, F., {Hauschildt}, P.~H., {Alexander}, D.~R., {Tamanai}, A., \&
  {Schweitzer}, A. 2001, \apj, 556, 357, \dodoi{10.1086/321547}

\bibitem[{{Allard} {et~al.}(2011){Allard}, {Homeier}, \&
  {Freytag}}]{2011ASPC..448...91A}
{Allard}, F., {Homeier}, D., \& {Freytag}, B. 2011, Astronomical Society of the
  Pacific Conference Series, Vol. 448, {Model Atmospheres From Very Low Mass
  Stars to Brown Dwarfs}, ed. C.~{Johns-Krull}, M.~K. {Browning}, \& A.~A.
  {West}, 91

\bibitem[{{Allard} {et~al.}(1999){Allard}, {Royer}, {Kielkopf}, \&
  {Feautrier}}]{1999PhRvA..60.1021A}
{Allard}, N.~F., {Royer}, A., {Kielkopf}, J.~F., \& {Feautrier}, N. 1999, \pra,
  60, 1021, \dodoi{10.1103/PhysRevA.60.1021}

\bibitem[{{Allard} {et~al.}(2007{\natexlab{b}}){Allard}, {Spiegelman}, \&
  {Kielkopf}}]{2007A&A...465.1085A}
{Allard}, N.~F., {Spiegelman}, F., \& {Kielkopf}, J.~F. 2007{\natexlab{b}},
  \aap, 465, 1085, \dodoi{10.1051/0004-6361:20066616}

\bibitem[{{Allard} {et~al.}(2016){Allard}, {Spiegelman}, \&
  {Kielkopf}}]{2016A&A...589A..21A}
---. 2016, \aap, 589, A21, \dodoi{10.1051/0004-6361/201628270}

\bibitem[{{Allende Prieto} {et~al.}(2004){Allende Prieto}, {Barklem},
  {Lambert}, \& {Cunha}}]{2004AandA...420..183A}
{Allende Prieto}, C., {Barklem}, P.~S., {Lambert}, D.~L., \& {Cunha}, K. 2004,
  \aap, 420, 183, \dodoi{10.1051/0004-6361:20035801}

\bibitem[{{Astropy Collaboration} {et~al.}(2013){Astropy Collaboration},
  {Robitaille}, {Tollerud}, {Greenfield}, {Droettboom}, {Bray}, {Aldcroft},
  {Davis}, {Ginsburg}, {Price-Whelan}, {Kerzendorf}, {Conley}, {Crighton},
  {Barbary}, {Muna}, {Ferguson}, {Grollier}, {Parikh}, {Nair}, {Unther},
  {Deil}, {Woillez}, {Conseil}, {Kramer}, {Turner}, {Singer}, {Fox}, {Weaver},
  {Zabalza}, {Edwards}, {Azalee Bostroem}, {Burke}, {Casey}, {Crawford},
  {Dencheva}, {Ely}, {Jenness}, {Labrie}, {Lim}, {Pierfederici}, {Pontzen},
  {Ptak}, {Refsdal}, {Servillat}, \& {Streicher}}]{2013A&A...558A..33A}
{Astropy Collaboration}, {Robitaille}, T.~P., {Tollerud}, E.~J., {et~al.} 2013,
  \aap, 558, A33, \dodoi{10.1051/0004-6361/201322068}

\bibitem[{{Astropy Collaboration} {et~al.}(2018){Astropy Collaboration},
  {Price-Whelan}, {Sip{\H o}cz}, {G{\"u}nther}, {Lim}, {Crawford}, {Conseil},
  {Shupe}, {Craig}, {Dencheva}, {Ginsburg}, {VanderPlas}, {Bradley},
  {P{\'e}rez-Su{\'a}rez}, {de Val-Borro}, {Aldcroft}, {Cruz}, {Robitaille},
  {Tollerud}, {Ardelean}, {Babej}, {Bach}, {Bachetti}, {Bakanov}, {Bamford},
  {Barentsen}, {Barmby}, {Baumbach}, {Berry}, {Biscani}, {Boquien}, {Bostroem},
  {Bouma}, {Brammer}, {Bray}, {Breytenbach}, {Buddelmeijer}, {Burke},
  {Calderone}, {Cano Rodr{\'{\i}}guez}, {Cara}, {Cardoso}, {Cheedella},
  {Copin}, {Corrales}, {Crichton}, {D'Avella}, {Deil}, {Depagne}, {Dietrich},
  {Donath}, {Droettboom}, {Earl}, {Erben}, {Fabbro}, {Ferreira}, {Finethy},
  {Fox}, {Garrison}, {Gibbons}, {Goldstein}, {Gommers}, {Greco}, {Greenfield},
  {Groener}, {Grollier}, {Hagen}, {Hirst}, {Homeier}, {Horton}, {Hosseinzadeh},
  {Hu}, {Hunkeler}, {Ivezi{\'c}}, {Jain}, {Jenness}, {Kanarek}, {Kendrew},
  {Kern}, {Kerzendorf}, {Khvalko}, {King}, {Kirkby}, {Kulkarni}, {Kumar},
  {Lee}, {Lenz}, {Littlefair}, {Ma}, {Macleod}, {Mastropietro}, {McCully},
  {Montagnac}, {Morris}, {Mueller}, {Mumford}, {Muna}, {Murphy}, {Nelson},
  {Nguyen}, {Ninan}, {N{\"o}the}, {Ogaz}, {Oh}, {Parejko}, {Parley}, {Pascual},
  {Patil}, {Patil}, {Plunkett}, {Prochaska}, {Rastogi}, {Reddy Janga},
  {Sabater}, {Sakurikar}, {Seifert}, {Sherbert}, {Sherwood-Taylor}, {Shih},
  {Sick}, {Silbiger}, {Singanamalla}, {Singer}, {Sladen}, {Sooley},
  {Sornarajah}, {Streicher}, {Teuben}, {Thomas}, {Tremblay}, {Turner},
  {Terr{\'o}n}, {van Kerkwijk}, {de la Vega}, {Watkins}, {Weaver}, {Whitmore},
  {Woillez}, {Zabalza}, \& {Astropy Contributors}}]{2018AJ....156..123A}
{Astropy Collaboration}, {Price-Whelan}, A.~M., {Sip{\H o}cz}, B.~M., {et~al.}
  2018, \aj, 156, 123, \dodoi{10.3847/1538-3881/aabc4f}

\bibitem[{{Bailer-Jones} {et~al.}(2018){Bailer-Jones}, {Rybizki}, {Fouesneau},
  {Mantelet}, \& {Andrae}}]{2018AJ....156...58B}
{Bailer-Jones}, C.~A.~L., {Rybizki}, J., {Fouesneau}, M., {Mantelet}, G., \&
  {Andrae}, R. 2018, \aj, 156, 58, \dodoi{10.3847/1538-3881/aacb21}

\bibitem[{{Baliunas} {et~al.}(1996){Baliunas}, {Sokoloff}, \&
  {Soon}}]{1996ApJ...457L..99B}
{Baliunas}, S., {Sokoloff}, D., \& {Soon}, W. 1996, \apjl, 457, L99,
  \dodoi{10.1086/309891}

\bibitem[{{Barnes} \& {Fortney}(2003)}]{2003ApJ...588..545B}
{Barnes}, J.~W., \& {Fortney}, J.~J. 2003, \apj, 588, 545,
  \dodoi{10.1086/373893}

\bibitem[{{Barnes}(2003)}]{2003ApJ...586..464B}
{Barnes}, S.~A. 2003, \apj, 586, 464, \dodoi{10.1086/367639}

\bibitem[{{Barnes}(2007)}]{2007ApJ...669.1167B}
---. 2007, \apj, 669, 1167, \dodoi{10.1086/519295}

\bibitem[{{Beatty} {et~al.}(2018){Beatty}, {Morley}, {Curtis}, {Burrows},
  {Davenport}, \& {Montet}}]{2018AJ....156..168B}
{Beatty}, T.~G., {Morley}, C.~V., {Curtis}, J.~L., {et~al.} 2018, \aj, 156,
  168, \dodoi{10.3847/1538-3881/aad697}

\bibitem[{{Bertelli} {et~al.}(2008){Bertelli}, {Girardi}, {Marigo}, \&
  {Nasi}}]{2008AandA...484..815B}
{Bertelli}, G., {Girardi}, L., {Marigo}, P., \& {Nasi}, E. 2008, \aap, 484,
  815, \dodoi{10.1051/0004-6361:20079165}

\bibitem[{{Bertelli} {et~al.}(2009){Bertelli}, {Nasi}, {Girardi}, \&
  {Marigo}}]{2009AandA...508..355B}
{Bertelli}, G., {Nasi}, E., {Girardi}, L., \& {Marigo}, P. 2009, \aap, 508,
  355, \dodoi{10.1051/0004-6361/200912093}

\bibitem[{{Best} {et~al.}(2021){Best}, {Liu}, {Magnier}, \&
  {Dupuy}}]{2021AJ....161...42B}
{Best}, W. M.~J., {Liu}, M.~C., {Magnier}, E.~A., \& {Dupuy}, T.~J. 2021, \aj,
  161, 42, \dodoi{10.3847/1538-3881/abc893}

\bibitem[{{Beuzit} {et~al.}(2004){Beuzit}, {S{\'e}gransan}, {Forveille},
  {Udry}, {Delfosse}, {Mayor}, {Perrier}, {Hainaut}, {Roddier}, {Roddier}, \&
  {Mart{\'\i}n}}]{2004A&A...425..997B}
{Beuzit}, J.~L., {S{\'e}gransan}, D., {Forveille}, T., {et~al.} 2004, \aap,
  425, 997, \dodoi{10.1051/0004-6361:20048006}

\bibitem[{{Bonfanti} {et~al.}(2016){Bonfanti}, {Ortolani}, \&
  {Nascimbeni}}]{2016AandA...585A...5B}
{Bonfanti}, A., {Ortolani}, S., \& {Nascimbeni}, V. 2016, \aap, 585, A5,
  \dodoi{10.1051/0004-6361/201527297}

\bibitem[{{Bonnefoy} {et~al.}(2014){Bonnefoy}, {Chauvin}, {Lagrange}, {Rojo},
  {Allard}, {Pinte}, {Dumas}, \& {Homeier}}]{2014A&A...562A.127B}
{Bonnefoy}, M., {Chauvin}, G., {Lagrange}, A.~M., {et~al.} 2014, \aap, 562,
  A127, \dodoi{10.1051/0004-6361/201118270}

\bibitem[{{Bonnefoy} {et~al.}(2018){Bonnefoy}, {Perraut}, {Lagrange},
  {Delorme}, {Vigan}, {Line}, {Rodet}, {Ginski}, {Mourard}, {Marleau},
  {Samland}, {Tremblin}, {Ligi}, {Cantalloube}, {Molli{\`e}re}, {Charnay},
  {Kuzuhara}, {Janson}, {Morley}, {Homeier}, {D'Orazi}, {Klahr}, {Mordasini},
  {Lavie}, {Baudino}, {Beust}, {Peretti}, {Musso Bartucci}, {Mesa},
  {B{\'e}zard}, {Boccaletti}, {Galicher}, {Hagelberg}, {Desidera}, {Biller},
  {Maire}, {Allard}, {Borgniet}, {Lannier}, {Meunier}, {Desort}, {Alecian},
  {Chauvin}, {Langlois}, {Henning}, {Mugnier}, {Mouillet}, {Gratton}, {Brandt},
  {Mc Elwain}, {Beuzit}, {Tamura}, {Hori}, {Brandner}, {Buenzli}, {Cheetham},
  {Cudel}, {Feldt}, {Kasper}, {Keppler}, {Kopytova}, {Meyer}, {Perrot},
  {Rouan}, {Salter}, {Schmidt}, {Sissa}, {Zurlo}, {Wildi}, {Blanchard}, {De
  Caprio}, {Delboulb{\'e}}, {Maurel}, {Moulin}, {Pavlov}, {Rabou}, {Ramos},
  {Roelfsema}, {Rousset}, {Stadler}, {Rigal}, \& {Weber}}]{2018A&A...618A..63B}
{Bonnefoy}, M., {Perraut}, K., {Lagrange}, A.~M., {et~al.} 2018, \aap, 618,
  A63, \dodoi{10.1051/0004-6361/201832942}

\bibitem[{{Bowler} {et~al.}(2017){Bowler}, {Liu}, {Mawet}, {Ngo}, {Malo},
  {Mace}, {McLane}, {Lu}, {Tristan}, {Hinkley}, {Hillenbrand}, {Shkolnik},
  {Benneke}, \& {Best}}]{2017AJ....153...18B}
{Bowler}, B.~P., {Liu}, M.~C., {Mawet}, D., {et~al.} 2017, \aj, 153, 18,
  \dodoi{10.3847/1538-3881/153/1/18}

\bibitem[{{Bowler} {et~al.}(2018){Bowler}, {Dupuy}, {Endl}, {Cochran},
  {MacQueen}, {Fulton}, {Petigura}, {Howard}, {Hirsch}, {Kratter}, {Crepp},
  {Biller}, {Johnson}, \& {Wittenmyer}}]{2018AJ....155..159B}
{Bowler}, B.~P., {Dupuy}, T.~J., {Endl}, M., {et~al.} 2018, \aj, 155, 159,
  \dodoi{10.3847/1538-3881/aab2a6}

\bibitem[{{Brandt} {et~al.}(2019){Brandt}, {Dupuy}, \&
  {Bowler}}]{2019AJ....158..140B}
{Brandt}, T.~D., {Dupuy}, T.~J., \& {Bowler}, B.~P. 2019, \aj, 158, 140,
  \dodoi{10.3847/1538-3881/ab04a8}

\bibitem[{{Brandt} {et~al.}(2020){Brandt}, {Dupuy}, {Bowler}, {Bardalez
  Gagliuffi}, {Faherty}, {Brandt}, \& {Michalik}}]{2020AJ....160..196B}
{Brandt}, T.~D., {Dupuy}, T.~J., {Bowler}, B.~P., {et~al.} 2020, \aj, 160, 196,
  \dodoi{10.3847/1538-3881/abb45e}

\bibitem[{{Bressan} {et~al.}(2012){Bressan}, {Marigo}, {Girardi}, {Salasnich},
  {Dal Cero}, {Rubele}, \& {Nanni}}]{2012MNRAS.427..127B}
{Bressan}, A., {Marigo}, P., {Girardi}, L., {et~al.} 2012, \mnras, 427, 127,
  \dodoi{10.1111/j.1365-2966.2012.21948.x}

\bibitem[{{Burgasser}(2007)}]{2007ApJ...658..617B}
{Burgasser}, A.~J. 2007, \apj, 658, 617, \dodoi{10.1086/511176}

\bibitem[{{Burgasser}(2014)}]{2014ASInC..11....7B}
{Burgasser}, A.~J. 2014, in Astronomical Society of India Conference Series,
  Vol.~11, Astronomical Society of India Conference Series, 7--16

\bibitem[{{Burgasser} {et~al.}(2006){Burgasser}, {Burrows}, \&
  {Kirkpatrick}}]{2006ApJ...639.1095B}
{Burgasser}, A.~J., {Burrows}, A., \& {Kirkpatrick}, J.~D. 2006, \apj, 639,
  1095, \dodoi{10.1086/499344}

\bibitem[{{Burgasser} {et~al.}(2004){Burgasser}, {McElwain}, {Kirkpatrick},
  {Cruz}, {Tinney}, \& {Reid}}]{2004AJ....127.2856B}
{Burgasser}, A.~J., {McElwain}, M.~W., {Kirkpatrick}, J.~D., {et~al.} 2004,
  \aj, 127, 2856, \dodoi{10.1086/383549}

\bibitem[{{Burgasser} {et~al.}(2000){Burgasser}, {Kirkpatrick}, {Cutri},
  {McCallon}, {Kopan}, {Gizis}, {Liebert}, {Reid}, {Brown}, {Monet}, {Dahn},
  {Beichman}, \& {Skrutskie}}]{2000ApJ...531L..57B}
{Burgasser}, A.~J., {Kirkpatrick}, J.~D., {Cutri}, R.~M., {et~al.} 2000, \apjl,
  531, L57, \dodoi{10.1086/312522}

\bibitem[{{Burgasser} {et~al.}(2010){Burgasser}, {Simcoe}, {Bochanski},
  {Saumon}, {Mamajek}, {Cushing}, {Marley}, {McMurtry}, {Pipher}, \&
  {Forrest}}]{2010ApJ...725.1405B}
{Burgasser}, A.~J., {Simcoe}, R.~A., {Bochanski}, J.~J., {et~al.} 2010, \apj,
  725, 1405, \dodoi{10.1088/0004-637X/725/2/1405}

\bibitem[{{Burningham} {et~al.}(2009){Burningham}, {Pinfield}, {Leggett},
  {Tinney}, {Liu}, {Homeier}, {West}, {Day-Jones}, {Huelamo}, {Dupuy}, {Zhang},
  {Murray}, {Lodieu}, {Barrado Y Navascu{\'e}s}, {Folkes}, {Galvez-Ortiz},
  {Jones}, {Lucas}, {Morales Calderon}, \& {Tamura}}]{2009MNRAS.395.1237B}
{Burningham}, B., {Pinfield}, D.~J., {Leggett}, S.~K., {et~al.} 2009, \mnras,
  395, 1237, \dodoi{10.1111/j.1365-2966.2009.14620.x}

\bibitem[{{Burningham} {et~al.}(2011){Burningham}, {Leggett}, {Homeier},
  {Saumon}, {Lucas}, {Pinfield}, {Tinney}, {Allard}, {Marley}, {Jones},
  {Murray}, {Ishii}, {Day-Jones}, {Gomes}, \& {Zhang}}]{2011MNRAS.414.3590B}
{Burningham}, B., {Leggett}, S.~K., {Homeier}, D., {et~al.} 2011, \mnras, 414,
  3590, \dodoi{10.1111/j.1365-2966.2011.18664.x}

\bibitem[{{Burrows} {et~al.}(2001){Burrows}, {Hubbard}, {Lunine}, \&
  {Liebert}}]{2001RvMP...73..719B}
{Burrows}, A., {Hubbard}, W.~B., {Lunine}, J.~I., \& {Liebert}, J. 2001,
  Reviews of Modern Physics, 73, 719, \dodoi{10.1103/RevModPhys.73.719}

\bibitem[{{Burrows} \& {Liebert}(1993)}]{1993RvMP...65..301B}
{Burrows}, A., \& {Liebert}, J. 1993, Reviews of Modern Physics, 65, 301,
  \dodoi{10.1103/RevModPhys.65.301}

\bibitem[{{Burrows} {et~al.}(2000){Burrows}, {Marley}, \&
  {Sharp}}]{2000ApJ...531..438B}
{Burrows}, A., {Marley}, M.~S., \& {Sharp}, C.~M. 2000, \apj, 531, 438,
  \dodoi{10.1086/308462}

\bibitem[{{Burrows} {et~al.}(2006){Burrows}, {Sudarsky}, \&
  {Hubeny}}]{2006ApJ...640.1063B}
{Burrows}, A., {Sudarsky}, D., \& {Hubeny}, I. 2006, \apj, 640, 1063,
  \dodoi{10.1086/500293}

\bibitem[{{Burrows} {et~al.}(2003){Burrows}, {Sudarsky}, \&
  {Lunine}}]{2003ApJ...596..587B}
{Burrows}, A., {Sudarsky}, D., \& {Lunine}, J.~I. 2003, \apj, 596, 587,
  \dodoi{10.1086/377709}

\bibitem[{{Burrows} \& {Volobuyev}(2003)}]{2003ApJ...583..985B}
{Burrows}, A., \& {Volobuyev}, M. 2003, \apj, 583, 985, \dodoi{10.1086/345412}

\bibitem[{{Burrows} {et~al.}(1997){Burrows}, {Marley}, {Hubbard}, {Lunine},
  {Guillot}, {Saumon}, {Freedman}, {Sudarsky}, \&
  {Sharp}}]{1997ApJ...491..856B}
{Burrows}, A., {Marley}, M., {Hubbard}, W.~B., {et~al.} 1997, \apj, 491, 856,
  \dodoi{10.1086/305002}

\bibitem[{{Carmichael} {et~al.}(2020){Carmichael}, {Quinn}, {Mustill}, {Huang},
  {Zhou}, {Persson}, {Nielsen}, {Collins}, {Ziegler}, {Collins}, {Rodriguez},
  {Shporer}, {Brahm}, {Mann}, {Bouchy}, {Fridlund}, {Stassun}, {Hellier},
  {Seidel}, {Stalport}, {Udry}, {Pepe}, {Ireland}, {{\v{Z}}erjal},
  {Brice{\~n}o}, {Law}, {Jord{\'a}n}, {Espinoza}, {Henning}, {Sarkis}, \&
  {Latham}}]{2020AJ....160...53C}
{Carmichael}, T.~W., {Quinn}, S.~N., {Mustill}, A.~J., {et~al.} 2020, \aj, 160,
  53, \dodoi{10.3847/1538-3881/ab9b84}

\bibitem[{{Casagrande} {et~al.}(2007){Casagrande}, {Flynn}, {Portinari},
  {Girardi}, \& {Jimenez}}]{2007MNRAS.382.1516C}
{Casagrande}, L., {Flynn}, C., {Portinari}, L., {Girardi}, L., \& {Jimenez}, R.
  2007, \mnras, 382, 1516, \dodoi{10.1111/j.1365-2966.2007.12512.x}

\bibitem[{{Casagrande} {et~al.}(2011){Casagrande}, {Sch{\"o}nrich}, {Asplund},
  {Cassisi}, {Ram{\'\i}rez}, {Mel{\'e}ndez}, {Bensby}, \&
  {Feltzing}}]{2011AandA...530A.138C}
{Casagrande}, L., {Sch{\"o}nrich}, R., {Asplund}, M., {et~al.} 2011, \aap, 530,
  A138, \dodoi{10.1051/0004-6361/201016276}

\bibitem[{{Chandrasekhar}(1939)}]{1939isss.book.....C}
{Chandrasekhar}, S. 1939, {An introduction to the study of stellar structure}

\bibitem[{{Chauvin} {et~al.}(2017){Chauvin}, {Desidera}, {Lagrange}, {Vigan},
  {Gratton}, {Langlois}, {Bonnefoy}, {Beuzit}, {Feldt}, {Mouillet}, {Meyer},
  {Cheetham}, {Biller}, {Boccaletti}, {D'Orazi}, {Galicher}, {Hagelberg},
  {Maire}, {Mesa}, {Olofsson}, {Samland}, {Schmidt}, {Sissa}, {Bonavita},
  {Charnay}, {Cudel}, {Daemgen}, {Delorme}, {Janin-Potiron}, {Janson},
  {Keppler}, {Le Coroller}, {Ligi}, {Marleau}, {Messina}, {Molli{\`e}re},
  {Mordasini}, {M{\"u}ller}, {Peretti}, {Perrot}, {Rodet}, {Rouan}, {Zurlo},
  {Dominik}, {Henning}, {Menard}, {Schmid}, {Turatto}, {Udry}, {Vakili}, {Abe},
  {Antichi}, {Baruffolo}, {Baudoz}, {Baudrand}, {Blanchard}, {Bazzon}, {Buey},
  {Carbillet}, {Carle}, {Charton}, {Cascone}, {Claudi}, {Costille}, {Deboulbe},
  {De Caprio}, {Dohlen}, {Fantinel}, {Feautrier}, {Fusco}, {Gigan}, {Giro},
  {Gisler}, {Gluck}, {Hubin}, {Hugot}, {Jaquet}, {Kasper}, {Madec}, {Magnard},
  {Martinez}, {Maurel}, {Le Mignant}, {M{\"o}ller-Nilsson}, {Llored}, {Moulin},
  {Orign{\'e}}, {Pavlov}, {Perret}, {Petit}, {Pragt}, {Puget}, {Rabou},
  {Ramos}, {Rigal}, {Rochat}, {Roelfsema}, {Rousset}, {Roux}, {Salasnich},
  {Sauvage}, {Sevin}, {Soenke}, {Stadler}, {Suarez}, {Weber}, {Wildi},
  {Antoniucci}, {Augereau}, {Baudino}, {Brandner}, {Engler}, {Girard}, {Gry},
  {Kral}, {Kopytova}, {Lagadec}, {Milli}, {Moutou}, {Schlieder},
  {Szul{\'a}gyi}, {Thalmann}, \& {Wahhaj}}]{2017A&A...605L...9C}
{Chauvin}, G., {Desidera}, S., {Lagrange}, A.~M., {et~al.} 2017, \aap, 605, L9,
  \dodoi{10.1051/0004-6361/201731152}

\bibitem[{{Cheetham} {et~al.}(2018){Cheetham}, {S{\'e}gransan}, {Peretti},
  {Delisle}, {Hagelberg}, {Beuzit}, {Forveille}, {Marmier}, {Udry}, \&
  {Wildi}}]{2018A&A...614A..16C}
{Cheetham}, A., {S{\'e}gransan}, D., {Peretti}, S., {et~al.} 2018, \aap, 614,
  A16, \dodoi{10.1051/0004-6361/201630136}

\bibitem[{{Cottaar} {et~al.}(2014){Cottaar}, {Covey}, {Meyer}, {Nidever},
  {Stassun}, {Foster}, {Tan}, {Chojnowski}, {da Rio}, {Flaherty}, {Frinchaboy},
  {Skrutskie}, {Majewski}, {Wilson}, \& {Zasowski}}]{2014ApJ...794..125C}
{Cottaar}, M., {Covey}, K.~R., {Meyer}, M.~R., {et~al.} 2014, \apj, 794, 125,
  \dodoi{10.1088/0004-637X/794/2/125}

\bibitem[{{Cushing} {et~al.}(2004){Cushing}, {Vacca}, \&
  {Rayner}}]{2004PASP..116..362C}
{Cushing}, M.~C., {Vacca}, W.~D., \& {Rayner}, J.~T. 2004, \pasp, 116, 362,
  \dodoi{10.1086/382907}

\bibitem[{{Cushing} {et~al.}(2008){Cushing}, {Marley}, {Saumon}, {Kelly},
  {Vacca}, {Rayner}, {Freedman}, {Lodders}, \& {Roellig}}]{2008ApJ...678.1372C}
{Cushing}, M.~C., {Marley}, M.~S., {Saumon}, D., {et~al.} 2008, \apj, 678,
  1372, \dodoi{10.1086/526489}

\bibitem[{{Cushing} {et~al.}(2011){Cushing}, {Kirkpatrick}, {Gelino},
  {Griffith}, {Skrutskie}, {Mainzer}, {Marsh}, {Beichman}, {Burgasser},
  {Prato}, {Simcoe}, {Marley}, {Saumon}, {Freedman}, {Eisenhardt}, \&
  {Wright}}]{2011ApJ...743...50C}
{Cushing}, M.~C., {Kirkpatrick}, J.~D., {Gelino}, C.~R., {et~al.} 2011, \apj,
  743, 50, \dodoi{10.1088/0004-637X/743/1/50}

\bibitem[{{Cutri} \& {et al.}(2014)}]{2014yCat.2328....0C}
{Cutri}, R.~M., \& {et al.} 2014, VizieR Online Data Catalog, 2328

\bibitem[{{Czekala} {et~al.}(2015){Czekala}, {Andrews}, {Mandel}, {Hogg}, \&
  {Green}}]{2015ApJ...812..128C}
{Czekala}, I., {Andrews}, S.~M., {Mandel}, K.~S., {Hogg}, D.~W., \& {Green},
  G.~M. 2015, \apj, 812, 128, \dodoi{10.1088/0004-637X/812/2/128}

\bibitem[{{Deacon} {et~al.}(2014){Deacon}, {Liu}, {Magnier}, {Aller}, {Best},
  {Dupuy}, {Bowler}, {Mann}, {Redstone}, {Burgett}, {Chambers}, {Draper},
  {Flewelling}, {Hodapp}, {Kaiser}, {Kudritzki}, {Morgan}, {Metcalfe}, {Price},
  {Tonry}, \& {Wainscoat}}]{2014ApJ...792..119D}
{Deacon}, N.~R., {Liu}, M.~C., {Magnier}, E.~A., {et~al.} 2014, \apj, 792, 119,
  \dodoi{10.1088/0004-637X/792/2/119}

\bibitem[{{Del Burgo} {et~al.}(2009){Del Burgo}, {Mart{\'\i}n}, {Zapatero
  Osorio}, \& {Hauschildt}}]{2009AandA...501.1059D}
{Del Burgo}, C., {Mart{\'\i}n}, E.~L., {Zapatero Osorio}, M.~R., \&
  {Hauschildt}, P.~H. 2009, \aap, 501, 1059,
  \dodoi{10.1051/0004-6361/200810752}

\bibitem[{{Delgado Mena} {et~al.}(2010){Delgado Mena}, {Israelian},
  {Gonz{\'a}lez Hern{\'a}ndez}, {Bond}, {Santos}, {Udry}, \&
  {Mayor}}]{2010ApJ...725.2349D}
{Delgado Mena}, E., {Israelian}, G., {Gonz{\'a}lez Hern{\'a}ndez}, J.~I.,
  {et~al.} 2010, \apj, 725, 2349, \dodoi{10.1088/0004-637X/725/2/2349}

\bibitem[{{Demarque} {et~al.}(2008){Demarque}, {Guenther}, {Li}, {Mazumdar}, \&
  {Straka}}]{2008ApandSS.316...31D}
{Demarque}, P., {Guenther}, D.~B., {Li}, L.~H., {Mazumdar}, A., \& {Straka},
  C.~W. 2008, \apss, 316, 31, \dodoi{10.1007/s10509-007-9698-y}

\bibitem[{{Demarque} {et~al.}(2004){Demarque}, {Woo}, {Kim}, \&
  {Yi}}]{2004ApJS..155..667D}
{Demarque}, P., {Woo}, J.-H., {Kim}, Y.-C., \& {Yi}, S.~K. 2004, \apjs, 155,
  667, \dodoi{10.1086/424966}

\bibitem[{{Deming} {et~al.}(2013){Deming}, {Wilkins}, {McCullough}, {Burrows},
  {Fortney}, {Agol}, {Dobbs-Dixon}, {Madhusudhan}, {Crouzet}, {Desert},
  {Gilliland}, {Haynes}, {Knutson}, {Line}, {Magic}, {Mand ell}, {Ranjan},
  {Charbonneau}, {Clampin}, {Seager}, \& {Showman}}]{2013ApJ...774...95D}
{Deming}, D., {Wilkins}, A., {McCullough}, P., {et~al.} 2013, \apj, 774, 95,
  \dodoi{10.1088/0004-637X/774/2/95}

\bibitem[{{Donahue}(1993)}]{1993PhDT.........3D}
{Donahue}, R.~A. 1993, PhD thesis, New Mexico State University, University
  Park.

\bibitem[{{Dupuy} \& {Liu}(2017)}]{2017ApJS..231...15D}
{Dupuy}, T.~J., \& {Liu}, M.~C. 2017, \apjs, 231, 15,
  \dodoi{10.3847/1538-4365/aa5e4c}

\bibitem[{{Dupuy} {et~al.}(2009){Dupuy}, {Liu}, \&
  {Ireland}}]{2009ApJ...692..729D}
{Dupuy}, T.~J., {Liu}, M.~C., \& {Ireland}, M.~J. 2009, \apj, 692, 729,
  \dodoi{10.1088/0004-637X/692/1/729}

\bibitem[{{Dupuy} {et~al.}(2014){Dupuy}, {Liu}, \&
  {Ireland}}]{2014ApJ...790..133D}
---. 2014, \apj, 790, 133, \dodoi{10.1088/0004-637X/790/2/133}

\bibitem[{{Dupuy} {et~al.}(2019){Dupuy}, {Liu}, {Best}, {Mann}, {Tucker},
  {Zhang}, {Baraffe}, {Chabrier}, {Forveille}, {Metchev}, {Tremblin}, {Do},
  {Payne}, {Shappee}, {Bond}, {Cetre}, {Chun}, {Delorme}, {Jovanovic},
  {Lilley}, {Mawet}, {Ragland}, {Wetherell}, \&
  {Wizinowich}}]{2019AJ....158..174D}
{Dupuy}, T.~J., {Liu}, M.~C., {Best}, W. M.~J., {et~al.} 2019, \aj, 158, 174,
  \dodoi{10.3847/1538-3881/ab3cd1}

\bibitem[{{Duquennoy} \& {Mayor}(1988)}]{1988A&A...200..135D}
{Duquennoy}, A., \& {Mayor}, M. 1988, \aap, 200, 135

\bibitem[{{Faherty} {et~al.}(2016){Faherty}, {Riedel}, {Cruz}, {Gagne},
  {Filippazzo}, {Lambrides}, {Fica}, {Weinberger}, {Thorstensen}, {Tinney},
  {Baldassare}, {Lemonier}, \& {Rice}}]{2016ApJS..225...10F}
{Faherty}, J.~K., {Riedel}, A.~R., {Cruz}, K.~L., {et~al.} 2016, \apjs, 225,
  10, \dodoi{10.3847/0067-0049/225/1/10}

\bibitem[{{Faherty} {et~al.}(2020){Faherty}, {Goodman}, {Caselden}, {Colin},
  {Kuchner}, {Meisner}, {Gagn{\'e}}, {Schneider}, {Gonzales}, {Bardalez
  Gagliuffi}, {Logsdon}, {Allers}, \& {Burgasser}}]{2020ApJ...889..176F}
{Faherty}, J.~K., {Goodman}, S., {Caselden}, D., {et~al.} 2020, \apj, 889, 176,
  \dodoi{10.3847/1538-4357/ab5303}

\bibitem[{{Fegley} \& {Lodders}(1994)}]{1994Icar..110..117F}
{Fegley}, Bruce, J., \& {Lodders}, K. 1994, \icarus, 110, 117,
  \dodoi{10.1006/icar.1994.1111}

\bibitem[{{Feltzing} \& {Gustafsson}(1998)}]{1998AandAS..129..237F}
{Feltzing}, S., \& {Gustafsson}, B. 1998, \aaps, 129, 237,
  \dodoi{10.1051/aas:1998400}

\bibitem[{{Fernandes} {et~al.}(2011){Fernandes}, {Vaz}, \&
  {Vicente}}]{2011AandA...532A..20F}
{Fernandes}, J.~M., {Vaz}, A.~I.~F., \& {Vicente}, L.~N. 2011, \aap, 532, A20,
  \dodoi{10.1051/0004-6361/200811182}

\bibitem[{{Fischer} {et~al.}(2003){Fischer}, {Butler}, {Marcy}, {Vogt}, \&
  {Henry}}]{2003ApJ...590.1081F}
{Fischer}, D.~A., {Butler}, R.~P., {Marcy}, G.~W., {Vogt}, S.~S., \& {Henry},
  G.~W. 2003, \apj, 590, 1081, \dodoi{10.1086/375027}

\bibitem[{{Foreman-Mackey} {et~al.}(2013){Foreman-Mackey}, {Hogg}, {Lang}, \&
  {Goodman}}]{2013PASP..125..306F}
{Foreman-Mackey}, D., {Hogg}, D.~W., {Lang}, D., \& {Goodman}, J. 2013, \pasp,
  125, 306, \dodoi{10.1086/670067}

\bibitem[{{Fortney} {et~al.}(2008{\natexlab{a}}){Fortney}, {Lodders}, {Marley},
  \& {Freedman}}]{2008ApJ...678.1419F}
{Fortney}, J.~J., {Lodders}, K., {Marley}, M.~S., \& {Freedman}, R.~S.
  2008{\natexlab{a}}, \apj, 678, 1419, \dodoi{10.1086/528370}

\bibitem[{{Fortney} {et~al.}(2005){Fortney}, {Marley}, {Lodders}, {Saumon}, \&
  {Freedman}}]{2005ApJ...627L..69F}
{Fortney}, J.~J., {Marley}, M.~S., {Lodders}, K., {Saumon}, D., \& {Freedman},
  R. 2005, \apjl, 627, L69, \dodoi{10.1086/431952}

\bibitem[{{Fortney} {et~al.}(2008{\natexlab{b}}){Fortney}, {Marley}, {Saumon},
  \& {Lodders}}]{2008ApJ...683.1104F}
{Fortney}, J.~J., {Marley}, M.~S., {Saumon}, D., \& {Lodders}, K.
  2008{\natexlab{b}}, \apj, 683, 1104, \dodoi{10.1086/589942}

\bibitem[{{Fortney} {et~al.}(2013){Fortney}, {Mordasini}, {Nettelmann},
  {Kempton}, {Greene}, \& {Zahnle}}]{2013ApJ...775...80F}
{Fortney}, J.~J., {Mordasini}, C., {Nettelmann}, N., {et~al.} 2013, \apj, 775,
  80, \dodoi{10.1088/0004-637X/775/1/80}

\bibitem[{{Gagn{\'e}} {et~al.}(2018){Gagn{\'e}}, {Mamajek}, {Malo}, {Riedel},
  {Rodriguez}, {Lafreni{\`e}re}, {Faherty}, {Roy-Loubier}, {Pueyo}, {Robin}, \&
  {Doyon}}]{2018ApJ...856...23G}
{Gagn{\'e}}, J., {Mamajek}, E.~E., {Malo}, L., {et~al.} 2018, \apj, 856, 23,
  \dodoi{10.3847/1538-4357/aaae09}

\bibitem[{{Gaia Collaboration} {et~al.}(2016){Gaia Collaboration}, {Prusti},
  {de Bruijne}, {Brown}, {Vallenari}, {Babusiaux}, {Bailer-Jones}, {Bastian},
  {Biermann}, {Evans}, {Eyer}, {Jansen}, {Jordi}, {Klioner}, {Lammers},
  {Lindegren}, {Luri}, {Mignard}, {Milligan}, {Panem}, {Poinsignon},
  {Pourbaix}, {Randich}, {Sarri}, {Sartoretti}, {Siddiqui}, {Soubiran},
  {Valette}, {van Leeuwen}, {Walton}, {Aerts}, {Arenou}, {Cropper}, {Drimmel},
  {H{\o}g}, {Katz}, {Lattanzi}, {O'Mullane}, {Grebel}, {Holland}, {Huc},
  {Passot}, {Bramante}, {Cacciari}, {Casta{\~n}eda}, {Chaoul}, {Cheek}, {De
  Angeli}, {Fabricius}, {Guerra}, {Hern{\'a}ndez}, {Jean-Antoine-Piccolo},
  {Masana}, {Messineo}, {Mowlavi}, {Nienartowicz}, {Ord{\'o}{\~n}ez-Blanco},
  {Panuzzo}, {Portell}, {Richards}, {Riello}, {Seabroke}, {Tanga},
  {Th{\'e}venin}, {Torra}, {Els}, {Gracia-Abril}, {Comoretto},
  {Garcia-Reinaldos}, {Lock}, {Mercier}, {Altmann}, {Andrae}, {Astraatmadja},
  {Bellas-Velidis}, {Benson}, {Berthier}, {Blomme}, {Busso}, {Carry},
  {Cellino}, {Clementini}, {Cowell}, {Creevey}, {Cuypers}, {Davidson}, {De
  Ridder}, {de Torres}, {Delchambre}, {Dell'Oro}, {Ducourant}, {Fr{\'e}mat},
  {Garc{\'\i}a-Torres}, {Gosset}, {Halbwachs}, {Hambly}, {Harrison}, {Hauser},
  {Hestroffer}, {Hodgkin}, {Huckle}, {Hutton}, {Jasniewicz}, {Jordan},
  {Kontizas}, {Korn}, {Lanzafame}, {Manteiga}, {Moitinho}, {Muinonen},
  {Osinde}, {Pancino}, {Pauwels}, {Petit}, {Recio-Blanco}, {Robin}, {Sarro},
  {Siopis}, {Smith}, {Smith}, {Sozzetti}, {Thuillot}, {van Reeven}, {Viala},
  {Abbas}, {Abreu Aramburu}, {Accart}, {Aguado}, {Allan}, {Allasia},
  {Altavilla}, {{\'A}lvarez}, {Alves}, {Anderson}, {Andrei}, {Anglada Varela},
  {Antiche}, {Antoja}, {Ant{\'o}n}, {Arcay}, {Atzei}, {Ayache}, {Bach},
  {Baker}, {Balaguer-N{\'u}{\~n}ez}, {Barache}, {Barata}, {Barbier}, {Barblan},
  {Baroni}, {Barrado y Navascu{\'e}s}, {Barros}, {Barstow}, {Becciani},
  {Bellazzini}, {Bellei}, {Bello Garc{\'\i}a}, {Belokurov}, {Bendjoya},
  {Berihuete}, {Bianchi}, {Bienaym{\'e}}, {Billebaud}, {Blagorodnova},
  {Blanco-Cuaresma}, {Boch}, {Bombrun}, {Borrachero}, {Bouquillon}, {Bourda},
  {Bouy}, {Bragaglia}, {Breddels}, {Brouillet}, {Br{\"u}semeister},
  {Bucciarelli}, {Budnik}, {Burgess}, {Burgon}, {Burlacu}, {Busonero}, {Buzzi},
  {Caffau}, {Cambras}, {Campbell}, {Cancelliere}, {Cantat-Gaudin}, {Carlucci},
  {Carrasco}, {Castellani}, {Charlot}, {Charnas}, {Charvet}, {Chassat},
  {Chiavassa}, {Clotet}, {Cocozza}, {Collins}, {Collins}, {Costigan}, {Crifo},
  {Cross}, {Crosta}, {Crowley}, {Dafonte}, {Damerdji}, {Dapergolas}, {David},
  {David}, {De Cat}, {de Felice}, {de Laverny}, {De Luise}, {De March}, {de
  Martino}, {de Souza}, {Debosscher}, {del Pozo}, {Delbo}, {Delgado},
  {Delgado}, {di Marco}, {Di Matteo}, {Diakite}, {Distefano}, {Dolding}, {Dos
  Anjos}, {Drazinos}, {Dur{\'a}n}, {Dzigan}, {Ecale}, {Edvardsson}, {Enke},
  {Erdmann}, {Escolar}, {Espina}, {Evans}, {Eynard Bontemps}, {Fabre},
  {Fabrizio}, {Faigler}, {Falc{\~a}o}, {Farr{\`a}s Casas}, {Faye}, {Federici},
  {Fedorets}, {Fern{\'a}ndez-Hern{\'a}ndez}, {Fernique}, {Fienga}, {Figueras},
  {Filippi}, {Findeisen}, {Fonti}, {Fouesneau}, {Fraile}, {Fraser}, {Fuchs},
  {Furnell}, {Gai}, {Galleti}, {Galluccio}, {Garabato}, {Garc{\'\i}a-Sedano},
  {Gar{\'e}}, {Garofalo}, {Garralda}, {Gavras}, {Gerssen}, {Geyer}, {Gilmore},
  {Girona}, {Giuffrida}, {Gomes}, {Gonz{\'a}lez-Marcos},
  {Gonz{\'a}lez-N{\'u}{\~n}ez}, {Gonz{\'a}lez-Vidal}, {Granvik}, {Guerrier},
  {Guillout}, {Guiraud}, {G{\'u}rpide}, {Guti{\'e}rrez-S{\'a}nchez}, {Guy},
  {Haigron}, {Hatzidimitriou}, {Haywood}, {Heiter}, {Helmi}, {Hobbs},
  {Hofmann}, {Holl}, {Holland }, {Hunt}, {Hypki}, {Icardi}, {Irwin}, {Jevardat
  de Fombelle}, {Jofr{\'e}}, {Jonker}, {Jorissen}, {Julbe}, {Karampelas},
  {Kochoska}, {Kohley}, {Kolenberg}, {Kontizas}, {Koposov}, {Kordopatis},
  {Koubsky}, {Kowalczyk}, {Krone-Martins}, {Kudryashova}, {Kull}, {Bachchan},
  {Lacoste-Seris}, {Lanza}, {Lavigne}, {Le Poncin-Lafitte}, {Lebreton},
  {Lebzelter}, {Leccia}, {Leclerc}, {Lecoeur-Taibi}, {Lemaitre}, {Lenhardt},
  {Leroux}, {Liao}, {Licata}, {Lindstr{\o}m}, {Lister}, {Livanou}, {Lobel},
  {L{\"o}ffler}, {L{\'o}pez}, {Lopez-Lozano}, {Lorenz}, {Loureiro},
  {MacDonald}, {Magalh{\~a}es Fernandes}, {Managau}, {Mann}, {Mantelet},
  {Marchal}, {Marchant}, {Marconi}, {Marie}, {Marinoni}, {Marrese},
  {Marschalk{\'o}}, {Marshall}, {Mart{\'\i}n-Fleitas}, {Martino}, {Mary},
  {Matijevi{\v{c}}}, {Mazeh}, {McMillan}, {Messina}, {Mestre}, {Michalik},
  {Millar}, {Miranda}, {Molina}, {Molinaro}, {Molinaro}, {Moln{\'a}r},
  {Moniez}, {Montegriffo}, {Monteiro}, {Mor}, {Mora}, {Morbidelli}, {Morel},
  {Morgenthaler}, {Morley}, {Morris}, {Mulone}, {Muraveva}, {Musella},
  {Narbonne}, {Nelemans}, {Nicastro}, {Noval}, {Ord{\'e}novic},
  {Ordieres-Mer{\'e}}, {Osborne}, {Pagani}, {Pagano}, {Pailler}, {Palacin},
  {Palaversa}, {Parsons}, {Paulsen}, {Pecoraro}, {Pedrosa}, {Pentik{\"a}inen},
  {Pereira}, {Pichon}, {Piersimoni}, {Pineau}, {Plachy}, {Plum}, {Poujoulet},
  {Pr{\v{s}}a}, {Pulone}, {Ragaini}, {Rago}, {Rambaux}, {Ramos-Lerate},
  {Ranalli}, {Rauw}, {Read}, {Regibo}, {Renk}, {Reyl{\'e}}, {Ribeiro},
  {Rimoldini}, {Ripepi}, {Riva}, {Rixon}, {Roelens}, {Romero-G{\'o}mez},
  {Rowell}, {Royer}, {Rudolph}, {Ruiz-Dern}, {Sadowski}, {Sagrist{\`a}
  Sell{\'e}s}, {Sahlmann}, {Salgado}, {Salguero}, {Sarasso}, {Savietto},
  {Schnorhk}, {Schultheis}, {Sciacca}, {Segol}, {Segovia}, {Segransan},
  {Serpell}, {Shih}, {Smareglia}, {Smart}, {Smith}, {Solano}, {Solitro},
  {Sordo}, {Soria Nieto}, {Souchay}, {Spagna}, {Spoto}, {Stampa}, {Steele},
  {Steidelm{\"u}ller}, {Stephenson}, {Stoev}, {Suess}, {S{\"u}veges}, {Surdej},
  {Szabados}, {Szegedi-Elek}, {Tapiador}, {Taris}, {Tauran}, {Taylor},
  {Teixeira}, {Terrett}, {Tingley}, {Trager}, {Turon}, {Ulla}, {Utrilla},
  {Valentini}, {van Elteren}, {Van Hemelryck}, {van Leeuwen}, {Varadi},
  {Vecchiato}, {Veljanoski}, {Via}, {Vicente}, {Vogt}, {Voss}, {Votruba},
  {Voutsinas}, {Walmsley}, {Weiler}, {Weingrill}, {Werner}, {Wevers},
  {Whitehead}, {Wyrzykowski}, {Yoldas}, {{\v{Z}}erjal}, {Zucker}, {Zurbach},
  {Zwitter}, {Alecu}, {Allen}, {Allende Prieto}, {Amorim},
  {Anglada-Escud{\'e}}, {Arsenijevic}, {Azaz}, {Balm}, {Beck}, {Bernstein},
  {Bigot}, {Bijaoui}, {Blasco}, {Bonfigli}, {Bono}, {Boudreault}, {Bressan},
  {Brown}, {Brunet}, {Bunclark}, {Buonanno}, {Butkevich}, {Carret}, {Carrion},
  {Chemin}, {Ch{\'e}reau}, {Corcione}, {Darmigny}, {de Boer}, {de Teodoro}, {de
  Zeeuw}, {Delle Luche}, {Domingues}, {Dubath}, {Fodor}, {Fr{\'e}zouls},
  {Fries}, {Fustes}, {Fyfe}, {Gallardo}, {Gallegos}, {Gardiol}, {Gebran},
  {Gomboc}, {G{\'o}mez}, {Grux}, {Gueguen}, {Heyrovsky}, {Hoar}, {Iannicola},
  {Isasi Parache}, {Janotto}, {Joliet}, {Jonckheere}, {Keil}, {Kim},
  {Klagyivik}, {Klar}, {Knude}, {Kochukhov}, {Kolka}, {Kos}, {Kutka}, {Lainey},
  {LeBouquin}, {Liu}, {Loreggia}, {Makarov}, {Marseille}, {Martayan},
  {Martinez-Rubi}, {Massart}, {Meynadier}, {Mignot}, {Munari}, {Nguyen},
  {Nordlander}, {Ocvirk}, {O'Flaherty}, {Olias Sanz}, {Ortiz}, {Osorio},
  {Oszkiewicz}, {Ouzounis}, {Palmer}, {Park}, {Pasquato}, {Peltzer}, {Peralta},
  {P{\'e}turaud}, {Pieniluoma}, {Pigozzi}, {Poels}, {Prat}, {Prod'homme},
  {Raison}, {Rebordao}, {Risquez}, {Rocca-Volmerange}, {Rosen}, {Ruiz-Fuertes},
  {Russo}, {Sembay}, {Serraller Vizcaino}, {Short}, {Siebert}, {Silva},
  {Sinachopoulos}, {Slezak}, {Soffel}, {Sosnowska}, {Strai{\v{z}}ys}, {ter
  Linden}, {Terrell}, {Theil}, {Tiede}, {Troisi}, {Tsalmantza}, {Tur},
  {Vaccari}, {Vachier}, {Valles}, {Van Hamme}, {Veltz}, {Virtanen}, {Wallut},
  {Wichmann}, {Wilkinson}, {Ziaeepour}, \& {Zschocke}}]{2016AandA...595A...1G}
{Gaia Collaboration}, {Prusti}, T., {de Bruijne}, J.~H.~J., {et~al.} 2016,
  \aap, 595, A1, \dodoi{10.1051/0004-6361/201629272}

\bibitem[{{Gaia Collaboration} {et~al.}(2018){Gaia Collaboration}, {Brown},
  {Vallenari}, {Prusti}, {de Bruijne}, {Babusiaux}, {Bailer-Jones}, {Biermann},
  {Evans}, {Eyer}, \& et~al.}]{2018AandA...616A...1G}
{Gaia Collaboration}, {Brown}, A.~G.~A., {Vallenari}, A., {et~al.} 2018, \aap,
  616, A1, \dodoi{10.1051/0004-6361/201833051}

\bibitem[{{Gaidos} \& {Mann}(2014)}]{2014ApJ...791...54G}
{Gaidos}, E., \& {Mann}, A.~W. 2014, \apj, 791, 54,
  \dodoi{10.1088/0004-637X/791/1/54}

\bibitem[{{Gaidos}(1998)}]{1998PASP..110.1259G}
{Gaidos}, E.~J. 1998, \pasp, 110, 1259, \dodoi{10.1086/316251}

\bibitem[{{Geballe} {et~al.}(2009){Geballe}, {Saumon}, {Golimowski}, {Leggett},
  {Marley}, \& {Noll}}]{2009ApJ...695..844G}
{Geballe}, T.~R., {Saumon}, D., {Golimowski}, D.~A., {et~al.} 2009, \apj, 695,
  844, \dodoi{10.1088/0004-637X/695/2/844}

\bibitem[{{Geballe} {et~al.}(2001){Geballe}, {Saumon}, {Leggett}, {Knapp},
  {Marley}, \& {Lodders}}]{2001ApJ...556..373G}
{Geballe}, T.~R., {Saumon}, D., {Leggett}, S.~K., {et~al.} 2001, \apj, 556,
  373, \dodoi{10.1086/321575}

\bibitem[{{Ghezzi} {et~al.}(2010){Ghezzi}, {Cunha}, {Smith}, {de Ara{\'u}jo},
  {Schuler}, \& {de la Reza}}]{2010ApJ...720.1290G}
{Ghezzi}, L., {Cunha}, K., {Smith}, V.~V., {et~al.} 2010, \apj, 720, 1290,
  \dodoi{10.1088/0004-637X/720/2/1290}

\bibitem[{{Goldman} {et~al.}(2010){Goldman}, {Marsat}, {Henning}, {Clemens}, \&
  {Greiner}}]{2010MNRAS.405.1140G}
{Goldman}, B., {Marsat}, S., {Henning}, T., {Clemens}, C., \& {Greiner}, J.
  2010, \mnras, 405, 1140, \dodoi{10.1111/j.1365-2966.2010.16524.x}

\bibitem[{{Goodman} \& {Weare}(2010)}]{2010CAMCS...5...65G}
{Goodman}, J., \& {Weare}, J. 2010, Communications in Applied Mathematics and
  Computational Science, 5, 65, \dodoi{10.2140/camcos.2010.5.65}

\bibitem[{{Gordon} \& {McBride}(1994)}]{gordon94}
{Gordon}, S., \& {McBride}, B.~J. 1994, NASA Reference Publ.~1311

\bibitem[{{Gray}(2008)}]{2008oasp.book.....G}
{Gray}, D.~F. 2008, {The Observation and Analysis of Stellar Photospheres}

\bibitem[{{Gray} {et~al.}(2003){Gray}, {Corbally}, {Garrison}, {McFadden}, \&
  {Robinson}}]{2003AJ....126.2048G}
{Gray}, R.~O., {Corbally}, C.~J., {Garrison}, R.~F., {McFadden}, M.~T., \&
  {Robinson}, P.~E. 2003, \aj, 126, 2048, \dodoi{10.1086/378365}

\bibitem[{{Gully-Santiago} {et~al.}(2017){Gully-Santiago}, {Herczeg},
  {Czekala}, {Somers}, {Grankin}, {Covey}, {Donati}, {Alencar}, {Hussain},
  {Shappee}, {Mace}, {Lee}, {Holoien}, {Jose}, \& {Liu}}]{2017ApJ...836..200G}
{Gully-Santiago}, M.~A., {Herczeg}, G.~J., {Czekala}, I., {et~al.} 2017, \apj,
  836, 200, \dodoi{10.3847/1538-4357/836/2/200}

\bibitem[{{Habib} {et~al.}(2007){Habib}, {Heitmann}, {Higdon}, {Nakhleh}, \&
  {Williams}}]{2007PhRvD..76h3503H}
{Habib}, S., {Heitmann}, K., {Higdon}, D., {Nakhleh}, C., \& {Williams}, B.
  2007, \prd, 76, 083503, \dodoi{10.1103/PhysRevD.76.083503}

\bibitem[{{Heitmann} {et~al.}(2009){Heitmann}, {Higdon}, {White}, {Habib},
  {Williams}, {Lawrence}, \& {Wagner}}]{2009ApJ...705..156H}
{Heitmann}, K., {Higdon}, D., {White}, M., {et~al.} 2009, \apj, 705, 156,
  \dodoi{10.1088/0004-637X/705/1/156}

\bibitem[{{Hempelmann} {et~al.}(1995){Hempelmann}, {Schmitt}, {Schultz},
  {Ruediger}, \& {Stepien}}]{1995AandA...294..515H}
{Hempelmann}, A., {Schmitt}, J.~H.~M.~M., {Schultz}, M., {Ruediger}, G., \&
  {Stepien}, K. 1995, \aap, 294, 515

\bibitem[{{Henry} {et~al.}(1996){Henry}, {Soderblom}, {Donahue}, \&
  {Baliunas}}]{1996AJ....111..439H}
{Henry}, T.~J., {Soderblom}, D.~R., {Donahue}, R.~A., \& {Baliunas}, S.~L.
  1996, \aj, 111, 439, \dodoi{10.1086/117796}

\bibitem[{{Hewett} {et~al.}(2006){Hewett}, {Warren}, {Leggett}, \&
  {Hodgkin}}]{2006MNRAS.367..454H}
{Hewett}, P.~C., {Warren}, S.~J., {Leggett}, S.~K., \& {Hodgkin}, S.~T. 2006,
  \mnras, 367, 454, \dodoi{10.1111/j.1365-2966.2005.09969.x}

\bibitem[{Hunter(2007)}]{Hunter:2007}
Hunter, J.~D. 2007, Computing in Science \& Engineering, 9, 90,
  \dodoi{10.1109/MCSE.2007.55}

\bibitem[{{Ivezi{\'c}} {et~al.}(2014){Ivezi{\'c}}, {Connelly}, {Vand erPlas},
  \& {Gray}}]{2014sdmm.book.....I}
{Ivezi{\'c}}, {\v{Z}}., {Connelly}, A.~J., {Vand erPlas}, J.~T., \& {Gray}, A.
  2014, {Statistics, Data Mining, and Machine Learning in Astronomy}

\bibitem[{{James}(1964)}]{1964ApJ...140..552J}
{James}, R.~A. 1964, \apj, 140, 552, \dodoi{10.1086/147949}

\bibitem[{{Jarrett} {et~al.}(2011){Jarrett}, {Cohen}, {Masci}, {Wright},
  {Stern}, {Benford}, {Blain}, {Carey}, {Cutri}, \&
  {Eisenhardt}}]{2011ApJ...735..112J}
{Jarrett}, T.~H., {Cohen}, M., {Masci}, F., {et~al.} 2011, \apj, 735, 112,
  \dodoi{10.1088/0004-637X/735/2/112}

\bibitem[{Jones {et~al.}(2001)Jones, Oliphant, Peterson, {et~al.}}]{scipy}
Jones, E., Oliphant, T., Peterson, P., {et~al.} 2001, {SciPy}: Open source
  scientific tools for {Python}.
\newblock \url{http://www.scipy.org/}

\bibitem[{{Jose} {et~al.}(2020){Jose}, {Biller}, {Albert}, {Dubber}, {Allers},
  {Herczeg}, {Liu}, {Pearson}, {Lalchand}, {Chen}, {Bonnefoy}, {Artigau},
  {Delorme}, {Chiang}, {Zhang}, \& {Oasa}}]{2020ApJ...892..122J}
{Jose}, J., {Biller}, B.~A., {Albert}, L., {et~al.} 2020, \apj, 892, 122,
  \dodoi{10.3847/1538-4357/ab74dd}

\bibitem[{{Kiraga} \& {Stepien}(2007)}]{2007AcA....57..149K}
{Kiraga}, M., \& {Stepien}, K. 2007, \actaa, 57, 149.
\newblock \doarXiv{0707.2577}

\bibitem[{{Kirkpatrick} {et~al.}(2019){Kirkpatrick}, {Martin}, {Smart},
  {Cayago}, {Beichman}, {Marocco}, {Gelino}, {Faherty}, {Cushing}, \&
  {Schneider}}]{2019ApJS..240...19K}
{Kirkpatrick}, J.~D., {Martin}, E.~C., {Smart}, R.~L., {et~al.} 2019, \apjs,
  240, 19, \dodoi{10.3847/1538-4365/aaf6af}

\bibitem[{{Kreidberg} {et~al.}(2014){Kreidberg}, {Bean}, {D{\'e}sert},
  {Benneke}, {Deming}, {Stevenson}, {Seager}, {Berta-Thompson}, {Seifahrt}, \&
  {Homeier}}]{2014Natur.505...69K}
{Kreidberg}, L., {Bean}, J.~L., {D{\'e}sert}, J.-M., {et~al.} 2014, \nat, 505,
  69, \dodoi{10.1038/nature12888}

\bibitem[{{Lawrence} {et~al.}(2007){Lawrence}, {Warren}, {Almaini}, {Edge},
  {Hambly}, {Jameson}, {Lucas}, {Casali}, {Adamson}, {Dye}, {Emerson},
  {Foucaud}, {Hewett}, {Hirst}, {Hodgkin}, {Irwin}, {Lodieu}, {McMahon},
  {Simpson}, {Smail}, {Mortlock}, \& {Folger}}]{2007MNRAS.379.1599L}
{Lawrence}, A., {Warren}, S.~J., {Almaini}, O., {et~al.} 2007, \mnras, 379,
  1599, \dodoi{10.1111/j.1365-2966.2007.12040.x}

\bibitem[{{Lawrence} {et~al.}(2012){Lawrence}, {Warren}, {Almaini}, {Edge},
  {Hambly}, {Jameson}, {Lucas}, {Casali}, {Adamson}, {Dye}, {Emerson},
  {Foucaud}, {Hewett}, {Hirst}, {Hodgkin}, {Irwin}, {Lodieu}, {McMahon},
  {Simpson}, {Smail}, {Mortlock}, \& {Folger}}]{2012yCat.2314....0L}
---. 2012, VizieR Online Data Catalog, II/314

\bibitem[{{Leggett} {et~al.}(2007){Leggett}, {Marley}, {Freedman}, {Saumon},
  {Liu}, {Geballe}, {Golimowski}, \& {Stephens}}]{2007ApJ...667..537L}
{Leggett}, S.~K., {Marley}, M.~S., {Freedman}, R., {et~al.} 2007, \apj, 667,
  537, \dodoi{10.1086/519948}

\bibitem[{{Leggett} {et~al.}(2017){Leggett}, {Tremblin}, {Esplin}, {Luhman}, \&
  {Morley}}]{2017ApJ...842..118L}
{Leggett}, S.~K., {Tremblin}, P., {Esplin}, T.~L., {Luhman}, K.~L., \&
  {Morley}, C.~V. 2017, \apj, 842, 118, \dodoi{10.3847/1538-4357/aa6fb5}

\bibitem[{{Leggett} {et~al.}(2010){Leggett}, {Burningham}, {Saumon}, {Marley},
  {Warren}, {Smart}, {Jones}, {Lucas}, {Pinfield}, \&
  {Tamura}}]{2010ApJ...710.1627L}
{Leggett}, S.~K., {Burningham}, B., {Saumon}, D., {et~al.} 2010, \apj, 710,
  1627, \dodoi{10.1088/0004-637X/710/2/1627}

\bibitem[{{Lenzuni} {et~al.}(1991){Lenzuni}, {Chernoff}, \&
  {Salpeter}}]{1991ApJS...76..759L}
{Lenzuni}, P., {Chernoff}, D.~F., \& {Salpeter}, E.~E. 1991, \apjs, 76, 759,
  \dodoi{10.1086/191580}

\bibitem[{{Ligi} {et~al.}(2016){Ligi}, {Creevey}, {Mourard}, {Crida},
  {Lagrange}, {Nardetto}, {Perraut}, {Schultheis}, {Tallon-Bosc}, \& {ten
  Brummelaar}}]{2016AandA...586A..94L}
{Ligi}, R., {Creevey}, O., {Mourard}, D., {et~al.} 2016, \aap, 586, A94,
  \dodoi{10.1051/0004-6361/201527054}

\bibitem[{{Lindegren}(2018)}]{Lindegren2018}
{Lindegren}, L. 2018, Gaia Technical Note: GAIA-C3-TN-LU-LL-124-01

\bibitem[{{Line} {et~al.}(2015){Line}, {Teske}, {Burningham}, {Fortney}, \&
  {Marley}}]{2015ApJ...807..183L}
{Line}, M.~R., {Teske}, J., {Burningham}, B., {Fortney}, J.~J., \& {Marley},
  M.~S. 2015, \apj, 807, 183, \dodoi{10.1088/0004-637X/807/2/183}

\bibitem[{{Line} {et~al.}(2016){Line}, {Stevenson}, {Bean}, {Desert},
  {Fortney}, {Kreidberg}, {Madhusudhan}, {Showman}, \&
  {Diamond-Lowe}}]{2016AJ....152..203L}
{Line}, M.~R., {Stevenson}, K.~B., {Bean}, J., {et~al.} 2016, \aj, 152, 203,
  \dodoi{10.3847/0004-6256/152/6/203}

\bibitem[{{Line} {et~al.}(2017){Line}, {Marley}, {Liu}, {Burningham}, {Morley},
  {Hinkel}, {Teske}, {Fortney}, {Freedman}, \& {Lupu}}]{2017ApJ...848...83L}
{Line}, M.~R., {Marley}, M.~S., {Liu}, M.~C., {et~al.} 2017, \apj, 848, 83,
  \dodoi{10.3847/1538-4357/aa7ff0}

\bibitem[{{Liu} {et~al.}(2016){Liu}, {Dupuy}, \&
  {Allers}}]{2016ApJ...833...96L}
{Liu}, M.~C., {Dupuy}, T.~J., \& {Allers}, K.~N. 2016, \apj, 833, 96,
  \dodoi{10.3847/1538-4357/833/1/96}

\bibitem[{{Liu} {et~al.}(2008){Liu}, {Dupuy}, \&
  {Ireland}}]{2008ApJ...689..436L}
{Liu}, M.~C., {Dupuy}, T.~J., \& {Ireland}, M.~J. 2008, \apj, 689, 436,
  \dodoi{10.1086/591837}

\bibitem[{{Liu} {et~al.}(2007){Liu}, {Leggett}, \&
  {Chiu}}]{2007ApJ...660.1507L}
{Liu}, M.~C., {Leggett}, S.~K., \& {Chiu}, K. 2007, \apj, 660, 1507,
  \dodoi{10.1086/512662}

\bibitem[{{Liu} {et~al.}(2011){Liu}, {Delorme}, {Dupuy}, {Bowler}, {Albert},
  {Artigau}, {Reyl{\'e}}, {Forveille}, \& {Delfosse}}]{2011ApJ...740..108L}
{Liu}, M.~C., {Delorme}, P., {Dupuy}, T.~J., {et~al.} 2011, \apj, 740, 108,
  \dodoi{10.1088/0004-637X/740/2/108}

\bibitem[{{Lodders}(1999)}]{1999ApJ...519..793L}
{Lodders}, K. 1999, \apj, 519, 793, \dodoi{10.1086/307387}

\bibitem[{{Lodieu}(2013)}]{2013MNRAS.431.3222L}
{Lodieu}, N. 2013, \mnras, 431, 3222, \dodoi{10.1093/mnras/stt402}

\bibitem[{{Luhman} \& {Mamajek}(2012)}]{2012ApJ...758...31L}
{Luhman}, K.~L., \& {Mamajek}, E.~E. 2012, \apj, 758, 31,
  \dodoi{10.1088/0004-637X/758/1/31}

\bibitem[{{Luhman} {et~al.}(2009){Luhman}, {Mamajek}, {Allen}, \&
  {Cruz}}]{2009ApJ...703..399L}
{Luhman}, K.~L., {Mamajek}, E.~E., {Allen}, P.~R., \& {Cruz}, K.~L. 2009, \apj,
  703, 399, \dodoi{10.1088/0004-637X/703/1/399}

\bibitem[{{Luhman} {et~al.}(2007){Luhman}, {Patten}, {Marengo}, {Schuster},
  {Hora}, {Ellis}, {Stauffer}, {Sonnett}, {Winston}, {Gutermuth}, {Megeath},
  {Backman}, {Henry}, {Werner}, \& {Fazio}}]{2007ApJ...654..570L}
{Luhman}, K.~L., {Patten}, B.~M., {Marengo}, M., {et~al.} 2007, \apj, 654, 570,
  \dodoi{10.1086/509073}

\bibitem[{{MacDonald} {et~al.}(2018){MacDonald}, {Marley}, {Fortney}, \&
  {Lewis}}]{2018ApJ...858...69M}
{MacDonald}, R.~J., {Marley}, M.~S., {Fortney}, J.~J., \& {Lewis}, N.~K. 2018,
  \apj, 858, 69, \dodoi{10.3847/1538-4357/aabb05}

\bibitem[{{Mainzer} {et~al.}(2011){Mainzer}, {Cushing}, {Skrutskie}, {Gelino},
  {Kirkpatrick}, {Jarrett}, {Masci}, {Marley}, {Saumon}, {Wright}, {Beaton},
  {Dietrich}, {Eisenhardt}, {Garnavich}, {Kuhn}, {Leisawitz}, {Marsh},
  {McLean}, {Padgett}, \& {Rueff}}]{2011ApJ...726...30M}
{Mainzer}, A., {Cushing}, M.~C., {Skrutskie}, M., {et~al.} 2011, \apj, 726, 30,
  \dodoi{10.1088/0004-637X/726/1/30}

\bibitem[{{Mamajek} \& {Hillenbrand}(2008)}]{2008ApJ...687.1264M}
{Mamajek}, E.~E., \& {Hillenbrand}, L.~A. 2008, \apj, 687, 1264,
  \dodoi{10.1086/591785}

\bibitem[{{Manjavacas} {et~al.}(2016){Manjavacas}, {Goldman}, {Alcal{\'a}},
  {Zapatero-Osorio}, {B{\'e}jar}, {Homeier}, {Bonnefoy}, {Smart}, {Henning}, \&
  {Allard}}]{2016MNRAS.455.1341M}
{Manjavacas}, E., {Goldman}, B., {Alcal{\'a}}, J.~M., {et~al.} 2016, \mnras,
  455, 1341, \dodoi{10.1093/mnras/stv2048}

\bibitem[{{Marley} {et~al.}(1999){Marley}, {Gelino}, {Stephens}, {Lunine}, \&
  {Freedman}}]{1999ApJ...513..879M}
{Marley}, M.~S., {Gelino}, C., {Stephens}, D., {Lunine}, J.~I., \& {Freedman},
  R. 1999, \apj, 513, 879, \dodoi{10.1086/306881}

\bibitem[{{Marley} \& {McKay}(1999)}]{1999Icar..138..268M}
{Marley}, M.~S., \& {McKay}, C.~P. 1999, \icarus, 138, 268,
  \dodoi{10.1006/icar.1998.6071}

\bibitem[{{Marley} \& {Robinson}(2015)}]{2015ARA&A..53..279M}
{Marley}, M.~S., \& {Robinson}, T.~D. 2015, \araa, 53, 279,
  \dodoi{10.1146/annurev-astro-082214-122522}

\bibitem[{{Marley} {et~al.}(2012){Marley}, {Saumon}, {Cushing}, {Ackerman},
  {Fortney}, \& {Freedman}}]{2012ApJ...754..135M}
{Marley}, M.~S., {Saumon}, D., {Cushing}, M., {et~al.} 2012, \apj, 754, 135,
  \dodoi{10.1088/0004-637X/754/2/135}

\bibitem[{{Marley} {et~al.}(2017){Marley}, {Saumon}, {Fortney}, {Morley},
  {Lupu}, {Freedman}, \& {Visscher}}]{2017AAS...23031507M}
{Marley}, M.~S., {Saumon}, D., {Fortney}, J.~J., {et~al.} 2017, in American
  Astronomical Society Meeting Abstracts, Vol. 230, American Astronomical
  Society Meeting Abstracts \#230, 315.07

\bibitem[{{Marley} {et~al.}(1996){Marley}, {Saumon}, {Guillot}, {Freedman},
  {Hubbard}, {Burrows}, \& {Lunine}}]{1996Sci...272.1919M}
{Marley}, M.~S., {Saumon}, D., {Guillot}, T., {et~al.} 1996, Science, 272,
  1919, \dodoi{10.1126/science.272.5270.1919}

\bibitem[{{Marley} {et~al.}(2002){Marley}, {Seager}, {Saumon}, {Lodders},
  {Ackerman}, {Freedman}, \& {Fan}}]{2002ApJ...568..335M}
{Marley}, M.~S., {Seager}, S., {Saumon}, D., {et~al.} 2002, \apj, 568, 335,
  \dodoi{10.1086/338800}

\bibitem[{{Marois} {et~al.}(2008){Marois}, {Macintosh}, {Barman}, {Zuckerman},
  {Song}, {Patience}, {Lafreni{\`e}re}, \& {Doyon}}]{2008Sci...322.1348M}
{Marois}, C., {Macintosh}, B., {Barman}, T., {et~al.} 2008, Science, 322, 1348,
  \dodoi{10.1126/science.1166585}

\bibitem[{{McKay} {et~al.}(1989){McKay}, {Pollack}, \&
  {Courtin}}]{1989Icar...80...23M}
{McKay}, C.~P., {Pollack}, J.~B., \& {Courtin}, R. 1989, \icarus, 80, 23,
  \dodoi{10.1016/0019-1035(89)90160-7}

\bibitem[{{Mishenina} {et~al.}(2012){Mishenina}, {Soubiran}, {Kovtyukh},
  {Katsova}, \& {Livshits}}]{2012AandA...547A.106M}
{Mishenina}, T.~V., {Soubiran}, C., {Kovtyukh}, V.~V., {Katsova}, M.~M., \&
  {Livshits}, M.~A. 2012, \aap, 547, A106, \dodoi{10.1051/0004-6361/201118412}

\bibitem[{{Molli{\`e}re} {et~al.}(2015){Molli{\`e}re}, {van Boekel},
  {Dullemond}, {Henning}, \& {Mordasini}}]{2015ApJ...813...47M}
{Molli{\`e}re}, P., {van Boekel}, R., {Dullemond}, C., {Henning}, T., \&
  {Mordasini}, C. 2015, \apj, 813, 47, \dodoi{10.1088/0004-637X/813/1/47}

\bibitem[{{Montes} {et~al.}(2001){Montes}, {L{\'o}pez-Santiago}, {G{\'a}lvez},
  {Fern{\'a}ndez-Figueroa}, {De Castro}, \& {Cornide}}]{2001MNRAS.328...45M}
{Montes}, D., {L{\'o}pez-Santiago}, J., {G{\'a}lvez}, M.~C., {et~al.} 2001,
  \mnras, 328, 45, \dodoi{10.1046/j.1365-8711.2001.04781.x}

\bibitem[{{Montet} {et~al.}(2016){Montet}, {Johnson}, {Fortney}, \&
  {Desert}}]{2016ApJ...822L...6M}
{Montet}, B.~T., {Johnson}, J.~A., {Fortney}, J.~J., \& {Desert}, J.-M. 2016,
  \apjl, 822, L6, \dodoi{10.3847/2041-8205/822/1/L6}

\bibitem[{{Morel}(1997)}]{1997AandAS..124..597M}
{Morel}, P. 1997, \aaps, 124, 597, \dodoi{10.1051/aas:1997209}

\bibitem[{{Morley} {et~al.}(2012){Morley}, {Fortney}, {Marley}, {Visscher},
  {Saumon}, \& {Leggett}}]{2012ApJ...756..172M}
{Morley}, C.~V., {Fortney}, J.~J., {Marley}, M.~S., {et~al.} 2012, \apj, 756,
  172, \dodoi{10.1088/0004-637X/756/2/172}

\bibitem[{{Moses} {et~al.}(2013){Moses}, {Madhusudhan}, {Visscher}, \&
  {Freedman}}]{2013ApJ...763...25M}
{Moses}, J.~I., {Madhusudhan}, N., {Visscher}, C., \& {Freedman}, R.~S. 2013,
  \apj, 763, 25, \dodoi{10.1088/0004-637X/763/1/25}

\bibitem[{{Mugrauer} {et~al.}(2006){Mugrauer}, {Seifahrt}, {Neuh{\"a}user}, \&
  {Mazeh}}]{2006MNRAS.373L..31M}
{Mugrauer}, M., {Seifahrt}, A., {Neuh{\"a}user}, R., \& {Mazeh}, T. 2006,
  \mnras, 373, L31, \dodoi{10.1111/j.1745-3933.2006.00237.x}

\bibitem[{Oliphant(2006)}]{numpy}
Oliphant, T. 2006, {NumPy}: A guide to {NumPy}, USA: Trelgol Publishing.
\newblock \url{http://www.numpy.org/}

\bibitem[{{Parmentier} {et~al.}(2016){Parmentier}, {Fortney}, {Showman},
  {Morley}, \& {Marley}}]{2016ApJ...828...22P}
{Parmentier}, V., {Fortney}, J.~J., {Showman}, A.~P., {Morley}, C., \&
  {Marley}, M.~S. 2016, \apj, 828, 22, \dodoi{10.3847/0004-637X/828/1/22}

\bibitem[{P\'erez \& Granger(2007)}]{PER-GRA:2007}
P\'erez, F., \& Granger, B.~E. 2007, Computing in Science and Engineering, 9,
  21, \dodoi{10.1109/MCSE.2007.53}

\bibitem[{{Petigura} \& {Marcy}(2011)}]{2011ApJ...735...41P}
{Petigura}, E.~A., \& {Marcy}, G.~W. 2011, \apj, 735, 41,
  \dodoi{10.1088/0004-637X/735/1/41}

\bibitem[{{Phillips} {et~al.}(2020){Phillips}, {Tremblin}, {Baraffe},
  {Chabrier}, {Allard}, {Spiegelman}, {Goyal}, {Drummond}, \&
  {H{\'e}brard}}]{2020A&A...637A..38P}
{Phillips}, M.~W., {Tremblin}, P., {Baraffe}, I., {et~al.} 2020, \aap, 637,
  A38, \dodoi{10.1051/0004-6361/201937381}

\bibitem[{{Pietrinferni} {et~al.}(2004){Pietrinferni}, {Cassisi}, {Salaris}, \&
  {Castelli}}]{2004ApJ...612..168P}
{Pietrinferni}, A., {Cassisi}, S., {Salaris}, M., \& {Castelli}, F. 2004, \apj,
  612, 168, \dodoi{10.1086/422498}

\bibitem[{{Pietrinferni} {et~al.}(2006){Pietrinferni}, {Cassisi}, {Salaris}, \&
  {Castelli}}]{2006ApJ...642..797P}
---. 2006, \apj, 642, 797, \dodoi{10.1086/501344}

\bibitem[{{Pietrinferni} {et~al.}(2009){Pietrinferni}, {Cassisi}, {Salaris},
  {Percival}, \& {Ferguson}}]{2009ApJ...697..275P}
{Pietrinferni}, A., {Cassisi}, S., {Salaris}, M., {Percival}, S., \&
  {Ferguson}, J.~W. 2009, \apj, 697, 275, \dodoi{10.1088/0004-637X/697/1/275}

\bibitem[{{Piffl} {et~al.}(2014){Piffl}, {Scannapieco}, {Binney}, {Steinmetz},
  {Scholz}, {Williams}, {de Jong}, {Kordopatis}, {Matijevi{\v{c}}},
  {Bienaym{\'e}}, {Bland-Hawthorn}, {Boeche}, {Freeman}, {Gibson}, {Gilmore},
  {Grebel}, {Helmi}, {Munari}, {Navarro}, {Parker}, {Reid}, {Seabroke},
  {Watson}, {Wyse}, \& {Zwitter}}]{2014A&A...562A..91P}
{Piffl}, T., {Scannapieco}, C., {Binney}, J., {et~al.} 2014, \aap, 562, A91,
  \dodoi{10.1051/0004-6361/201322531}

\bibitem[{{Pinfield} {et~al.}(2006){Pinfield}, {Jones}, {Lucas}, {Kendall},
  {Folkes}, {Day-Jones}, {Chappelle}, \& {Steele}}]{2006MNRAS.368.1281P}
{Pinfield}, D.~J., {Jones}, H.~R.~A., {Lucas}, P.~W., {et~al.} 2006, \mnras,
  368, 1281, \dodoi{10.1111/j.1365-2966.2006.10213.x}

\bibitem[{{Pizzolato} {et~al.}(2003){Pizzolato}, {Maggio}, {Micela},
  {Sciortino}, \& {Ventura}}]{2003AandA...397..147P}
{Pizzolato}, N., {Maggio}, A., {Micela}, G., {Sciortino}, S., \& {Ventura}, P.
  2003, \aap, 397, 147, \dodoi{10.1051/0004-6361:20021560}

\bibitem[{{Prugniel} {et~al.}(2011){Prugniel}, {Vauglin}, \&
  {Koleva}}]{2011AandA...531A.165P}
{Prugniel}, P., {Vauglin}, I., \& {Koleva}, M. 2011, \aap, 531, A165,
  \dodoi{10.1051/0004-6361/201116769}

\bibitem[{{Ram{\'\i}rez} {et~al.}(2013){Ram{\'\i}rez}, {Allende Prieto}, \&
  {Lambert}}]{2013ApJ...764...78R}
{Ram{\'\i}rez}, I., {Allende Prieto}, C., \& {Lambert}, D.~L. 2013, \apj, 764,
  78, \dodoi{10.1088/0004-637X/764/1/78}

\bibitem[{{Rayner} {et~al.}(2003){Rayner}, {Toomey}, {Onaka}, {Denault},
  {Stahlberger}, {Vacca}, {Cushing}, \& {Wang}}]{2003PASP..115..362R}
{Rayner}, J.~T., {Toomey}, D.~W., {Onaka}, P.~M., {et~al.} 2003, \pasp, 115,
  362, \dodoi{10.1086/367745}

\bibitem[{{Rice} {et~al.}(2010){Rice}, {Barman}, {Mclean}, {Prato}, \&
  {Kirkpatrick}}]{2010ApJS..186...63R}
{Rice}, E.~L., {Barman}, T., {Mclean}, I.~S., {Prato}, L., \& {Kirkpatrick},
  J.~D. 2010, \apjs, 186, 63, \dodoi{10.1088/0067-0049/186/1/63}

\bibitem[{{Rocha-Pinto} {et~al.}(2004){Rocha-Pinto}, {Flynn}, {Scalo},
  {H{\"a}nninen}, {Maciel}, \& {Hensler}}]{2004AandA...423..517R}
{Rocha-Pinto}, H.~J., {Flynn}, C., {Scalo}, J., {et~al.} 2004, \aap, 423, 517,
  \dodoi{10.1051/0004-6361:20035617}

\bibitem[{{Salasnich} {et~al.}(2000){Salasnich}, {Girardi}, {Weiss}, \&
  {Chiosi}}]{2000AandA...361.1023S}
{Salasnich}, B., {Girardi}, L., {Weiss}, A., \& {Chiosi}, C. 2000, \aap, 361,
  1023.
\newblock \doarXiv{astro-ph/0007388}

\bibitem[{{Samland} {et~al.}(2017{\natexlab{a}}){Samland}, {Molli{\`e}re},
  {Bonnefoy}, {Maire}, {Cantalloube}, {Cheetham}, {Mesa}, {Gratton}, {Biller},
  \& {Wahhaj}}]{2017A&A...603A..57S}
{Samland}, M., {Molli{\`e}re}, P., {Bonnefoy}, M., {et~al.} 2017{\natexlab{a}},
  \aap, 603, A57, \dodoi{10.1051/0004-6361/201629767}

\bibitem[{{Samland} {et~al.}(2017{\natexlab{b}}){Samland}, {Molli{\`e}re},
  {Bonnefoy}, {Maire}, {Cantalloube}, {Cheetham}, {Mesa}, {Gratton}, {Biller},
  {Wahhaj}, {Bouwman}, {Brandner}, {Melnick}, {Carson}, {Janson}, {Henning},
  {Homeier}, {Mordasini}, {Langlois}, {Quanz}, {van Boekel}, {Zurlo},
  {Schlieder}, {Avenhaus}, {Beuzit}, {Boccaletti}, {Bonavita}, {Chauvin},
  {Claudi}, {Cudel}, {Desidera}, {Feldt}, {Fusco}, {Galicher}, {Kopytova},
  {Lagrange}, {Le Coroller}, {Martinez}, {Moeller-Nilsson}, {Mouillet},
  {Mugnier}, {Perrot}, {Sevin}, {Sissa}, {Vigan}, \&
  {Weber}}]{2017AandA...603A..57S}
---. 2017{\natexlab{b}}, \aap, 603, A57, \dodoi{10.1051/0004-6361/201629767}

\bibitem[{{Santos} {et~al.}(2004){Santos}, {Israelian}, \&
  {Mayor}}]{2004AandA...415.1153S}
{Santos}, N.~C., {Israelian}, G., \& {Mayor}, M. 2004, \aap, 415, 1153,
  \dodoi{10.1051/0004-6361:20034469}

\bibitem[{{Santos} {et~al.}(2005){Santos}, {Israelian}, {Mayor}, {Bento},
  {Almeida}, {Sousa}, \& {Ecuvillon}}]{2005AandA...437.1127S}
{Santos}, N.~C., {Israelian}, G., {Mayor}, M., {et~al.} 2005, \aap, 437, 1127,
  \dodoi{10.1051/0004-6361:20052895}

\bibitem[{{Saumon} {et~al.}(2000){Saumon}, {Geballe}, {Leggett}, {Marley},
  {Freedman}, {Lodders}, {Fegley}, \& {Sengupta}}]{2000ApJ...541..374S}
{Saumon}, D., {Geballe}, T.~R., {Leggett}, S.~K., {et~al.} 2000, \apj, 541,
  374, \dodoi{10.1086/309410}

\bibitem[{{Saumon} \& {Marley}(2008)}]{2008ApJ...689.1327S}
{Saumon}, D., \& {Marley}, M.~S. 2008, \apj, 689, 1327, \dodoi{10.1086/592734}

\bibitem[{{Saumon} {et~al.}(2012){Saumon}, {Marley}, {Abel}, {Frommhold}, \&
  {Freedman}}]{2012ApJ...750...74S}
{Saumon}, D., {Marley}, M.~S., {Abel}, M., {Frommhold}, L., \& {Freedman},
  R.~S. 2012, \apj, 750, 74, \dodoi{10.1088/0004-637X/750/1/74}

\bibitem[{{Saumon} {et~al.}(2006){Saumon}, {Marley}, {Cushing}, {Leggett},
  {Roellig}, {Lodders}, \& {Freedman}}]{2006ApJ...647..552S}
{Saumon}, D., {Marley}, M.~S., {Cushing}, M.~C., {et~al.} 2006, \apj, 647, 552,
  \dodoi{10.1086/505419}

\bibitem[{{Saumon} {et~al.}(2003){Saumon}, {Marley}, {Lodders}, \&
  {Freedman}}]{2003IAUS..211..345S}
{Saumon}, D., {Marley}, M.~S., {Lodders}, K., \& {Freedman}, R.~S. 2003, in IAU
  Symposium, Vol. 211, Brown Dwarfs, ed. E.~{Mart{\'\i}n}, 345

\bibitem[{{Scholz}(2010)}]{2010A&A...515A..92S}
{Scholz}, R.-D. 2010, \aap, 515, A92, \dodoi{10.1051/0004-6361/201014264}

\bibitem[{{Smith} {et~al.}(2007){Smith}, {Ruchti}, {Helmi}, {Wyse},
  {Fulbright}, {Freeman}, {Navarro}, {Seabroke}, {Steinmetz}, {Williams},
  {Bienaym{\'e}}, {Binney}, {Bland -Hawthorn}, {Dehnen}, {Gibson}, {Gilmore},
  {Grebel}, {Munari}, {Parker}, {Scholz}, {Siebert}, {Watson}, \&
  {Zwitter}}]{2007MNRAS.379..755S}
{Smith}, M.~C., {Ruchti}, G.~R., {Helmi}, A., {et~al.} 2007, \mnras, 379, 755,
  \dodoi{10.1111/j.1365-2966.2007.11964.x}

\bibitem[{{Soderblom} {et~al.}(1991){Soderblom}, {Duncan}, \&
  {Johnson}}]{1991ApJ...375..722S}
{Soderblom}, D.~R., {Duncan}, D.~K., \& {Johnson}, D. R.~H. 1991, \apj, 375,
  722, \dodoi{10.1086/170238}

\bibitem[{{Sorahana} \& {Yamamura}(2012)}]{2012ApJ...760..151S}
{Sorahana}, S., \& {Yamamura}, I. 2012, \apj, 760, 151,
  \dodoi{10.1088/0004-637X/760/2/151}

\bibitem[{{Spiegel} {et~al.}(2011){Spiegel}, {Burrows}, \&
  {Milsom}}]{2011ApJ...727...57S}
{Spiegel}, D.~S., {Burrows}, A., \& {Milsom}, J.~A. 2011, \apj, 727, 57,
  \dodoi{10.1088/0004-637X/727/1/57}

\bibitem[{{Stephens} {et~al.}(2009){Stephens}, {Leggett}, {Cushing}, {Marley},
  {Saumon}, {Geballe}, {Golimowski}, {Fan}, \& {Noll}}]{2009ApJ...702..154S}
{Stephens}, D.~C., {Leggett}, S.~K., {Cushing}, M.~C., {et~al.} 2009, \apj,
  702, 154, \dodoi{10.1088/0004-637X/702/1/154}

\bibitem[{{Takeda} {et~al.}(2007){Takeda}, {Ford}, {Sills}, {Rasio}, {Fischer},
  \& {Valenti}}]{2007ApJS..168..297T}
{Takeda}, G., {Ford}, E.~B., {Sills}, A., {et~al.} 2007, \apjs, 168, 297,
  \dodoi{10.1086/509763}

\bibitem[{{Taylor}(2005)}]{2005ASPC..347...29T}
{Taylor}, M.~B. 2005, in Astronomical Society of the Pacific Conference Series,
  Vol. 347, Astronomical Data Analysis Software and Systems XIV, ed.
  P.~{Shopbell}, M.~{Britton}, \& R.~{Ebert}, 29

\bibitem[{{Testi}(2009)}]{2009AandA...503..639T}
{Testi}, L. 2009, \aap, 503, 639, \dodoi{10.1051/0004-6361/200810699}

\bibitem[{{Thor{\'e}n} \& {Feltzing}(2000)}]{2000AandA...363..692T}
{Thor{\'e}n}, P., \& {Feltzing}, S. 2000, \aap, 363, 692.
\newblock \doarXiv{astro-ph/0010067}

\bibitem[{{Tremblin} {et~al.}(2015){Tremblin}, {Amundsen}, {Mourier},
  {Baraffe}, {Chabrier}, {Drummond}, {Homeier}, \&
  {Venot}}]{2015ApJ...804L..17T}
{Tremblin}, P., {Amundsen}, D.~S., {Mourier}, P., {et~al.} 2015, \apjl, 804,
  L17, \dodoi{10.1088/2041-8205/804/1/L17}

\bibitem[{{Triaud} {et~al.}(2020){Triaud}, {Burgasser}, {Burdanov}, {Kunovac
  Hod{\v{z}}i{\'c}}, {Alonso}, {Bardalez Gagliuffi}, {Delrez}, {Demory}, {de
  Wit}, {Ducrot}, {Hessman}, {Husser}, {Jehin}, {Pedersen}, {Queloz},
  {McCormac}, {Murray}, {Sebastian}, {Thompson}, {Van Grootel}, \&
  {Gillon}}]{2020NatAs...4..650T}
{Triaud}, A. H.~M.~J., {Burgasser}, A.~J., {Burdanov}, A., {et~al.} 2020,
  Nature Astronomy, 4, 650, \dodoi{10.1038/s41550-020-1018-2}

\bibitem[{{Tsuji}(2002)}]{2002ApJ...575..264T}
{Tsuji}, T. 2002, \apj, 575, 264, \dodoi{10.1086/341262}

\bibitem[{{Tsuji}(2005)}]{2005ApJ...621.1033T}
---. 2005, \apj, 621, 1033, \dodoi{10.1086/427747}

\bibitem[{{Tsuji} {et~al.}(1999){Tsuji}, {Ohnaka}, \&
  {Aoki}}]{1999ApJ...520L.119T}
{Tsuji}, T., {Ohnaka}, K., \& {Aoki}, W. 1999, \apj, 520, L119,
  \dodoi{10.1086/312161}

\bibitem[{{Valenti} \& {Fischer}(2005)}]{2005ApJS..159..141V}
{Valenti}, J.~A., \& {Fischer}, D.~A. 2005, \apjs, 159, 141,
  \dodoi{10.1086/430500}

\bibitem[{{Vican}(2012)}]{2012AJ....143..135V}
{Vican}, L. 2012, \aj, 143, 135, \dodoi{10.1088/0004-6256/143/6/135}

\bibitem[{{Visscher} {et~al.}(2010){Visscher}, {Lodders}, \&
  {Fegley}}]{2010ApJ...716.1060V}
{Visscher}, C., {Lodders}, K., \& {Fegley}, Bruce, J. 2010, \apj, 716, 1060,
  \dodoi{10.1088/0004-637X/716/2/1060}

\bibitem[{{Wright} {et~al.}(2004){Wright}, {Marcy}, {Butler}, \&
  {Vogt}}]{2004ApJS..152..261W}
{Wright}, J.~T., {Marcy}, G.~W., {Butler}, R.~P., \& {Vogt}, S.~S. 2004, \apjs,
  152, 261, \dodoi{10.1086/386283}

\bibitem[{{Zahnle} \& {Marley}(2014)}]{2014ApJ...797...41Z}
{Zahnle}, K.~J., \& {Marley}, M.~S. 2014, \apj, 797, 41,
  \dodoi{10.1088/0004-637X/797/1/41}

\bibitem[{{Zhang} {et~al.}(2021){Zhang}, {Liu}, {Best}, {Dupuy}, \&
  {Siverd}}]{2021arXiv210205045Z}
{Zhang}, Z., {Liu}, M.~C., {Best}, W. M.~J., {Dupuy}, T.~J., \& {Siverd}, R.~J.
  2021, arXiv e-prints, arXiv:2102.05045.
\newblock \doarXiv{2102.05045}

\bibitem[{{Zhang} {et~al.}(2018){Zhang}, {Liu}, {Best}, {Magnier}, {Aller},
  {Chambers}, {Draper}, {Flewelling}, {Hodapp}, {Kaiser}, {Kudritzki},
  {Metcalfe}, {Wainscoat}, \& {Waters}}]{2018ApJ...858...41Z}
{Zhang}, Z., {Liu}, M.~C., {Best}, W.~M.~J., {et~al.} 2018, \apj, 858, 41,
  \dodoi{10.3847/1538-4357/aab269}

\bibitem[{{Zhang} {et~al.}(2020){Zhang}, {Liu}, {Hermes}, {Magnier}, {Marley},
  {Tremblay}, {Tucker}, {Do}, {Payne}, \& {Shappee}}]{2020ApJ...891..171Z}
{Zhang}, Z., {Liu}, M.~C., {Hermes}, J.~J., {et~al.} 2020, \apj, 891, 171,
  \dodoi{10.3847/1538-4357/ab765c}

\end{thebibliography}

%%%%%%%%%%%%
%%%   TABLES   %%%
%%%%%%%%%%%%
\clearpage
\tabletypesize{\scriptsize}

%\global\pdfpageattr\expandafter{\the\pdfpageattr/Rotate 0} 
\begin{deluxetable}{l|c|c|l} 
\tablewidth{0pc} 
\tablecaption{Metallicity and Age of HD 3651A \label{tab:HD3651}} 
\tablehead{ \multicolumn{1}{l}{}  &  \multicolumn{1}{c}{Value\tablenotemark{a}}  &  \multicolumn{1}{c}{Reference}  &  \multicolumn{1}{l}{Notes\tablenotemark{b}}   } 
\startdata 
Metallicity (dex)  &  $0.13 \pm 0.06$  &  7  &  Spectroscopy ($R \approx 5.0 \times 10^{4}$)  \\ 
  &  $0.12 \pm 0.04$  &  10  &  Spectroscopy ($R \approx 5.7 \times 10^{4}$)  \\ 
  &  $0.12$  &  19  &  Spectroscopy ($R \approx 6.0 \times 10^{4}$)  \\ 
  &  $0.16$  &  21  &  Spectroscopy ($R \approx 5.0 \times 10^{4}$)  \\ 
  &  $0.15 \pm 0.05$  &  23  &  Spectroscopy ($R \approx 4.2 \times 10^{4}$)  \\ 
  &  $0.18 \pm 0.07$  &  24  &  Spectroscopy ($R \approx 4.5 \times 10^{4}$)  \\ 
  &  $0.16 \pm 0.04$  &  27  &  Spectroscopy ($R \approx 6.0 \times 10^{4}$)  \\ 
\hline 
\textbf{Adopted Metallicity (dex)} &  $\mathbf{0.1-0.2}$  &    &   \\ 
\hline \hline 
Isochrone Age (Gyr)  &  $3-12.5$  &  12  &  Y$^{2}$ (Yonsei-Yale) isochrones (8)  \\ 
  &  $5.13$  &  14  &  Padova isochrones (5)  \\ 
  &  $>11.8$  &  16  &  YERC (Yale Stellar Evolution Code) isochrones (17)  \\ 
  &  $6.059$  &  20  &  CESAM isochrones (4)  \\ 
  &  $6.9 \pm 2.8$  &  25  &  PARSEC (Padova and Trieste Stellar Evolutionary Code) isochrones (22)  \\ 
  &  $10.00 \pm 3.52$  &  26  &  PARSEC (Padova and Trieste Stellar Evolutionary Code) isochrones (22)  \\ 
  &  $10.2^{+2.3}_{-5.3}$  &  27  &  Y$^{2}$ (Yonsei-Yale) isochrones (8)  \\ 
age range  &  $3-13$  &    &    \\ 
\hline 
Gyrochronology Age (Gyr)  &  $6.10 \pm 0.99$  &  13  &  Rotation period $44$~days (3)  \\ 
  &  $7.3 \pm 1.8$  &  1  &  Rotation period $44$~days (3)  \\ 
age range  &  $6.1-7.3$  &    &    \\ 
\hline 
Stellar-Activity Age (Gyr)  &  $6.4 - 7.7$  &  18  &  $\log{R^{\prime}_{HK}} = - 4.991$ (3)  \\ 
  &  $8.3 \pm 2.1$  &  1  &  $\log{R^{\prime}_{HK}} = - 5.09$ (6,15)  \\ 
  &  $3.9 \pm 1.0$  &  1  &  $\log{R^{\prime}_{HK}} = - 4.85$ (9)  \\ 
  &  $7.0 \pm 1.8$  &  1  &  $\log{R^{\prime}_{HK}} = - 5.02$ (11)  \\ 
  &  $2.7 \pm 1.0$  &  1  &  $\log{R_{X}} \equiv \log{(L_{X}/L_{\rm bol})} = - 5.69$ (2)  \\ 
age range    &  $2.7-8.3$  &    &    \\ 
\hline 
\textbf{Adopted Age (Gyr)} &  $\mathbf{4.5 - 8.3}$  &    &     \\ 
\enddata 
\tablenotetext{a}{The gyrochronology and stellar-activity ages from this work are computed based on the rotation-age, $R^{\prime}_{\rm HK}$-age, and $R_{X}$-age relations from \cite{2008ApJ...687.1264M}. } 
\tablenotetext{b}{Notes contain (1) the method and spectral resolution used to compute the metallicity, and (2) the models, rotation periods, stellar-activity indices, and related references (indicated by the number in parenthesese) used to compute the object's ages. } 
\tablerefs{(1) This work, (2) \cite{1995AandA...294..515H}, (3) \cite{1996ApJ...457L..99B}, (4) \cite{1997AandAS..124..597M}, (5) \cite{2000AandA...361.1023S}, (6) \cite{2003AJ....126.2048G}, (7) \cite{2004AandA...420..183A}, (8) \cite{2004ApJS..155..667D}, (9) \cite{2004AandA...423..517R}, (10) \cite{2004AandA...415.1153S}, (11) \cite{2004ApJS..152..261W}, (12) \cite{2005ApJS..159..141V}, (13) \cite{2007ApJ...669.1167B}, (14) \cite{2007MNRAS.382.1516C}, (15) \cite{2007ApJ...660.1507L}, (16) \cite{2007ApJS..168..297T}, (17) \cite{2008ApandSS.316...31D}, (18) \cite{2008ApJ...687.1264M}, (19) \cite{2010ApJ...725.2349D}, (20) \cite{2011AandA...532A..20F}, (21) \cite{2011ApJ...735...41P}, (22) \cite{2012MNRAS.427..127B}, (23) \cite{2012AandA...547A.106M}, (24) \cite{2013ApJ...764...78R}, (25) \cite{2016AandA...585A...5B}, (26) \cite{2016AandA...586A..94L}, (27) \cite{2018AandA...614A..55A}} 
\end{deluxetable}

%\global\pdfpageattr\expandafter{\the\pdfpageattr/Rotate 0} 
\begin{deluxetable}{l|c|c|l} 
\tablewidth{0pc} 
\tablecaption{Metallicity and Age of GJ 570A \label{tab:GJ570}} 
\tablehead{ \multicolumn{1}{l}{}  &  \multicolumn{1}{c}{Value\tablenotemark{a}}  &  \multicolumn{1}{c}{Reference}  &  \multicolumn{1}{l}{Notes\tablenotemark{b}}   } 
\startdata 
Metallicity (dex)  &  $0.04 \pm 0.09$  &  4  &  Spectroscopy ($R \approx 1.0 \times 10^{5}$)  \\ 
  &  $0.04$  &  5  &  Spectroscopy ($R \approx 0.7-1.0 \times 10^{5}$)  \\ 
  &  $0.07 \pm 0.10$  &  9  &  Spectroscopy ($R \approx 4.8 \times 10^{4}$)  \\ 
  &  $0.03 \pm 0.03$  &  19  &  Spectroscopy ($R \approx 4.8 \times 10^{4}$)  \\ 
  &  $-0.04 \pm 0.05$  &  21  &  Spectroscopy ($R \approx 4.2 \times 10^{4}$)  \\ 
  &  $-0.05 \pm 0.17$  &  23  &  Spectroscopy ($R \approx 6.5-8.3 \times 10^{4}$)  \\ 
  &  $0.05 \pm 0.04$  &  24  &  Spectroscopy ($R \approx 6.0 \times 10^{4}$)  \\ 
\hline 
\textbf{Adopted Metallicity (dex)} &  $\mathbf{-0.05 - 0.05}$  &    &   \\ 
\hline \hline 
Isochrone Age (Gyr)  &  $3.3^{+8.3}_{-3.1}$  &  10  &  Y$^{2}$ (Yonsei-Yale) isochrones (7)  \\ 
  &  $<0.6$  &  14  &  YERC (Yale Stellar Evolution Code) isochrones (16)  \\ 
  &  $1^{+3}_{-1}$  &  19  &  Y$^{2}$ (Yonsei-Yale) isochrones (7)  \\ 
  &  $6.56^{+4.93}_{-4.57}$  &  20  &  BASTI isochrones (8,11,18)  \\ 
  &  $6.68^{+4.84}_{-4.68}$  &  20  &  Padova isochrones (15,17)  \\ 
  &  $7.92^{+4.89}_{-3.97}$  &  24  &  Y$^{2}$ (Yonsei-Yale) isochrones (7)  \\ 
age range  &  $1-8$  &    &    \\ 
\hline 
Gyrochronology Age (Gyr)  &  $3.69 \pm 0.56$  &  12  &  Rotation period $44.6$~d (6)  \\ 
  &  $5.30 \pm 1.24$  &  1  &  Rotation period $44.6$~d (6)  \\ 
age range  &  $3.7-5.3$  &    &    \\ 
\hline 
Stellar-Activity Age (Gyr)  &  $2.5 \pm 0.6$  &  1  &  $\log{R^{\prime}_{HK}} = - 4.75$ (2)  \\ 
  &  $0.6 \pm 0.1$  &  1  &  $\log{R^{\prime}_{HK}} = - 4.49$ (3)  \\ 
  &  $1.4 \pm 0.3$  &  1  &  $\log{R^{\prime}_{HK}} = - 4.63$ (22)  \\ 
  &  $1.7 \pm 0.7$  &  1  &  $\log{R_{X}} \equiv \log{(L_{X}/L_{\rm bol})} = - 5.38$ (6)  \\ 
  &  $1.4 \pm 0.6$  &  1  &  $\log{R_{X}} = - 5.27$ (13)  \\ 
age range    &  $0.6-2.5$  &    &    \\ 
\hline 
\textbf{Adopted Age (Gyr)} &  $\mathbf{1.4-5.2}$  &    &     \\ 
\enddata 
\tablenotetext{a}{The gyrochronology and stellar-activity ages from this work are computed based on the rotation-age, $R^{\prime}_{\rm HK}$-age, and $R_{X}$-age relations from \cite{2008ApJ...687.1264M}. } 
\tablenotetext{b}{Notes contain (1) the method and spectral resolution used to compute the metallicity, and (2) the models, rotation periods, stellar-activity indices, and related references (indicated by the number in parenthesese) used to compute the object's ages. } 
\tablerefs{(1) This work, (2) \cite{1991ApJ...375..722S}, (3) \cite{1996AJ....111..439H}, (4) \cite{1998AandAS..129..237F}, (5) \cite{2000AandA...363..692T}, (6) \cite{2003AandA...397..147P}, (7) \cite{2004ApJS..155..667D}, (8) \cite{2004ApJ...612..168P}, (9) \cite{2005AandA...437.1127S}, (10) \cite{2005ApJS..159..141V}, (11) \cite{2006ApJ...642..797P}, (12) \cite{2007ApJ...669.1167B}, (13) \cite{2007ApJ...660.1507L}, (14) \cite{2007ApJS..168..297T}, (15) \cite{2008AandA...484..815B}, (16) \cite{2008ApandSS.316...31D}, (17) \cite{2009AandA...508..355B}, (18) \cite{2009ApJ...697..275P}, (19) \cite{2010ApJ...720.1290G}, (20) \cite{2011AandA...530A.138C}, (21) \cite{2011AandA...531A.165P}, (22) \cite{2012AJ....143..135V}, (23) \cite{2015ApJ...807..183L}, (24) \cite{2018AandA...614A..55A}} 
\end{deluxetable}

%\global\pdfpageattr\expandafter{\the\pdfpageattr/Rotate 0} 
\begin{deluxetable}{lccc} 
\tablewidth{0pc} 
\setlength{\tabcolsep}{0.3in} 
\tablecaption{Astrometry and Photometry of HD~3651B, GJ~570D, and Ross~458C \label{tab:astrom_phot}} 
\tablehead{ \multicolumn{1}{l}{}  &  \multicolumn{1}{c}{HD 3651B}  &  \multicolumn{1}{c}{GJ 570D}  &  \multicolumn{1}{c}{Ross 458C}  } 
\startdata 
 Spectral Type  &  T7.5   &   T7.5  &   T8     \\ 
 R.A. (hh:mm:ss.ss; J2000)  &  00:39:18.91   &   14:57:15.84  &   13:00:41.64     \\ 
 Decl. (hh:mm:ss.ss; J2000)  &  +21:15:16.8   &   -21:22:08.1  &   +12:21:14.6     \\ 
 $\mu_{\alpha}\ \cos{\delta}$ (mas yr$^{-1}$)  &  $-462.06 \pm 0.11$   &   $1031.40 \pm 0.17$  &   $-632.15 \pm 0.50$     \\ 
 $\mu_{\delta}$ (mas yr$^{-1}$)  &  $-369.81 \pm 0.06$   &   $-1723.65 \pm 0.13$  &   $-36.02 \pm 0.19$     \\ 
 Parallax (mas)  &  $89.79 \pm 0.06$   &   $170.01 \pm 0.09$  &   $86.86 \pm 0.15$     \\ 
 Astrometry References &  5,7   &   5,7  &   5,7     \\ 
\hline 
 $Y_{\rm MKO}$ (mag)  &  $17.12 \pm 0.06$   &   $15.78 \pm 0.10$  &   $17.72 \pm 0.03$     \\ 
 $J_{\rm MKO}$ (mag)  &  $16.16 \pm 0.03$   &   $14.82 \pm 0.05$  &   $16.69 \pm 0.02$     \\ 
 $H_{\rm MKO}$ (mag)  &  $16.68 \pm 0.04$   &   $15.28 \pm 0.05$  &   $17.01 \pm 0.04$     \\ 
 $K_{\rm MKO}$ (mag)  &  $16.87 \pm 0.05$   &   $15.52 \pm 0.05$  &   $16.90 \pm 0.06$     \\ 
 MKO References &  1,9   &   2  &   3     \\ 
\hline 
 $W1$ (mag)  &  --   &   $14.93 \pm 0.04$  &   $16.04 \pm 0.06$     \\ 
 $W2$ (mag)  &  --   &   $12.13 \pm 0.02$  &   $13.85 \pm 0.04$     \\ 
 AllWISE Reference &  ...   &   4  &   4     \\ 
\hline 
 $[3.6]$ (mag)  &  $15.38 \pm 0.04$   &   $13.88 \pm 0.02$  &   $15.28 \pm 0.01$     \\ 
 $[4.5]$ (mag)  &  $13.62 \pm 0.02$   &   $12.15 \pm 0.02$  &   $13.77 \pm 0.01$     \\ 
 {\it Spitzer}/IRAC Reference &  2   &   8  &   6     \\ 
\enddata 
\tablerefs{(1) \cite{2007ApJ...654..570L}, (2) \cite{2010ApJ...710.1627L}, (3) \cite{2012yCat.2314....0L}, (4) \cite{2014yCat.2328....0C}, (5) \cite{2016AandA...595A...1G}, (6) \cite{2017ApJ...842..118L}, (7) \cite{2018AandA...616A...1G}, (8) \cite{2019ApJS..240...19K}, (9) \cite{2021AJ....161...42B}} 
\end{deluxetable}

%\global\pdfpageattr\expandafter{\the\pdfpageattr/Rotate 90} 
\begin{longrotatetable} 
\begin{deluxetable}{llcllllllcllclcccl} 
\tablewidth{0pc} 
\setlength{\tabcolsep}{0.01in} 
\tablecaption{Spectroscopically Inferred Physical Properties of HD~3651B, GJ~570D, and Ross~458C \label{tab:results}} 
\tablehead{ \multicolumn{1}{l}{}  &  \multicolumn{1}{l}{}  &  \multicolumn{1}{c}{}  &  \multicolumn{6}{c}{Fitted Parameters\tablenotemark{a}}  &  \multicolumn{1}{c}{}  &  \multicolumn{2}{c}{Derived Parameters\tablenotemark{b}}  &  \multicolumn{1}{c}{}  &  \multicolumn{3}{c}{Models\tablenotemark{c}}  &  \multicolumn{1}{c}{}  &  \multicolumn{1}{l}{}   \\ 
\cline{4-9} \cline{11-12} \cline{14-16} 
\multicolumn{1}{l}{Object}  &  \multicolumn{1}{l}{$\lambda\lambda$}  &  \multicolumn{1}{c}{}  &  \multicolumn{1}{l}{$T_{\rm eff}$}  &  \multicolumn{1}{l}{$\log{g}$}  &  \multicolumn{1}{l}{$Z$\tablenotemark{d}}  &  \multicolumn{1}{l}{$v_{r}$}  &  \multicolumn{1}{l}{$v\sin{i}$}  &  \multicolumn{1}{l}{$\log{\Omega}$}  &  \multicolumn{1}{c}{}  &  \multicolumn{1}{l}{$R$}  &  \multicolumn{1}{l}{$M$}  &  \multicolumn{1}{c}{}  &  \multicolumn{1}{l}{Grids}  &  \multicolumn{1}{c}{Cloudless?}  &  \multicolumn{1}{c}{Chem.Eq.?}  &  \multicolumn{1}{c}{}  &  \multicolumn{1}{l}{References}  \\ 
\multicolumn{1}{l}{}  &  \multicolumn{1}{l}{($\mu$m)}  &  \multicolumn{1}{c}{}  &  \multicolumn{1}{l}{(K)}  &  \multicolumn{1}{l}{(dex)}  &  \multicolumn{1}{l}{(dex)}  &  \multicolumn{1}{l}{(km s$^{-1}$)}  &  \multicolumn{1}{l}{(km s$^{-1}$)}  &  \multicolumn{1}{l}{(dex)}  &  \multicolumn{1}{c}{}  &  \multicolumn{1}{l}{($R_{\rm Jup}$)}  &  \multicolumn{1}{l}{($M_{\rm Jup}$)}  &  \multicolumn{1}{c}{}  &  \multicolumn{1}{c}{}  &  \multicolumn{1}{c}{}  &  \multicolumn{1}{c}{}  &  \multicolumn{1}{c}{}  &  \multicolumn{1}{l}{}   } 
\startdata 
 HD 3651B  &  $1.0-2.5$  &  &  $818^{+28\ (+19)}_{-28\ (-19)}$  &  $3.94^{+0.29\ (+0.20)}_{-0.28\ (-0.20)}$  &  $-0.22^{+0.16\ (+0.11)}_{-0.16\ (-0.10)}$  &  $318^{+202\ (+95)}_{-202\ (+96)}$  &  $27^{+27\ (+20)}_{-27\ (+18)}$  &  $-19.543^{+0.072\ (+0.051)}_{-0.070\ (+0.046)}$  &  &  $0.81^{+0.07}_{-0.06}$  &  $2.3^{+2.3}_{-1.1}$  &  &  26,28  &  Yes  &  Yes  &  &  This Work (Starfish)   \\ 
   &  $1.0-2.5$  &  &  $824^{+2}_{-3}$  &  $4.001^{+0.006}_{-0.116}$  &  $-0.141^{+0.017}_{-0.016}$  &  $252^{+188\ (+23)}_{-197\ (+29)}$  &  $28^{+17}_{-18}$  &  $-19.572^{+0.021\ (+0.006)}_{-0.018\ (+0.005)}$  &  &  $0.787^{+0.019}_{-0.016}$  &  $2.47^{+0.16}_{-0.54}$  &  &  26,28  &  Yes  &  Yes  &  &  This Work (Traditional)   \\ 
     &  $1.0-2.1$  &  &  $790 \pm 30$  &  $5.0 \pm 0.3$  &  $0.12 \pm 0.04$$^{\star}$  &  $\cdots$  &  $\cdots$  &  $\cdots$  &  &  $\cdots$  &  $\cdots$  &  &  $\cdots$\tablenotemark{e}  &    &    &  &  10   \\ 
     &  $0.7-2.5$  &  &  $820-830$  &  $5.4-5.5$  &  $\approx 0.2$  &  $\cdots$  &  $\cdots$  &  $\cdots$  &  &  $\cdots$  &  $\cdots$  &  &  1,3,6  &  Yes  &  Yes  &  &  11   \\ 
     &  $1.15-2.25$  &  &  $\approx 800$  &  $4.5-4.8$  &  $0$$^{\star}$  &  $\cdots$  &  $\cdots$  &  $\cdots$  &  &  $0.996-1.06$  &  $\cdots$  &  &  2,5  &  Yes  &  Yes  &  &  20   \\ 
     &  $1.15-2.25$  &  &  $850$  &  $4.5$  &  $0$$^{\star}$  &  $\cdots$  &  $\cdots$  &  $\cdots$  &  &  $1.07$  &  $\cdots$  &  &  16  &    &  Yes  &  &  20   \\ 
     &  $0.85-2.4$  &  &  $783 ^{+13}_{-12}$  &  $4.64 \pm 0.04$  &  $0.25 \pm 0.04$  &  $\cdots$  &  $\cdots$  &  $\cdots$  &  &  $0.89 \pm 0.03$  &  $\cdots$  &  &  23,27  &  Yes  &  Yes  &  &  27   \\ 
     &  $1.0-2.5$  &  &  $719 ^{+19}_{-25}$  &  $5.12^{+0.1}_{-0.2}$  &  $0.08^{+0.05}_{-0.06}$  &  $\cdots$  &  $\cdots$  &  $\cdots$  &  &  $1.10^{+0.1}_{-0.07}$  &  $\cdots$  &  &  $\cdots$\tablenotemark{f}  &    &    &  &  25   \\ 
     &    &  &    &    &    &    &    &    &  &    &   &  &    &    &   &  &     \\ 
 GJ 570D  &  $1.0-2.5$  &  &  $828^{+25\ (+16)}_{-26\ (-16)}$  &  $3.90^{+0.25\ (+0.16)}_{-0.25\ (-0.15)}$  &  $-0.33^{+0.14\ (+0.08)}_{-0.14\ (-0.08)}$  &  $431^{+213\ (+114)}_{-212\ (+115)}$  &  $26^{+25\ (+18)}_{-26\ (+17)}$  &  $-19.012^{+0.067\ (+0.041)}_{-0.066\ (+0.037)}$  &  &  $0.79^{+0.06}_{-0.06}$  &  $2.0^{+1.7}_{-0.9}$  &  &  26,28  &  Yes  &  Yes  &  &  This Work (Starfish)   \\ 
   &  $1.0-2.5$  &  &  $825^{+1}_{-1}$  &  $4.153^{+0.016}_{-0.014}$  &  $-0.269^{+0.009}_{-0.008}$  &  $118^{+180\ (+21)}_{-180\ (+22)}$  &  $35^{+23}_{-24}$  &  $-19.010^{+0.020\ (+0.004)}_{-0.020\ (+0.004)}$  &  &  $0.794^{+0.019}_{-0.018}$  &  $3.62^{+0.21}_{-0.19}$  &  &  26,28  &  Yes  &  Yes  &  &  This Work (Traditional)   \\ 
     &  $1.0-2.1$  &  &  $780-820$  &  $5.1$  &  $0$$^{\star}$  &  $\cdots$  &  $\cdots$  &  $\cdots$  &  &  $\cdots$  &  $\cdots$  &  &  $\cdots$\tablenotemark{e}  &    &    &  &  8   \\ 
     &  $0.7-14.5$  &  &  $800-820$  &  $5.09-5.23$  &  $0$$^{\star}$  &  $\cdots$  &  $\cdots$  &  $\cdots$  &  &  $\cdots$  &  $\cdots$  &  &  9  &  Yes  &    &  &  9,14   \\ 
     &  $1.147-1.347$  &  &  $948 \pm 53$  &  $4.5 \pm 0.5$  &  $0$$^{\star}$  &  $\cdots$  &  $32 \pm 8$  &  $\cdots$  &  &  $\cdots$  &  $\cdots$  &  &  2  &  Yes  &  Yes  &  &  13   \\ 
     &  $0.8-2.4$  &  &  $900$  &  $5.0$  &  $0$$^{\star}$  &  $\cdots$  &  $\cdots$  &  $\cdots$  &  &  $\cdots$  &  $\cdots$  &  &  2  &  Yes  &  Yes  &  &  15   \\ 
     &  $1.15-2.25$  &  &  $\approx 800$  &  $4.5-4.8$  &  $0$$^{\star}$  &  $\cdots$  &  $\cdots$  &  $\cdots$  &  &  $0.996-1.06$  &  $\cdots$  &  &  2,5  &  Yes  &  Yes  &  &  20   \\ 
     &  $1.15-2.25$  &  &  $900$  &  $5.0$  &  $0$$^{\star}$  &  $\cdots$  &  $\cdots$  &  $\cdots$  &  &  $0.903$  &  $\cdots$  &  &  16  &    &  Yes  &  &  20   \\ 
     &  $1.0-5.0$  &  &  $700$  &  $4.5$  &  $0$$^{\star}$  &  $\cdots$  &  $\cdots$  &  $\cdots$  &  &  $\cdots$  &  $\cdots$  &  &  4,7  &    &  Yes  &  &  22   \\ 
     &  $0.85-2.4$  &  &  $769 ^{+14}_{-13}$  &  $4.67 \pm 0.04$  &  $0.11 \pm 0.04$  &  $\cdots$  &  $\cdots$  &  $\cdots$  &  &  $0.94 \pm 0.04$  &  $\cdots$  &  &  23,27  &  Yes  &  Yes  &  &  27   \\ 
     &  $1.0-2.5$  &  &  $719 ^{+20}_{-22}$  &  $4.8^{+0.3}_{-0.3}$  &  $-0.15^{+0.07}_{-0.09}$  &  $\cdots$  &  $\cdots$  &  $\cdots$  &  &  $1.14^{+0.1}_{-0.09}$  &  $\cdots$  &  &  $\cdots$\tablenotemark{f}  &    &    &  &  25   \\ 
     &    &  &    &    &    &    &    &    &  &    &   &  &    &    &   &  &     \\ 
 Ross 458C  &  $1.0-2.5$  &  &  $804^{+30\ (+22)}_{-29\ (-21)}$  &  $4.09^{+0.31\ (+0.23)}_{-0.33\ (-0.28)}$  &  $0.23^{+0.20\ (+0.17)}_{-0.23\ (-0.20)}$  &  $276^{+217\ (+125)}_{-217\ (+119)}$  &  $28^{+30\ (+23)}_{-28\ (+19)}$  &  $-19.727^{+0.075\ (+0.054)}_{-0.076\ (+0.057)}$  &  &  $0.68^{+0.06}_{-0.06}$  &  $2.3^{+2.3}_{-1.2}$  &  &  26,28  &  Yes  &  Yes  &  &  This Work (Starfish)   \\ 
   &  $1.0-2.5$  &  &  $836^{+3}_{-3}$  &  $4.085^{+0.017}_{-0.014}$  &  $0.499^{+0.001}_{-0.002}$  &  $168^{+182\ (+24)}_{-183\ (+29)}$  &  $30^{+17}_{-21}$  &  $-19.805^{+0.021\ (+0.006)}_{-0.018\ (+0.006)}$  &  &  $0.622^{+0.016}_{-0.013}$  &  $1.90^{+0.12}_{-0.10}$  &  &  26,28  &  Yes  &  Yes  &  &  This Work (Traditional)   \\ 
     &  $0.85-2.35$  &  &  $760^{+70}_{-45}$  &  $4.2^{+0.3}_{-0.2}$  &  $\approx 0.3$  &  $\cdots$  &  $\cdots$  &  $\cdots$  &  &  $\cdots$  &  $\cdots$  &  &  12  &  Yes  &  Yes  &  &  17   \\ 
     &  $0.85-2.35$  &  &  $635^{+25}_{-35}$  &  $\approx 4.0$  &  $0$$^{\star}$  &  $\cdots$  &  $\cdots$  &  $\cdots$  &  &  $\cdots$  &  $\cdots$  &  &  12  &    &  Yes  &  &  17   \\ 
     &  $1.0-5.0$  &  &  $650-750$  &  $4.0$  &  $0$$^{\star}$  &  $\cdots$  &  $\cdots$  &  $\cdots$  &  &  $\cdots$  &  $\cdots$  &  &  12  &  Yes  &  Yes  &  &  19   \\ 
     &  $1.0-5.0$  &  &  $650-700$  &  $4.0$  &  $0.3$$^{\star}$  &  $\cdots$  &  $\cdots$  &  $\cdots$  &  &  $\cdots$  &  $\cdots$  &  &  12  &  Yes  &  Yes  &  &  19   \\ 
     &  $1.0-2.5$  &  &  $600$  &  $4.0$  &  $0$$^{\star}$  &  $\cdots$  &  $\cdots$  &  $\cdots$  &  &  $\cdots$  &  $\cdots$  &  &  12  &    &  Yes  &  &  19   \\ 
     &  $1.0-5.0$  &  &  $700-750$  &  $4.5-5.0$  &  $0$$^{\star}$  &  $\cdots$  &  $\cdots$  &  $\cdots$  &  &  $\cdots$  &  $\cdots$  &  &  18  &    &  Yes  &  &  19   \\ 
     &  $1.0-5.0$  &  &  $700-750$  &  $5.0$  &  $0.3$$^{\star}$  &  $\cdots$  &  $\cdots$  &  $\cdots$  &  &  $\cdots$  &  $\cdots$  &  &  18  &    &  Yes  &  &  19   \\ 
     &  $1.15-2.25$  &  &  $860-900$  &  $4.0-4.24$  &  $0$$^{\star}$  &  $\cdots$  &  $\cdots$  &  $\cdots$  &  &  $1.24-1.27$  &  $\cdots$  &  &  2,5  &  Yes  &  Yes  &  &  20   \\ 
     &  $1.15-2.25$  &  &  $850$  &  $4.5$  &  $0$$^{\star}$  &  $\cdots$  &  $\cdots$  &  $\cdots$  &  &  $1.07$  &  $\cdots$  &  &  16  &    &  Yes  &  &  20   \\ 
     &  $0.85-2.35$  &  &  $700$  &  $4.0$  &  $0$$^{\star}$  &  $\cdots$  &  $\cdots$  &  $\cdots$  &  &  $\cdots$  &  $\cdots$  &  &  21  &    &  Yes  &  &  21   \\ 
     &  $0.85-2.35$  &  &  $875$  &  $5.0$  &  $0.3$$^{\star}$  &  $\cdots$  &  $\cdots$  &  $\cdots$  &  &  $\cdots$  &  $\cdots$  &  &  24  &  Yes  &    &  &  24   \\ 
\enddata 
\tablenotetext{a}{Parameters derived from this work are shown as median and $1\sigma$ uncertainties, with systematic errors (Section~\ref{subsubsec:systematics}) already incorporated. Values inside parentheses are formal spectral-fitting uncertainties from Starfish before applying any systeamtic errors.}  
\tablenotetext{b}{Radii ($R$) is derived from the objects' parallaxes and fitted $\log{\Omega}$ and mass ($M$) is derved from $R$ and the fitted $\log{g}$. Age ($t$) is derived by interpolating the evolutionary models using the fitted $\{T_{\rm eff},\ \log{g},\ Z\}$ values.}  
\tablenotetext{c}{Grid models adopted by previous forward-modeling analyses and whether these models are generated with cloudless and/or chemical equilibrium assumptions (only shown if "Yes").}  
\tablenotetext{d}{The "$\star$" symbols mark metallicities that are assumed instead of fitted during the spectral fitting process.}  
\tablenotetext{e}{Parameters are derived from the empirical analysis of spectral indices by \cite{2006ApJ...639.1095B}.}  
\tablenotetext{f}{Parameters are derived from retrieval analysis.}  
\tablerefs{(1) \cite{2001ApJ...556..872A}, (2) \cite{2001ApJ...556..357A}, (3) \cite{2002ApJ...568..335M}, (4) \cite{2002ApJ...575..264T}, (5) \cite{2003ApJ...596..587B}, (6) \cite{2003IAUS..211..345S}, (7) \cite{2005ApJ...621.1033T}, (8) \cite{2006ApJ...639.1095B}, (9) \cite{2006ApJ...647..552S}, (10) \cite{2007ApJ...658..617B}, (11) \cite{2007ApJ...667..537L}, (12) \cite{2008ApJ...689.1327S}, (13) \cite{2009AandA...501.1059D}, (14) \cite{2009ApJ...695..844G}, (15) \cite{2009AandA...503..639T}, (16) \cite{2010HiA....15..756A}, (17) \cite{2010ApJ...725.1405B}, (18) \cite{2011ASPC..448...91A}, (19) \cite{2011MNRAS.414.3590B}, (20) \cite{2011ApJ...740..108L}, (21) \cite{2012ApJ...756..172M}, (22) \cite{2012ApJ...760..151S}, (23) \cite{2015ApJ...813...47M}, (24) \cite{2015ApJ...804L..17T}, (25) \cite{2017ApJ...848...83L}, (26) \cite{2017AAS...23031507M}, (27) \cite{2017AandA...603A..57S}, (28) Marley et al. (Submitted)} 
\end{deluxetable} 
\end{longrotatetable}

\begin{longrotatetable} 
\begin{deluxetable}{lcccc} 
\tablewidth{0pc} 
\setlength{\tabcolsep}{0.3in} 
\tablecaption{Spectroscopically Inferred Covariance Hyper-Parameters and $\epsilon_{J}$ of HD~3651B, GJ~570D, and Ross~458C \label{tab:hyper_params}} 
\tablehead{ \multicolumn{1}{l}{Object}  &  \multicolumn{1}{c}{$a_{N}$}  &  \multicolumn{1}{c}{$\ell$}  &  \multicolumn{1}{c}{$\log{a_{G}}$}  &  \multicolumn{1}{c}{$\epsilon_{J}$}   \\ 
\multicolumn{1}{l}{}  &  \multicolumn{1}{c}{}  &  \multicolumn{1}{c}{(km s$^{-1}$)}  &  \multicolumn{1}{c}{(dex)}  &  \multicolumn{1}{c}{}} 
\startdata 
 HD 3651B  &  $1.06^{+0.04}_{-0.04}$  &  $8114^{+751}_{-1015}$  &  $-34.64^{+0.11}_{-0.11}$  &  $0.026^{+0.004}_{-0.003}$ \\  
  GJ 570D  &  $1.04^{+0.05}_{-0.05}$  &  $2714^{+437}_{-372}$  &  $-33.79^{+0.08}_{-0.08}$  &  $0.020^{+0.002}_{-0.002}$ \\  
  Ross 458C  &  $0.93^{+0.04}_{-0.04}$  &  $8541^{+471}_{-747}$  &  $-34.67^{+0.09}_{-0.09}$  &  $0.042^{+0.005}_{-0.004}$ \\  
 \enddata 
\end{deluxetable} 
\end{longrotatetable}

\begin{longrotatetable} 
\begin{deluxetable}{lcccccccccccc} 
\tablewidth{0pc} 
\tablecaption{Atmospheric and Evolutionary Model Parameters of Benchmark Companions  \label{tab:comp_evo}} 
\tablehead{ \multicolumn{1}{l}{}  &  \multicolumn{1}{c}{}  &  \multicolumn{3}{c}{HD 3651B}  &  \multicolumn{1}{c}{}  &  \multicolumn{3}{c}{GJ 570D}   &   \multicolumn{1}{c}{}  &  \multicolumn{3}{c}{Ross 458C}   \\ 
\cline{3-5} \cline{7-9} \cline{11-13} 
\multicolumn{1}{l}{Parameter}  &  \multicolumn{1}{c}{}  &  \multicolumn{1}{c}{Atmospheric}  &  \multicolumn{1}{c}{Evolutionary}  &  \multicolumn{1}{c}{Atm. $-$ Evo.}  &  \multicolumn{1}{c}{}  &  \multicolumn{1}{c}{Atmospheric}  &  \multicolumn{1}{c}{Evolutionary}  &  \multicolumn{1}{c}{Atm. $-$ Evo.}  &  \multicolumn{1}{c}{}  &  \multicolumn{1}{c}{Atmospheric}  &  \multicolumn{1}{c}{Evolutionary}  &  \multicolumn{1}{c}{Atm. $-$ Evo.}   } 
\startdata 
Primary Star Age ($t_{\rm \star}$; Gyr)  &  &  \multicolumn{3}{c}{$4.5-8.3$}  &  &  \multicolumn{3}{c}{$1.4-5.2$}  &  &  \multicolumn{3}{c}{$0.15-0.8$}  \\ 
Primary Star Metallicity ($Z_{\rm \star}$; dex)  &  &  \multicolumn{3}{c}{$0.1-0.2$}  &  &  \multicolumn{3}{c}{$-0.05 - 0.05$}  &  &  \multicolumn{3}{c}{$0.16-0.32$}  \\ 
Effective Temperature ($T_{\rm eff}$; K)  &  &  $818^{+28}_{-28}$  &  $809^{+15}_{-16}$  &  $9^{+32}_{-32}$  &  &  $828^{+25}_{-26}$  &  $786^{+20}_{-20}$  &  $42^{+32}_{-32}$  &  &  $804^{+30}_{-29}$  &  $682^{+16}_{-17}$  &  $122^{+34}_{-33}$  \\ 
Surface Gravity ($\log{g}$; cgs)  &  &  $3.94^{+0.29}_{-0.28}$  &  $5.26^{+0.06}_{-0.06}$  &  $-1.31^{+0.29}_{-0.29}$  &  &  $3.90^{+0.25}_{-0.25}$  &  $5.04^{+0.13}_{-0.13}$  &  $-1.13^{+0.28}_{-0.28}$  &  &  $4.09^{+0.31}_{-0.33}$  &  $4.38^{+0.16}_{-0.17}$  &  $-0.29^{+0.34}_{-0.36}$  \\ 
Metallicity ($Z$; dex)  &  &  $-0.22^{+0.16}_{-0.16}$  &  $0.15^{+0.03}_{-0.03}$  &  $-0.37^{+0.17}_{-0.16}$  &  &  $-0.33^{+0.14}_{-0.14}$  &  $0.00^{+0.03}_{-0.03}$  &  $-0.33^{+0.14}_{-0.13}$  &  &  $0.23^{+0.20}_{-0.23}$  &  $0.25^{+0.06}_{-0.05}$  &  $-0.02^{+0.20}_{-0.23}$  \\ 
Radius ($R$; R$_{\rm Jup}$)  &  &  $0.81^{+0.07}_{-0.06}$  &  $0.81^{+0.02}_{-0.02}$  &  $0.00^{+0.07}_{-0.07}$  &  &  $0.79^{+0.06}_{-0.06}$  &  $0.89^{+0.05}_{-0.04}$  &  $-0.09^{+0.07}_{-0.07}$  &  &  $0.68^{+0.06}_{-0.06}$  &  $1.10^{+0.05}_{-0.05}$  &  $-0.42^{+0.08}_{-0.07}$  \\ 
Mass ($M$; M$_{\rm Jup}$)  &  &  $2.3^{+2.3}_{-1.1}$  &  $48.1^{+4.4}_{-4.0}$  &  $-45.2^{+4.2}_{-4.7}$  &  &  $2.0^{+1.7}_{-0.9}$  &  $34.6^{+7.5}_{-6.5}$  &  $-32.2^{+6.5}_{-7.6}$  &  &  $2.3^{+2.3}_{-1.2}$  &  $11.7^{+3.6}_{-3.0}$  &  $-9.0^{+3.3}_{-3.9}$  \\ 
Bolometric Luminosity\tablenotemark{a} ($\log{(L_{\rm bol}/L_{\odot})}$; dex)  &  &  $-5.56^{+0.04}_{-0.03}$  &  $-5.57^{+0.03}_{-0.03}$  &  $-0.01^{+0.04}_{-0.04}$  &  &  $-5.55^{+0.03}_{-0.03}$  &  $-5.54^{+0.03}_{-0.03}$  &  $0.01^{+0.04}_{-0.04}$  &  &  $-5.72^{+0.03}_{-0.03}$  &  $-5.60^{+0.03}_{-0.04}$  &  $0.11^{+0.05}_{-0.05}$  \\ 
\enddata 
\tablenotetext{a}{Atmospheric bolometric luminosities are computed by integrating the objects' $1.0-2.5$~$\mu$m SpeX data and fitted model spectra to shorter and longer wavelengths spanning $0.4-50$~$\mu$m, with these spectra scaled by the objects' measured parallaxes.} 
\end{deluxetable} 
\end{longrotatetable}

\vfill
\eject
\end{document}